%% file: MPFBench.tex
\documentclass{article} % For LaTeX2e
\usepackage{iclr2025_conference,times}

\input{math_commands.tex}

\usepackage{hyperref}       % hyperlinks
\usepackage{url}            % simple URL typesetting
\usepackage{booktabs}       % professional-quality tables
\usepackage{amsfonts}       % blackboard math symbols
\usepackage{nicefrac}       % compact symbols for 1/2, etc.
\usepackage{microtype}      % microtypography
\usepackage{xcolor}         % colors
\usepackage{algorithm}
\usepackage{algpseudocode}
\usepackage{natbib}
\usepackage[labelfont=bf,textfont=it,font=small]{caption}
\usepackage{pgfplots}
\pgfplotsset{compat=1.18}

\usepackage{amsmath}
\usepackage{subcaption}
\usepackage{graphicx}
\usepackage{array}
\usepackage{enumitem}
\usepackage{xspace}
\usepackage{amssymb}  % For checkmark symbol
\usepackage{amsfonts}  % For mathematical fonts
\usepackage{pifont}    % For more symbols
\usepackage{titlesec}
\usepackage{multirow}

\definecolor{darkgreen}{rgb}{0.0, 0.5, 0.0}

\titlespacing\section{0pt}{6pt plus 1pt minus 1pt}{3pt plus 1pt minus 1pt}
\titlespacing\subsection{0pt}{6pt plus 1pt minus 1pt}{3pt plus 1pt minus 1pt}
\titlespacing\subsubsection{0pt}{6pt plus 2pt minus 2pt}{3pt plus 2pt minus 2pt}

\hypersetup{colorlinks,
      linkcolor=blue,
      citecolor=blue,
      urlcolor=blue}

\newcommand{\cref}[2]{\hyperref[#2]{#1~\ref*{#2}}}
\newcommand{\colref}[2]{\hyperref[#2]{#1~\ref*{#2}}}
\newcommand{\eqnref}[1]{\colref{Equation}{#1}}
\newcommand{\figureref}[1]{\colref{Figure}{#1}}

\newcommand{\sectionref}[1]{\colref{Section}{#1}}
\newcommand{\tabref}[1]{\colref{Table}{#1}}
\newcommand{\coloredref}[2]{\hyperref[#2]{#1~\ref*{#2}}}
\newcommand{\coloredsubref}[3]{\hyperref[#2]{#1~\ref*{#2}{#3}}}
\newcolumntype{P}[1]{>{\centering\arraybackslash}p{#1}}

\newcommand{\MPFBench}{
  \href{https://baskargroup.github.io/mpf-bench/}{\textcolor{green!40!black}{MPF-Bench}}
}

\title{\MPFBench: A Large Scale Dataset for SciML of Multi-Phase-Flows: Droplet and Bubble Dynamics}

\author{
  \textbf{Mehdi Shadkhah}\textsuperscript{1},
  \textbf{Ronak Tali}\textsuperscript{1}\thanks{These authors contributed equally to this work.},
  \textbf{Ali Rabeh}\textsuperscript{1\fnsymbol{footnote}},
    \textbf{Cheng-Hau Yang}\textsuperscript{1\fnsymbol{footnote}}, 
    \textbf{Ethan Herron}\textsuperscript{1\fnsymbol{footnote}},\\
  \textbf{Abhisek Upadhyaya}\textsuperscript{2}, 
  \textbf{Adarsh Krishnamurthy}\textsuperscript{1},
  \textbf{Chinmay Hegde}\textsuperscript{2},
  \textbf{Aditya Balu}\textsuperscript{1}, \\
  \textbf{Baskar Ganapathysubramanian}\textsuperscript{1} \\
  \And
  \textsuperscript{1}Iowa State University\\
  \texttt{\{mehdish,rtali,arabeh,edherron,chenghau,}\\
  \texttt{adarsh,baditya,baskarg\}@iastate.edu}\\
  \And
  \textsuperscript{2}New York University\\
  \texttt{\{au2216,chinmay.h\}@nyu.edu} 
  }

\iclrfinalcopy % Uncomment for camera-ready version, but NOT for submission.
\begin{document}

\maketitle

\begin{abstract}
Multiphase fluid dynamics, such as falling droplets and rising bubbles, are critical to many industrial applications. However, simulating these phenomena efficiently is challenging due to the complexity of instabilities, wave patterns, and bubble breakup. This paper investigates the potential of scientific machine learning (SciML) to model these dynamics using neural operators and foundation models. We apply sequence-to-sequence techniques on a comprehensive dataset generated from 11,000 simulations, comprising 1 million time snapshots, produced with a well-validated Lattice Boltzmann method (LBM) framework. The results demonstrate the ability of machine learning models to capture transient dynamics and intricate fluid interactions, paving the way for more accurate and computationally efficient SciML-based solvers for multiphase applications.
\end{abstract}

% \section{Submission of conference papers to ICLR 2025}

% ICLR requires electronic submissions, processed by
% \url{https://openreview.net/}. See ICLR's website for more instructions.

% If your paper is ultimately accepted, the statement {\tt
%   {\textbackslash}iclrfinalcopy} should be inserted to adjust the
% format to the camera ready requirements.

% The format for the submissions is a variant of the NeurIPS format.
% Please read carefully the instructions below, and follow them
% faithfully.

\section{Introduction}
% \textcolor{red}{
% - We can talk about the importance of multiphase flow and its applications in two paragraphs with references. (Drug Delivery, Lab on Chip, Separation Processes, Nanofluidics, Coating Processes) \\
% }

Flow behavior in multiphase flow is crucial for many industrial and chemical applications. In drug delivery, two-phase flow can be used to create uniform drug-loaded microspheres or microcapsules. These microcapsules can provide controlled and sustained release of drugs, improving therapeutic outcomes~\citep{hernot2008microbubbles,sattari2020multiphase}. Two-phase flows are also essential for rapid diagnostics and biochemical applications in lab-on-a-chip technologies~\citep{haeberle2007microfluidic,mark2010microfluidic}. Discrete phase bubbles in microchannels, generated via T-junctions~(\cite{thorsen2001dynamic}), co-flowing systems~\citep{cramer2004drop}, or flow-focusing techniques~\citep{anna2003formation}, have a high surface-to-volume ratio, enhancing reaction efficiency and sensitivity. The shearing forces of the continuous phase precisely control bubble size and formation, which is crucial for device performance. By thoroughly understanding gas-liquid or liquid-liquid interactions, engineers can optimize mixing conditions~\citep{schwesinger1996modular,stroock2002chaotic,tice2003formation} to enhance reaction rates, improve product consistency, and reduce energy consumption. 
% In the oil industry, separators~\cite{sayda2007modeling,laleh2012design} separate different phases of fluids, such as oil, water, and gas. Understanding the dynamics of multi-phase flow helps optimize the design and operation of separators, ensuring efficient separation, improved purity, and increased yield. Additionally, understanding two-phase flow is crucial for coating processes~\cite{weinstein2004coating}. By precisely controlling how the liquid coating material interacts with the gaseous or liquid surface, we can achieve a even coating to improve product's durability.

% \textcolor{red}{
% - We can talk about phenomena involved in droplet and bubble dynamics +  their complexity (breakup, deformation, surface tension, and so on.) - mention references please \\
% }

Bubbles (\textit{lighter fluid volumes moving in a denser fluid medium}) and droplets (\textit{heavier fluid volumes moving in a lighter fluid medium}) play an integral role in applications such as drug delivery and lab-on-a-chip technologies. The dynamics of droplets and bubbles exhibit significant complexity, primarily due to phenomena such as breakup, deformation, and surface tension. Firstly, the breakup of droplets and bubbles is a highly nonlinear and complex process governed by factors such as viscosity ratio, density ratio, and surface tension. For example, for high inertia flows, the fast and irregular breakup results in smaller and widely-distributed droplets; at low Reynolds numbers, laminar flow leads to a more even breakup and larger droplets~\citep{eggers2008physics}. Secondly, droplets can be deformed by shear and pressure forces. Various studies have shown that the Capillary number~\citep{vananroye2008microconfined,liu2022enhanced}, Atwood number~\citep{fakhari2010investigation,singh2020role}, and Reynolds number~\citep{vontas2020droplet,xu2020droplet,seksinsky2021droplet} all have a significant impact on droplet deformation.

% \textcolor{red}{
% - bubble rise is a famous benchmark problem. Many people worked on it (Aland, Marand, ...) \\
% }

To better understand multiphase phenomena (both droplets and bubbles), researchers often perform a canonical simulation/experiment called the bubble rising case~\citep{Bhaga_Weber_1981, HUA2007769,Hysing2009, AMAYABOWER20101191, aland2012benchmark, Yuan2017, KHANWALE2023111874, rabeh2024modeling}, where a bubble is placed in a higher density fluid so that the bubble moves up due to buoyancy. Conversely, using a droplet of higher density causes the droplet to fall down due to gravity~\citep{YANG2021103561,JALAAL2012115}. This canonical study is essential since it provides insights into bubble dynamics and shape evolution, which are critical factors for optimizing industrial processes and improving numerical models in fluid dynamics research. Nonetheless, capturing the bubble-rising or droplet-falling phenomenon is a multiscale problem with forces acting at different scales, ranging from microscale molecular interactions to macroscale fluid dynamics. Therefore, high-fidelity simulations are essential to accurately resolve these interactions, particularly at the thin interfaces where precise capturing of surface tension and interfacial dynamics is critical.

Scientific Machine Learning (SciML) represents a powerful approach for addressing multiphase flow problems. SciML leverages the inherent physics to develop models that can learn from complex data and produce reliable predictions~\citep{karniadakis2021physicsinformed,hassan2023bubbleml,m2024pinn,rabeh2024geometry}. A key ingredient to training and accessing SciML solvers is a comprehensive dataset~\citep{tali2024flowbench}, which\MPFBench is an example of such a benchmark dataset. It includes wave patterns, bubble and droplet dynamics, and breakup. %It remains to be seen whether neural operators can be trained to predict bubble and droplet dynamics since most have not been tested on a dataset as complex as. Such phenomena are known to be chaotic and highly nonlinear by nature, therefore the extensive data provided by this benchmark is invaluable in guiding future research in SciM.
% \textcolor{red}{
% - We can talk about our dataset in the last paragraph, it has 10,000 simulations and more than 2 million time-series snapshots. We change density ratio, viscosity ratio, Reynolds number, and Bond number\\
% - You can explain the variety of phenomena that we capture here, breakup, deformation 
% - Get idea from the FlowBench paper to explain the dataset here.
% }

There are several approaches to using machine learning to solve scientific problems, including Physics-Informed Neural Networks (PINNs) \citep{raissi2019} and neural operators \citep{li2021, raonic2023, lu2020}. However, PINNs suffer from hard convergence and high generalization error \citep{rathore2024}. In this paper, we focus on using neural operators and foundation models which use supervised learning.\MPFBench has three major features:

\begin{itemize}[topsep=0pt, itemsep=0pt, left=0pt]
\item \textbf{Scientific machine learning evaluations}: We test our dataset on several neural operators and foundation models using the sequence-to-sequence time series concatenation technique. Our dataset serves as a good test for these models to evaluate their ability to learn multiscale physics data.
\item \textbf{Extensive amount of data}: Our dataset includes 11,000 simulations in 2D and 3D with over 1 million time-series snapshots. This extensive volume of data allows for robust training of SciML models, which will help in advancing the development of accurate and reliable SciML models for multiphase flow dynamics. 
\item \textbf{Multiphase simulations}: We conduct simulations of rising bubbles and falling droplets, solving the Navier-Stokes equations coupled with the Allen-Cahn equation. This approach captures considerable physical phenomena, including breakup and deformation.

%\item \textbf{Wide range of properties}: The dataset varies key parameters such as density ratio, viscosity ratio, Reynolds number, and Bond number. This diversity ensures the models trained on this data can potentially generalize across diverse flow conditions and interface dynamics. 

\end{itemize}

% \begin{itemize}[topsep=3pt, itemsep=0pt, left=0pt] 
% \item \textbf{Multiphysics simulations}:
% \textcolor{red}{
% - The rising bubble and falling droplet simulation and equation that we solve: here NS + Allen-Cahn
% }

% \item \textbf{Complex phenomena}:
% \textcolor{red}{
% - We capture complex phenomena: Surface Tension Effects - Buoyancy - Drag Force - Coalescence - Breakup - Capillary Waves - Ellipsoidal Deformation - Wake Formation - Diffusion - Convective Transport - Rayleigh-Taylor Instability - Oscillatory Motion - Wall Effects and ... 
% }
% \textcolor{red}{
% - Talk about dataset again why we chose these problems (rising bubble and falling droplet) \\
% - Why do we change the density ratio and dimensionless numbers?
% }

% \item \textbf{Wide range of properties (3 orders of magnitudes)}:
% \textcolor{red}{
% - Coalescence/breakup
% - Oscillatory motion
% - Capillary waves for droplet tails deforming into small bubbles
% -Diffusion due to viscosity ratio and Bond number 
% }

% \end{itemize}

\textbf{Our Contributions:} We summarize our main contributions below:
\begin{itemize}[left=0pt,topsep=0pt]
    \item Six neural operators and foundation models trained on our data i.e., predicting concentration, velocity, and pressure solution fields using previous time solutions as input to the models. To our knowledge, no study has evaluated the performance of neural operators and foundation models on multiphase flows. %The preliminary benchmark results are promising and gives us hope that the community will embrace this dataset and obtain superlative performance on this dataset using current and future SciML models.

    \item Our dataset features 11,000 simulations and over 1 million time-series snapshots, with variations in density ratio, viscosity ratio, Reynolds number, and Bond number. This extensive dataset encompasses many phenomena, ranging from subtle surface deformations in bubble oscillations to full bubble breakups driven by surface tension and density ratio variations. The richness and breadth of this dataset offer deep insights into the intricate dynamics of multiphase flows, making it a valuable resource for advancing research in this field. We provide our \href{https://figshare.com/s/bbb0c2463e8c8a24814a}{dataset} as a benchmark for others interested in developing and evaluating SciML models. Additional details can also be found in our \href{https://lobster-app-cbyg9.ondigitalocean.app/}{website}.

\end{itemize}

\section{Related Work} \label{related_work}

\begin{table}[ht]
\centering
\setlength\extrarowheight{2pt}
\caption{Comparison of public Multiphase Flow Datasets}
\begin{tabular}{>{\raggedright\arraybackslash}m{0.13\linewidth}%
>{\raggedleft\arraybackslash}m{0.1\linewidth}%
>{\raggedleft\arraybackslash}m{0.1\linewidth}%
>{\raggedright\arraybackslash}m{0.15\linewidth}%
>{\raggedright\arraybackslash}m{0.12\linewidth}%
>{\raggedright\arraybackslash}m{0.2\linewidth}}
\hline
\textbf{Name} & \textbf{Samples} & \textbf{Snapshots} & \textbf{Scope} & \textbf{Sources} & \textbf{Ranges of material properties} \\ \hline
% \textbf{Stanford Multiphase Flow Database (SMFD)} & 5659 & Gas-liquid flow in pipes of all inclinations & 15 different sources, 70\% lab, 30\% field & No mention. \\ \hline
\textbf{Flow Experiment Dataset} & 2904 & 2904 & Horizontal pipes, effects of density, viscosity, surface tension & Controlled lab environment & $\rho$: [1, 1.5] gm/cc,\newline  $\mu$: [1, 3.1] cP, \newline  \text{Surface tension} = [32, 70] mN/m \\ \hline
\textbf{BubbleML} & 79 & 7641 & pool boiling, flow boiling, and sub-cooled boiling & 2D and 3D Numerical simulations based on Flash-X & Re = 0.0042,\newline  $\rho^*$ = 0.0083,\newline  $\mu^*$ = 1, \newline Pr = 8.4, \newline  We = 1, \newline Fr = [1, 100] \\ \hline
\textbf{\MPFBench} & 11000 & $>$ 1 million & Droplet and bubble dynamics & 2D and 3D Simulations using LBM & $\rho^* : [10, 1000]$, \newline  $\mu^* : [1, 100]$,  \newline Bo : $[10, 500]$, \newline  Re : $[10, 1000]$ \\ \hline
\label{tab:CompareTable}
\end{tabular}
\end{table}

The Stanford Multiphase Flow Database (SMFD) used in~\citep{chaari2018optimized}, the flow experiment dataset~\citep{al2021effects}, and the BubbleML dataset~\citep{hassan2023bubbleml} are resources for understanding multiphase flow dynamics. 

The SMFD features 5659 measurements across a range of gas and liquid properties, pipe characteristics, and operational conditions. This dataset, derived from laboratory and field sources, supports various pipe inclinations and flow patterns. SMFD covers different flow regimes, including stratified, slug, and annular flows. However, it does not appear publicly available, so we cannot identify the number of individual snapshots in this dataset.

The flow experiment dataset~\citep{al2021effects} focuses on the effects of density, viscosity, and surface tension on two-phase flow regimes and pressure drops in horizontal pipes. The 2904 measurements from air-liquid system experiments provide insights into fluid properties' influence on flow regimes and pressure drops. This dataset's development of flow regimes and pressure contour maps enhances the understanding of fluid behavior in horizontal two-phase flows.

Additionally, the BubbleML dataset~\citep{hassan2023bubbleml} is a data collection focused on multiphysics phase change phenomena generated through physics-driven simulations, providing ground truth information for various boiling scenarios, including nucleate pool boiling, flow boiling, and sub-cooled boiling. We summarize these and other databases alongside our dataset in \tabref{tab:CompareTable}.

% \begin{itemize}
%     \item \textcolor{red}{ We need to mention that these other benchmark problems describe different problems. Since multiphase flow is highly nonlinear and dependent on ICs/BCs.}

%     \item \textcolor{red}{ Range of material properties that other benchmarks are discussing}

%     \item \textcolor{red}{ Inverse the table just like in flowbench, remove boxes, include a range of properties (density/viscosity ratio, surface tension), include the number of data points/simulations available}

% \end{itemize}

\section{Multi-phase flow (MPF) Bench}
\label{gen_inst}

% \begin{itemize}
%     \item \textcolor{red}{ Add introductory paragraph just like in flowbench}
% \end{itemize}

We present the\MPFBench dataset, encompassing 5500 bubble rise and 5500 droplet flow simulations, with each simulation containing 100 time-snapshots, making it, to our knowledge, two orders of magnitude larger -- in terms of number of time-snapshots -- than any existing multiphase flow dataset. This dataset features 2D and 3D transient simulations, capturing a spectrum of flow behaviors influenced by surface tension and density/viscosity ratios.\MPFBench includes scenarios from bubble oscillations with minor surface deformations to complete bubble breakup, offering a comprehensive resource for studying bubble rise and droplet fall dynamics.

\subsection{Problem Definition: Initial and Boundary Conditions, and outputs}\label{subsec:problem}
% \begin{itemize}
%     \item \textcolor{red}{u,v,p,$\rho$,c}
%     \item \textcolor{red}{free-slip, periodic on top/bottom}
%     \item \textcolor{red}{Nice figure with labels of IC and BC}

% \end{itemize}

\begin{figure}[ht]
    \centering
    \includegraphics[width=0.95\textwidth]{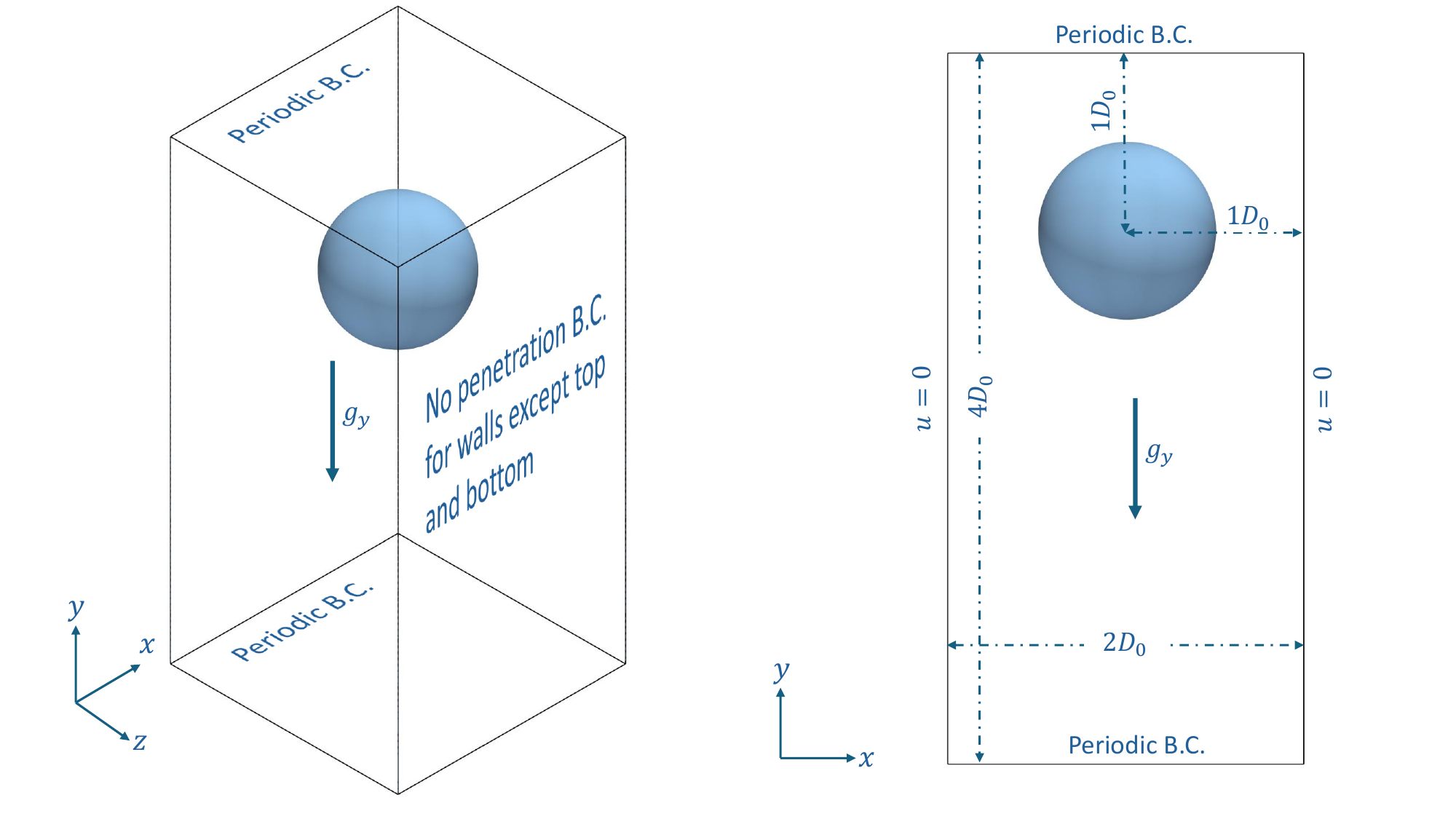}
    \caption{Boundary conditions for the simulation of a falling droplet. The left panel illustrates the 3D case, while the right panel illustrates the 2D case.}
    \label{fig:falling_droplet_bc}
\end{figure}

We consider 2D and 3D simulations of bubble rise and droplet fall simulations using the lattice Boltzmann method. The domain sizes for 2D and 3D are \([256, 512]\) and \([128, 256, 128]\) lattice units, respectively. For 2D simulations, the bubble is initially centered at \((64, 64)\) and the droplet is centered at \((128, 384)\). In 3D, the bubble is centered at \((64, 64, 64)\) while the droplet is centered at \((128, 384, 64)\). The initial diameter \(D_0\) for both problems is set to 128 lattice units in 2D and 64 lattice units in 3D. The boundary conditions are set to free-slip on the side walls and periodic at the top and bottom as illustrated in \figureref{fig:falling_droplet_bc}. This problem is driven mainly by the density and viscosity ratio of the two phases in addition to the Reynolds and Bond numbers. The Reynolds number measures the ratio of inertial forces to viscous forces, while the Bond number measures the ratio of gravitational forces to surface tension forces. Below is the definition of these four dimensionless numbers:
\begin{equation}
\rho^* = \frac{\rho_h}{\rho_l}, \quad
\mu^* = \frac{\mu_h}{\mu_l}, \quad
\text{Re}_h = \frac{\sqrt{g_y \rho_h (\rho_h - \rho_l) D^3}}{\mu_h}, \quad
\text{Bo} = \frac{g_y (\rho_h - \rho_l) D^2}{\sigma}
\end{equation}

where \(h\) and \(l\) indices refer to the heavy and light fluids, respectively. We have selected random, dimensionless numbers uniformly to ensure the entire defined range is covered. The outputs of the simulations are the interface indicator (\(c\)), velocity components (\(u, v, w\)), pressure (\(p\)), and density (\(\rho\)), which provide insights into the dynamics of multiphase flow and the interactions between the phases.

\begin{figure}[ht]
    \centering
    \begin{subfigure}[t]{0.49\textwidth}
        \centering
        \includegraphics[width=\linewidth,trim=0 0 0 0,clip]{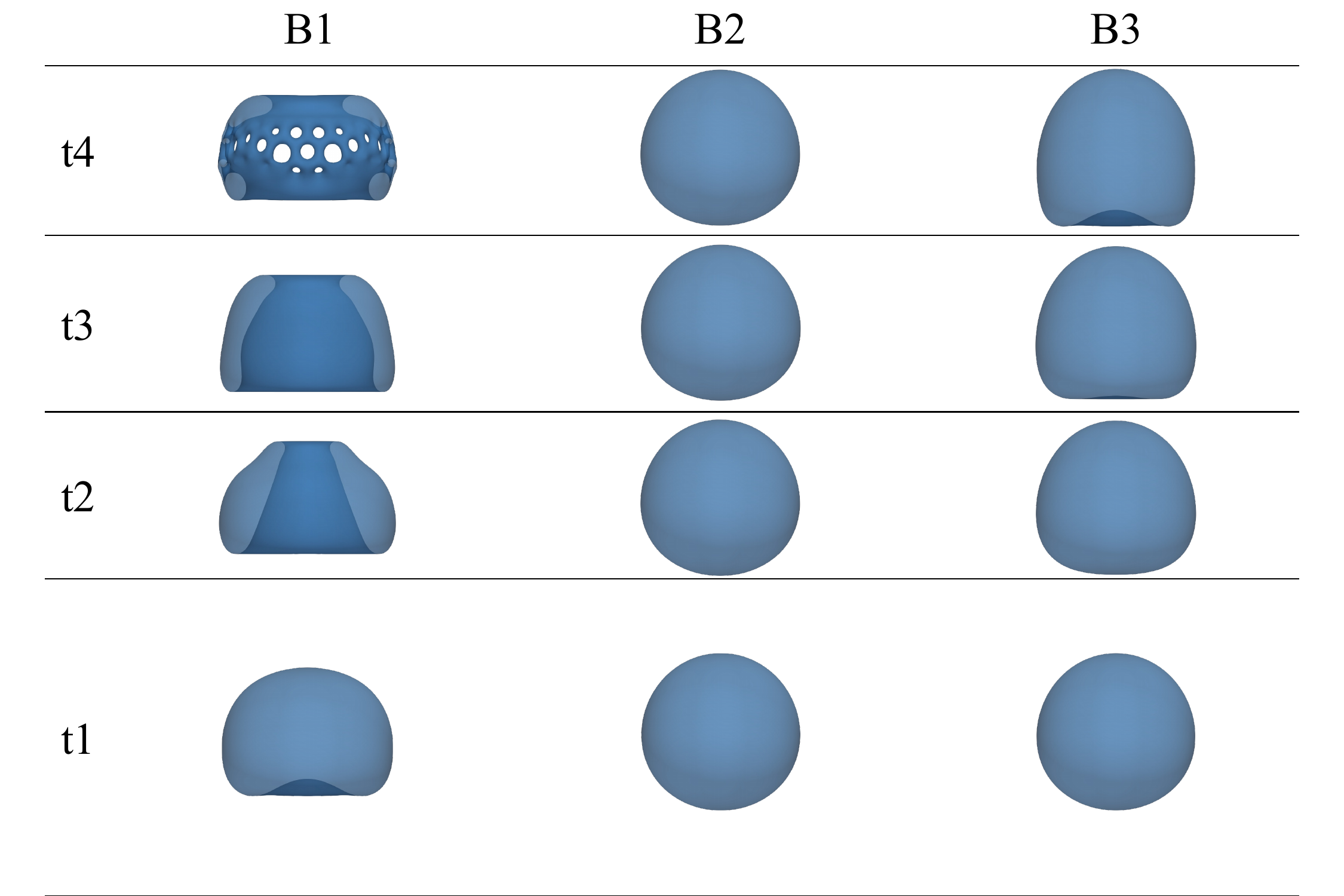} 
        \caption{Rising Bubble}
        \label{3D_contours_bubble}
    \end{subfigure}
    \begin{subfigure}[t]{0.49\textwidth}
        \centering
        \includegraphics[width=\linewidth,trim=0 0 0 0,clip]{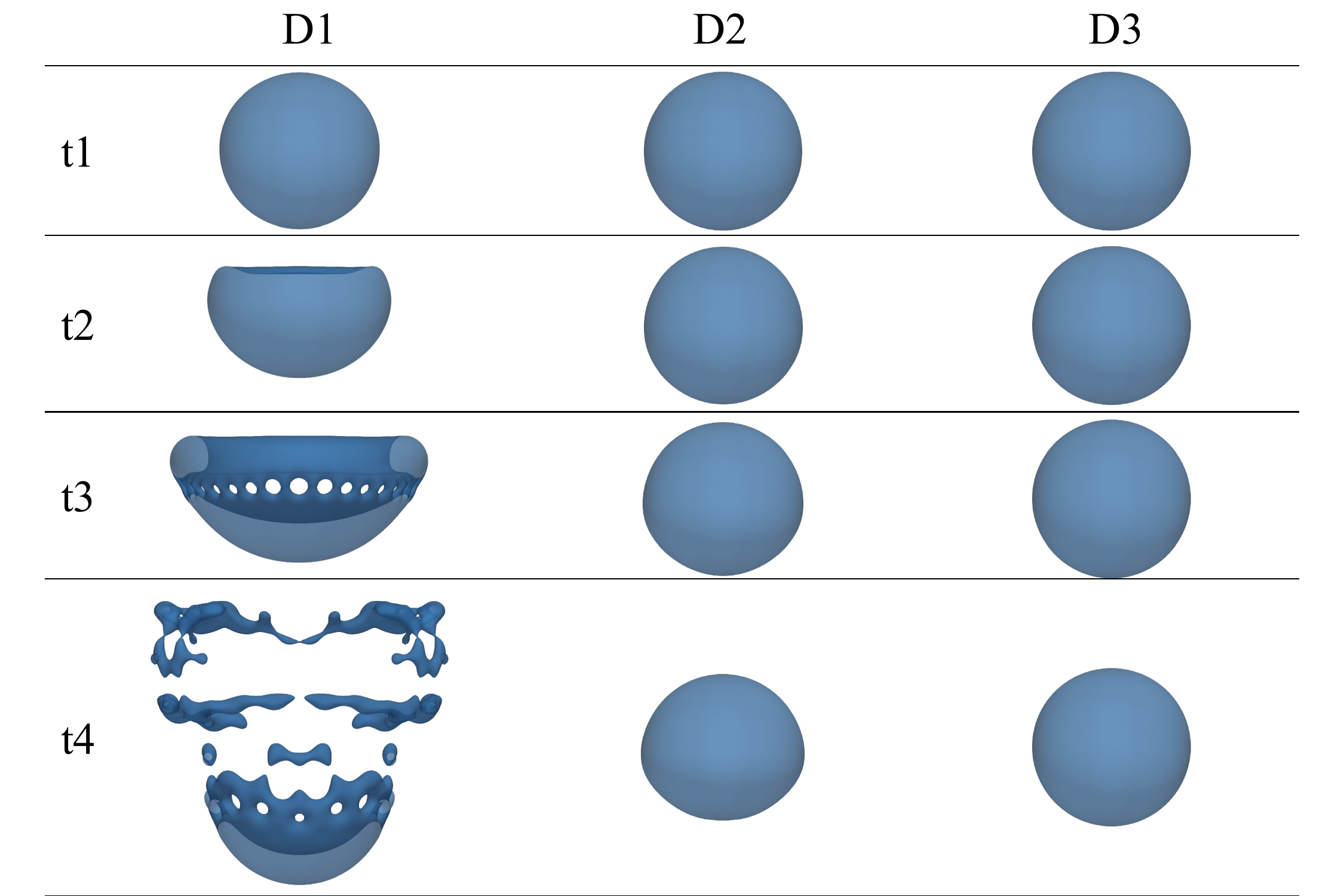} 
        \caption{Falling Droplet}
        \label{3D_contours_drop}
    \end{subfigure}
    \caption{(a) Snapshot of a 3D rising bubble and (b) snapshot of a 3D falling droplet. The properties of the fluids for each case are detailed in \tabref{tab:properties}.  \label{fig:3D-shape}}
    \label{3D_contours}
\end{figure}

\begin{figure}[t!]
    \centering

    \begin{subfigure}[t]{0.45\textwidth}
        \centering
        \includegraphics[width=\linewidth,trim=0 30 300 0 ,clip]{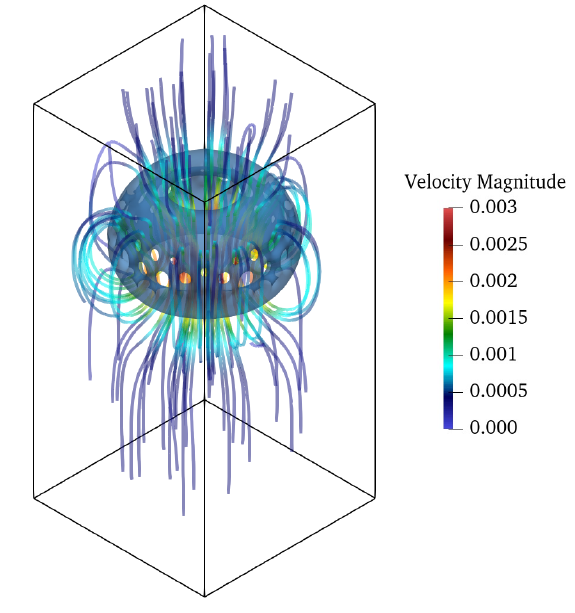} 
        \caption{Rising Bubble (B1)}
        \label{3D-streamline-B1}
    \end{subfigure}
    \begin{subfigure}[t]{0.45\textwidth}
        \centering
        \includegraphics[width=\linewidth,trim=300 30 0 0 ,clip]{Figures/3D/StreamLine3D-two.pdf} 
        \caption{Falling Droplet (D1)}
        \label{3D-streamline-D1}
    \end{subfigure}    
    \caption{Streamlines of a 3D rising bubble (a) and a 3D falling droplet (b), with colors indicating the magnitude of velocity. The properties of the fluids for each case are detailed in \tabref{tab:properties}.    \label{tab:3D-LIC}}
    \label{3D-streamline}
\end{figure}

\begin{figure}[t!]
    \begin{subfigure}{0.4\linewidth}
        \centering
        \setlength\extrarowheight{3pt}
        \begin{tabular}{>{\centering\arraybackslash}m{0.05\linewidth}>{\centering\arraybackslash}m{0.28\linewidth}>{\centering\arraybackslash}m{0.28\linewidth}>{\centering\arraybackslash}m{0.28\linewidth}}
        \hline
         & B1 & B2 & B3 \\
        \hline
        % [trim={left bottom right top},clip]
        t5 &
        \includegraphics[width=\linewidth,trim=300 0 300 0,clip]{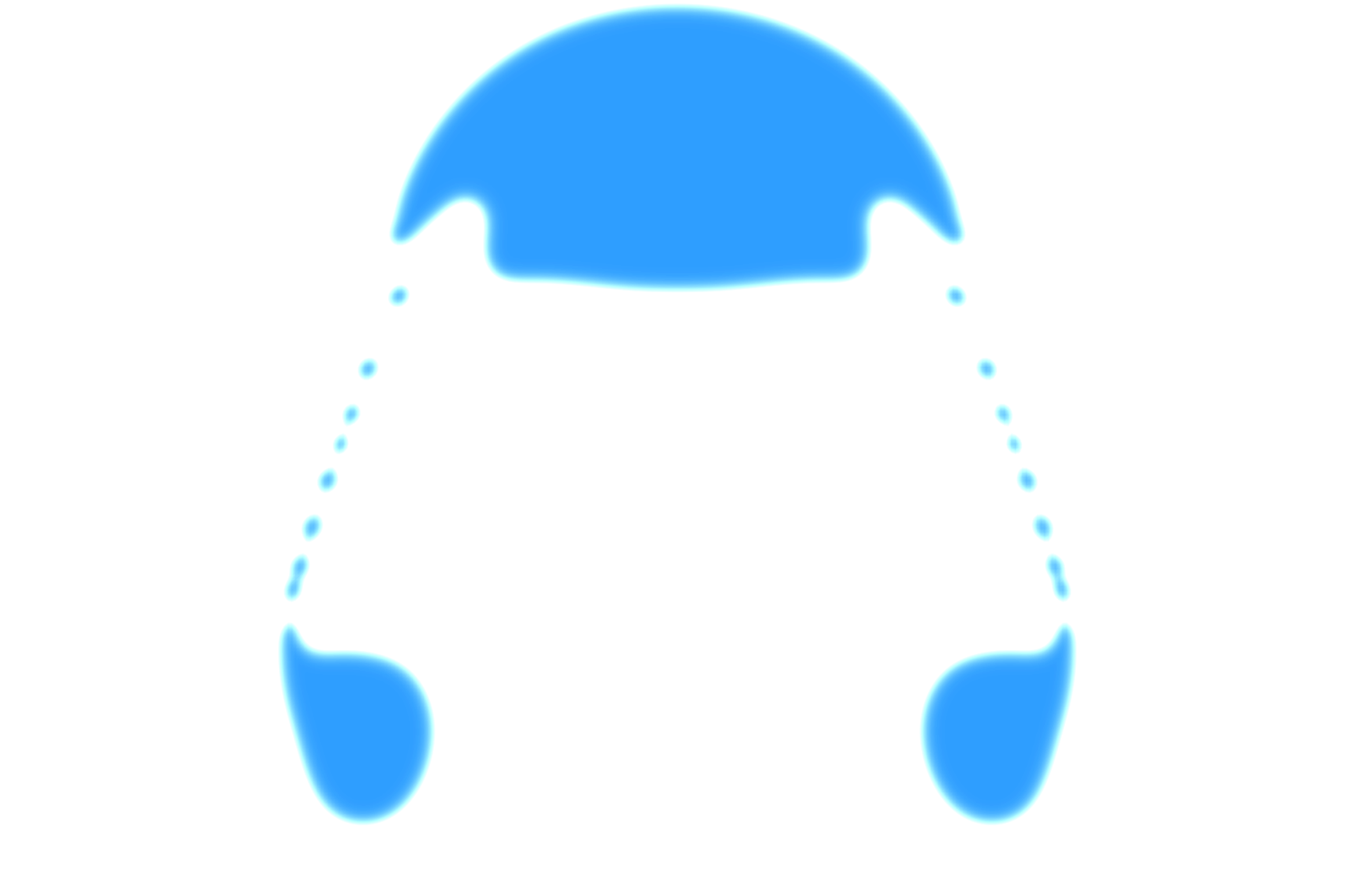} &
        \includegraphics[width=\linewidth,trim=300 0 300 0,clip]{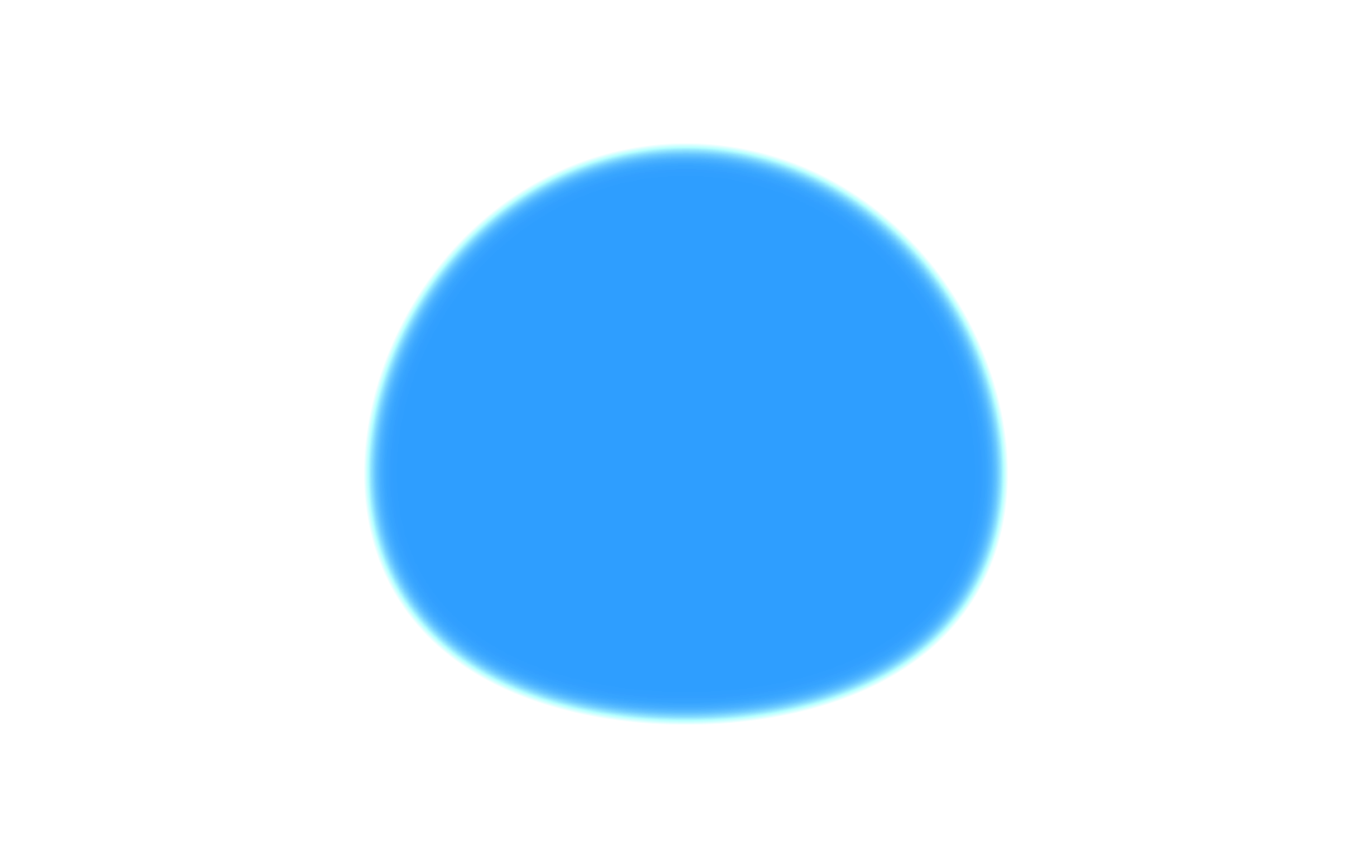} &
        \includegraphics[width=\linewidth,trim=300 0 300 0,clip]{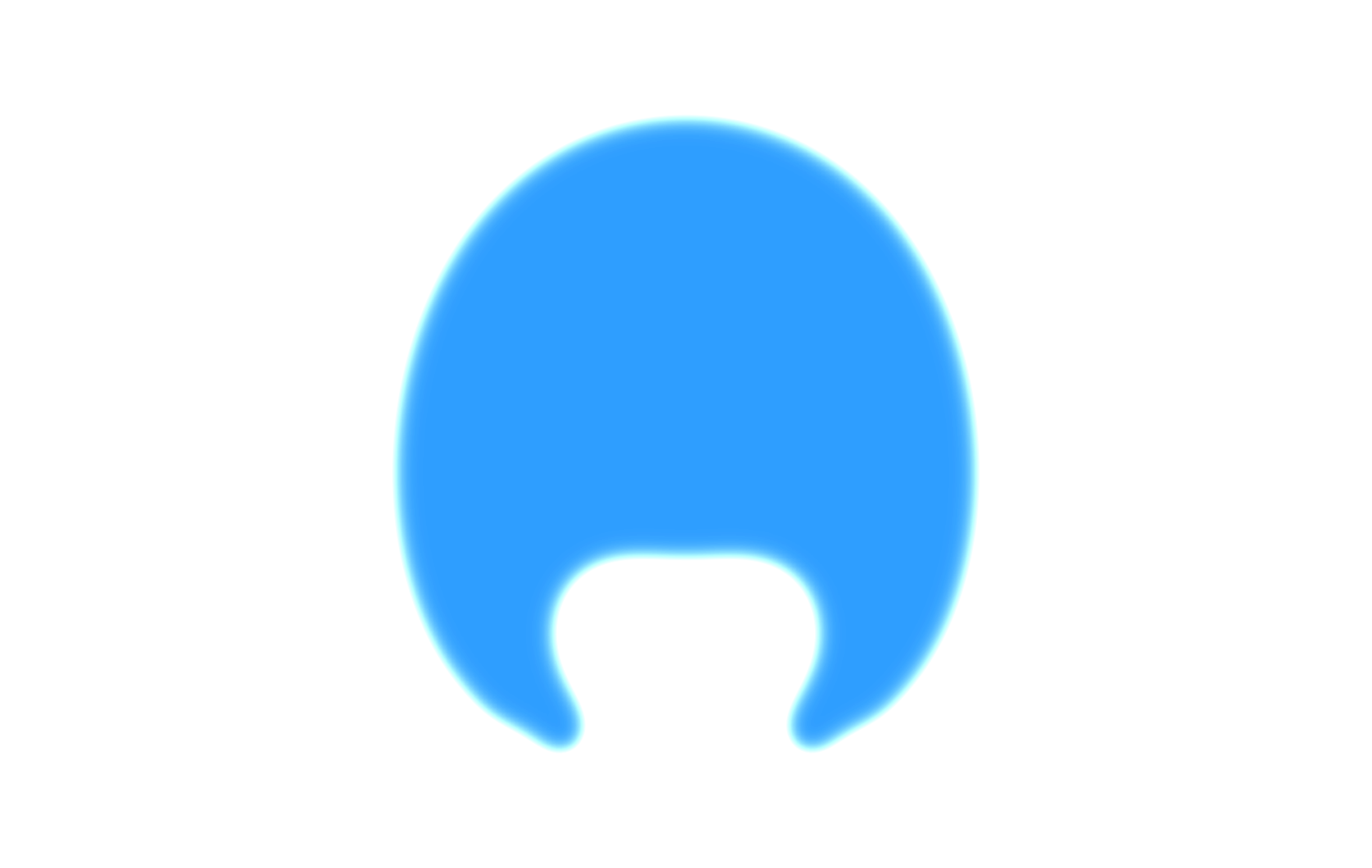} \\
        \hline
        t4 &
        \includegraphics[width=\linewidth,trim=300 170 300 170,clip]{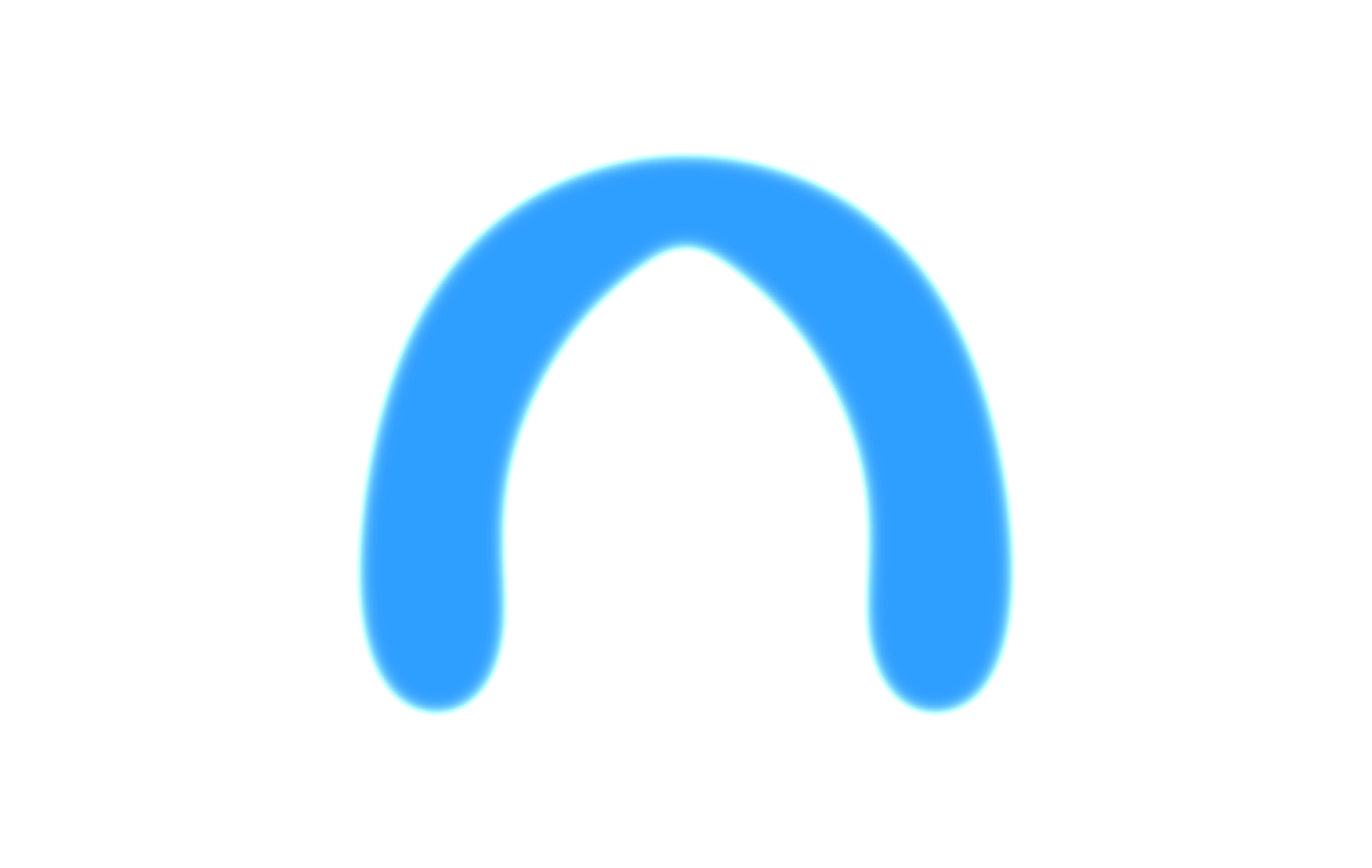} &
        \includegraphics[width=\linewidth,trim=300 170 300 170,clip]{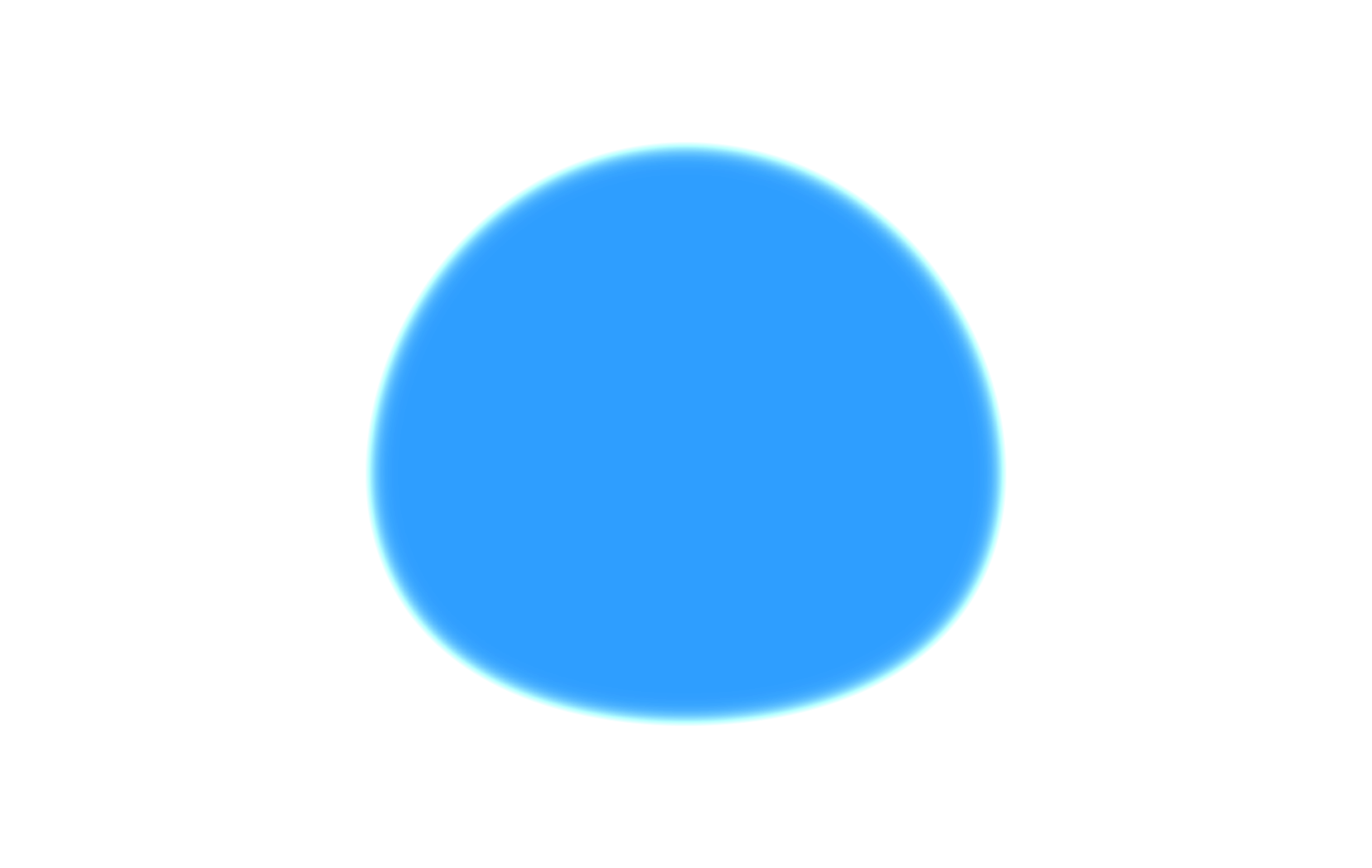} &
        \includegraphics[width=\linewidth,trim=300 170 300 170,clip]{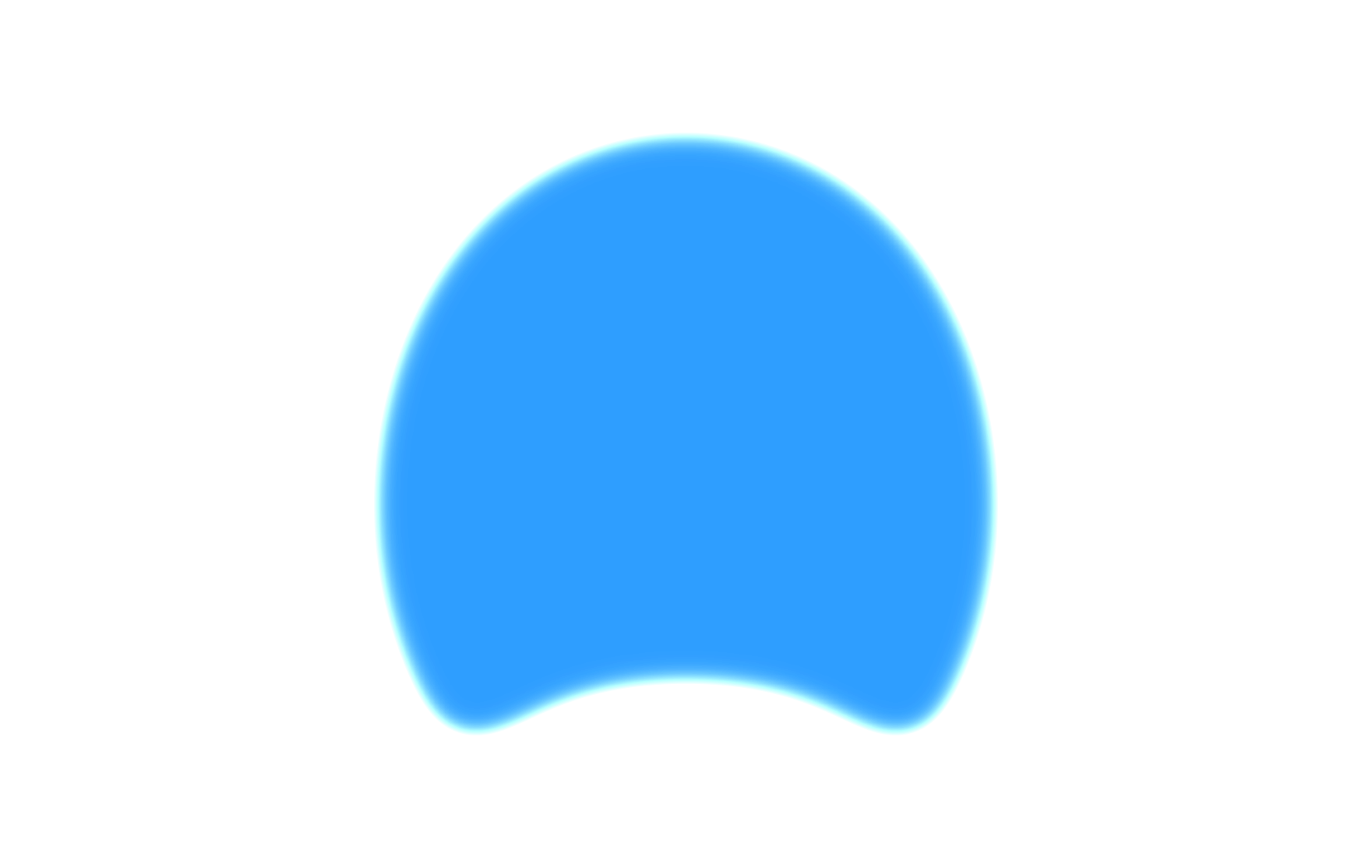} \\
        \hline
        t3 &
        \includegraphics[width=\linewidth,trim=300 170 300 170,clip]{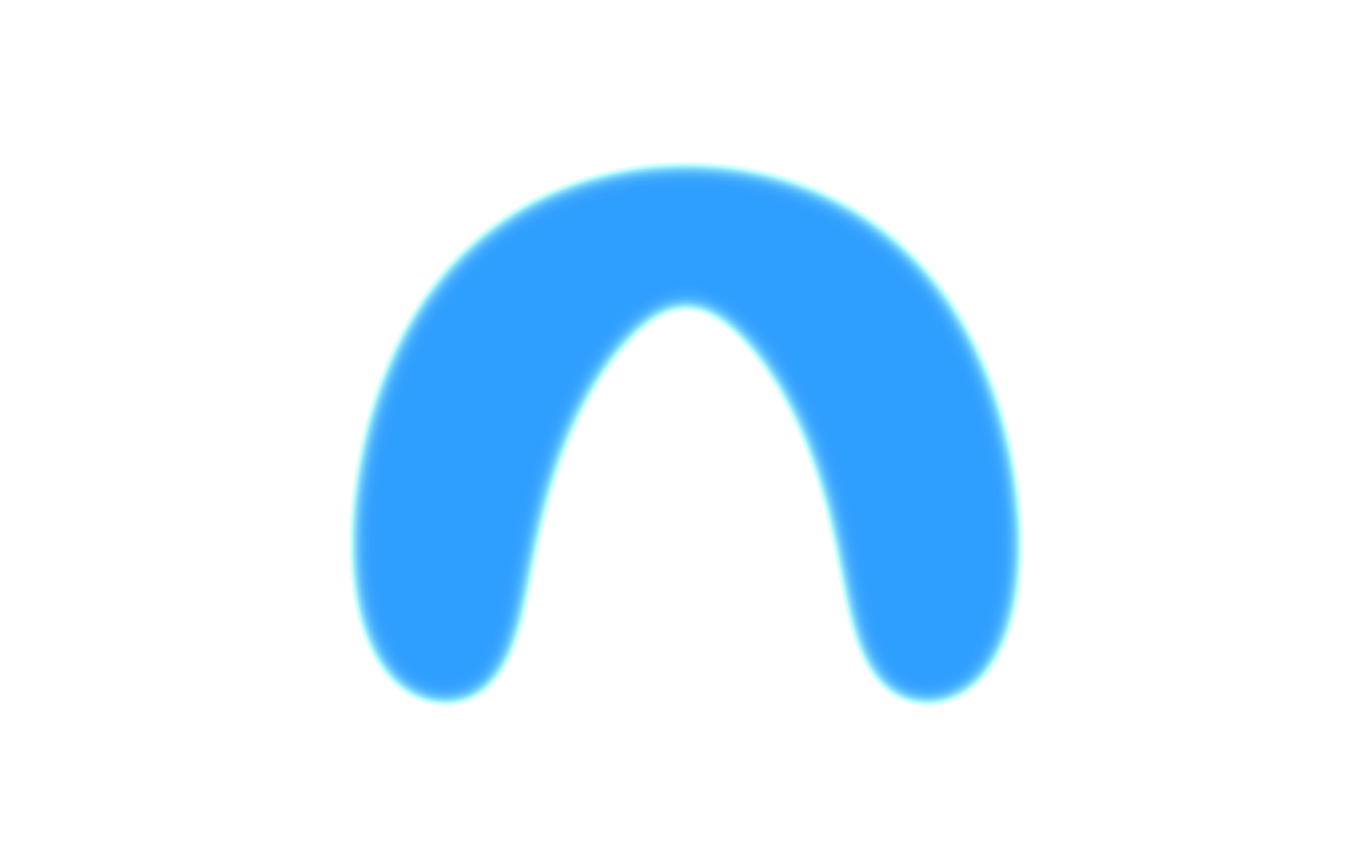} &
        \includegraphics[width=\linewidth,trim=300 170 300 170,clip]{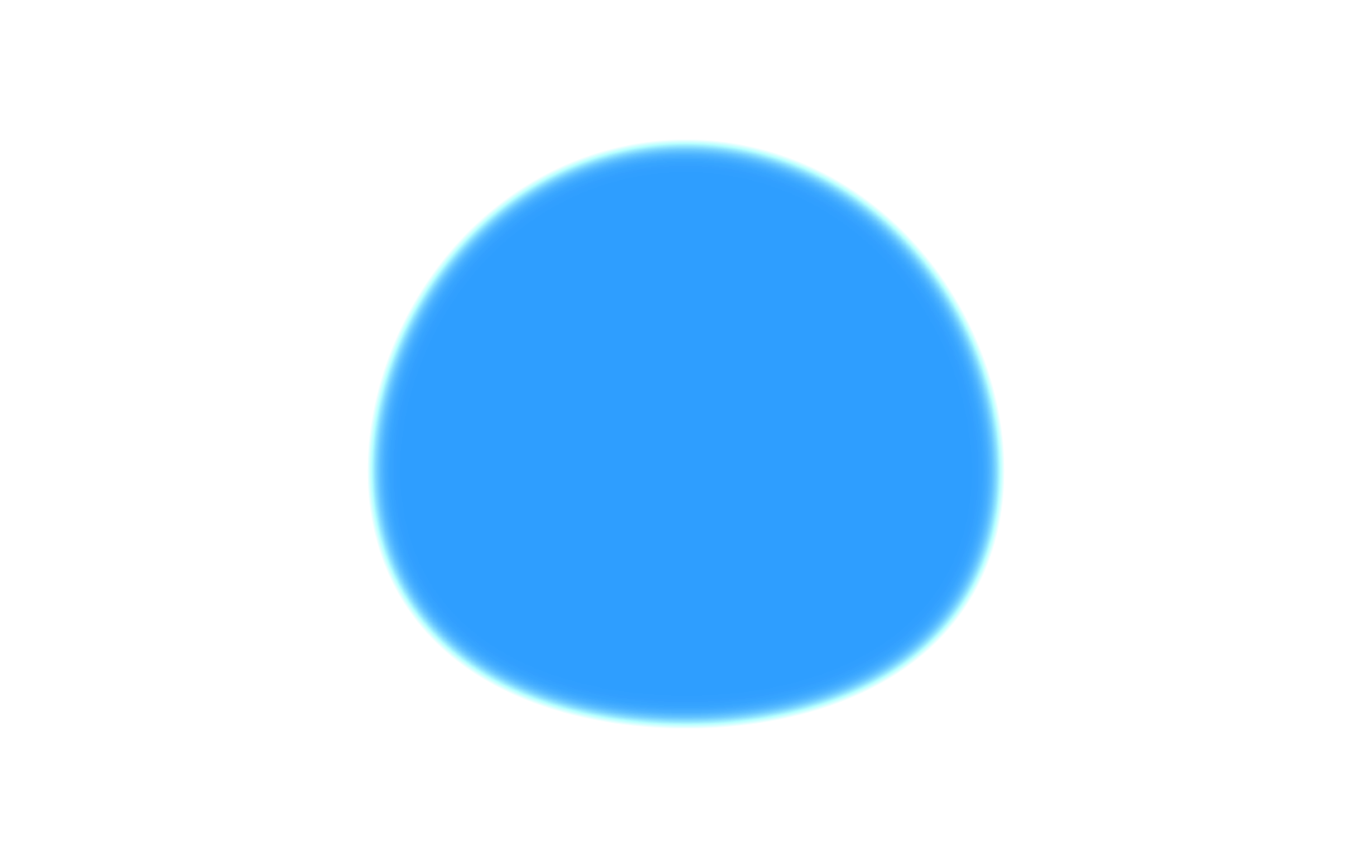} &
        \includegraphics[width=\linewidth,trim=300 170 300 170,clip]{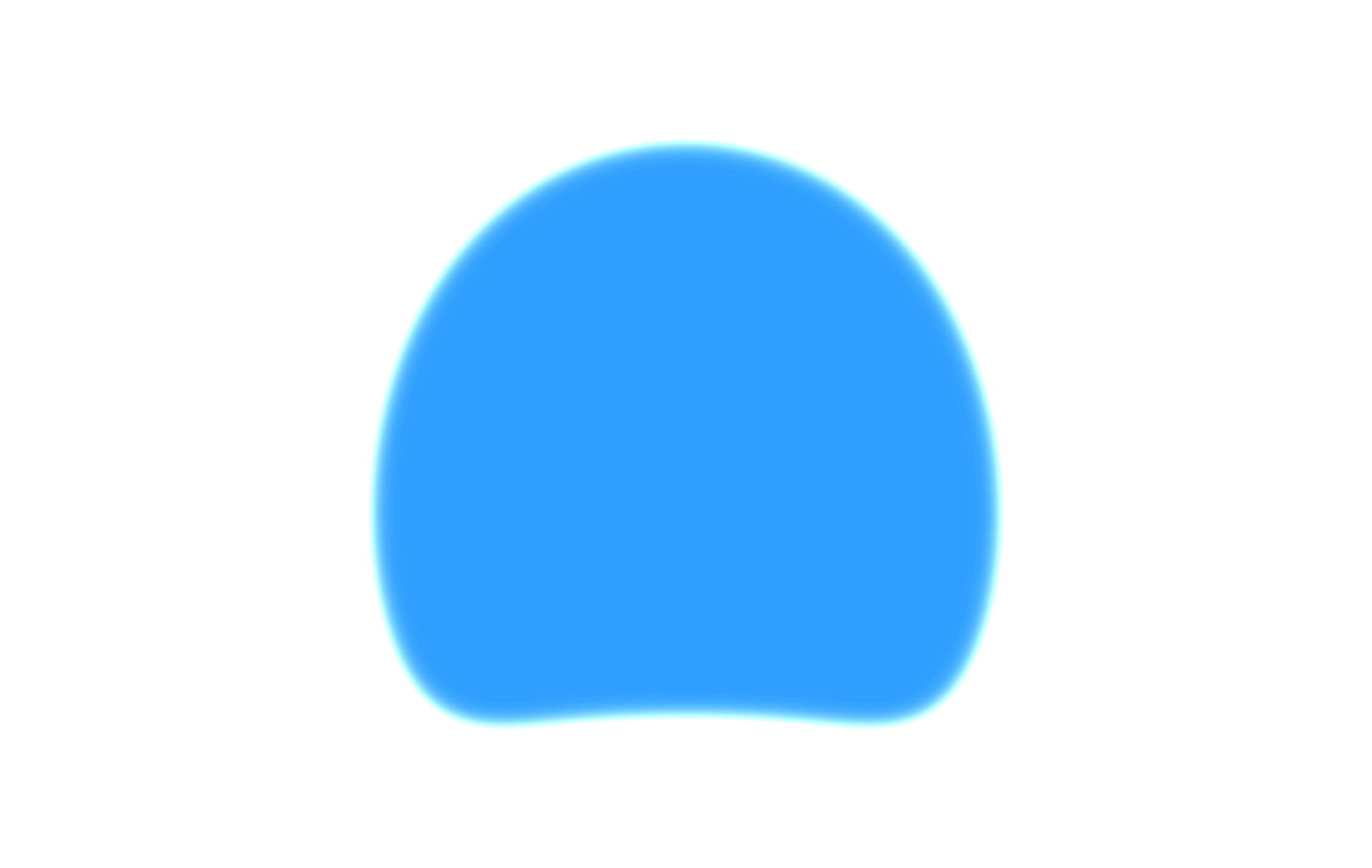} \\
        \hline
        t2 &
        \includegraphics[width=\linewidth,trim=300 170 300 170,clip]{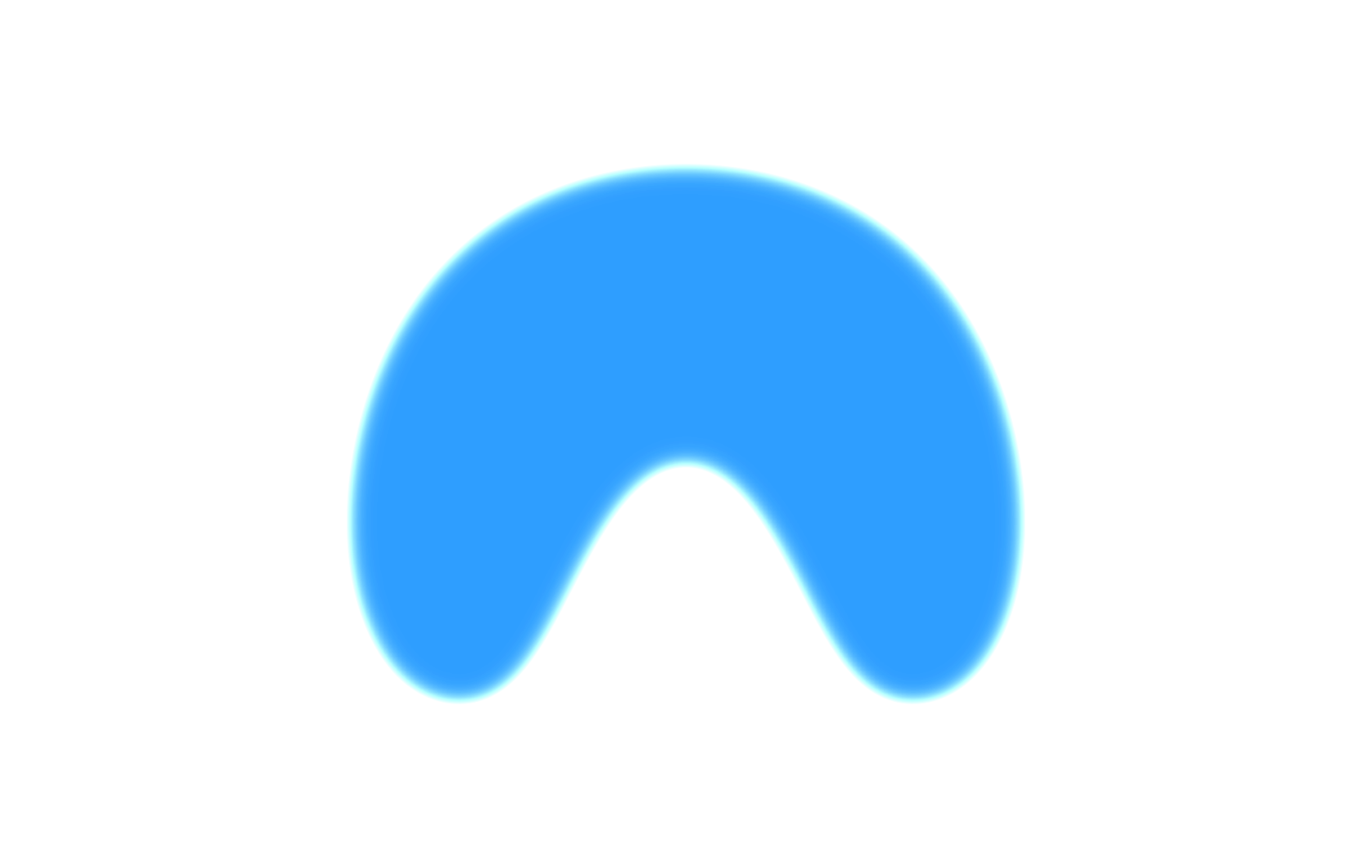} &
        \includegraphics[width=\linewidth,trim=300 170 300 170,clip]{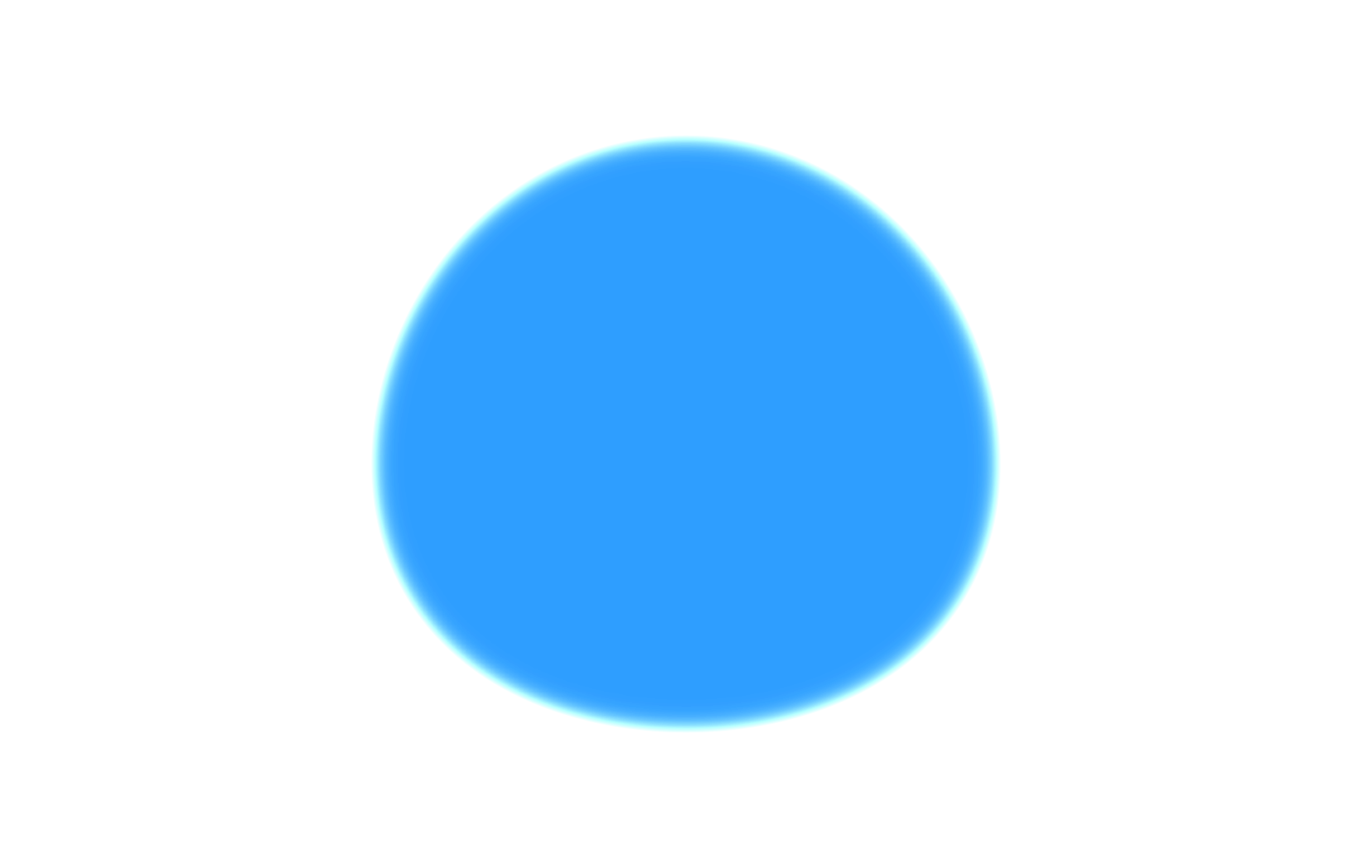} &
        \includegraphics[width=\linewidth,trim=300 170 300 170,clip]{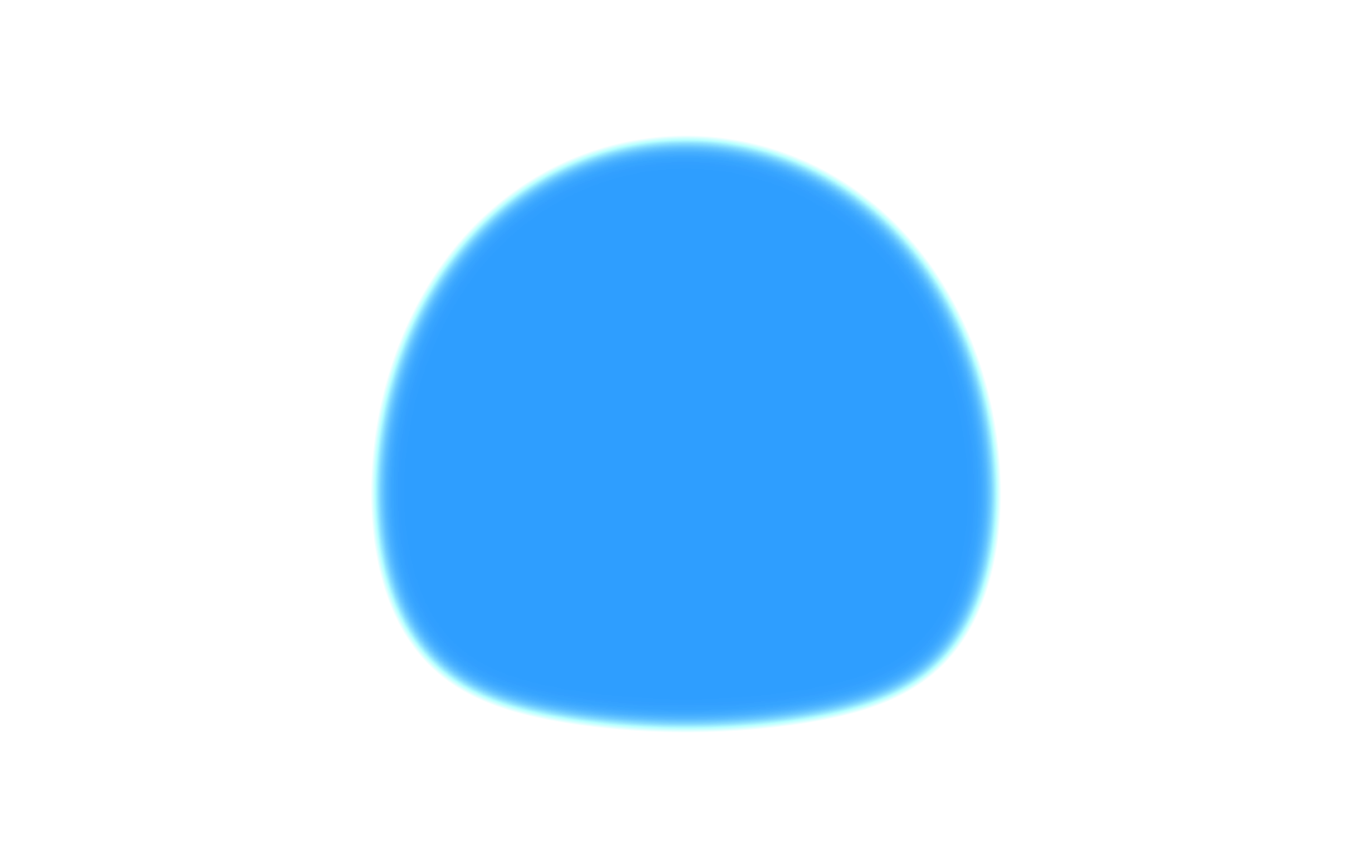} \\
        \hline
        t1 &
        \includegraphics[width=\linewidth,trim=300 170 300 170,clip]{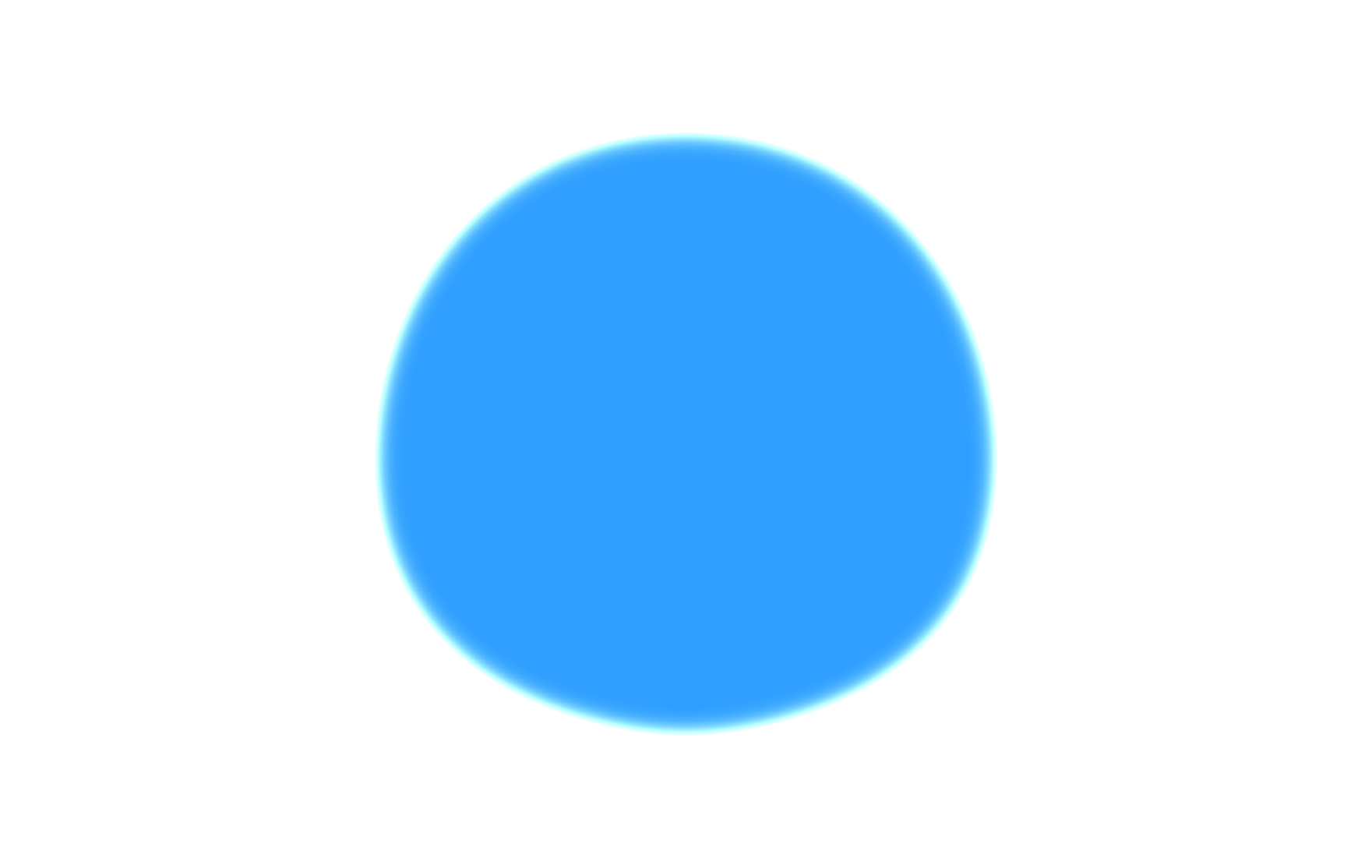} &
        \includegraphics[width=\linewidth,trim=300 170 300 170,clip]{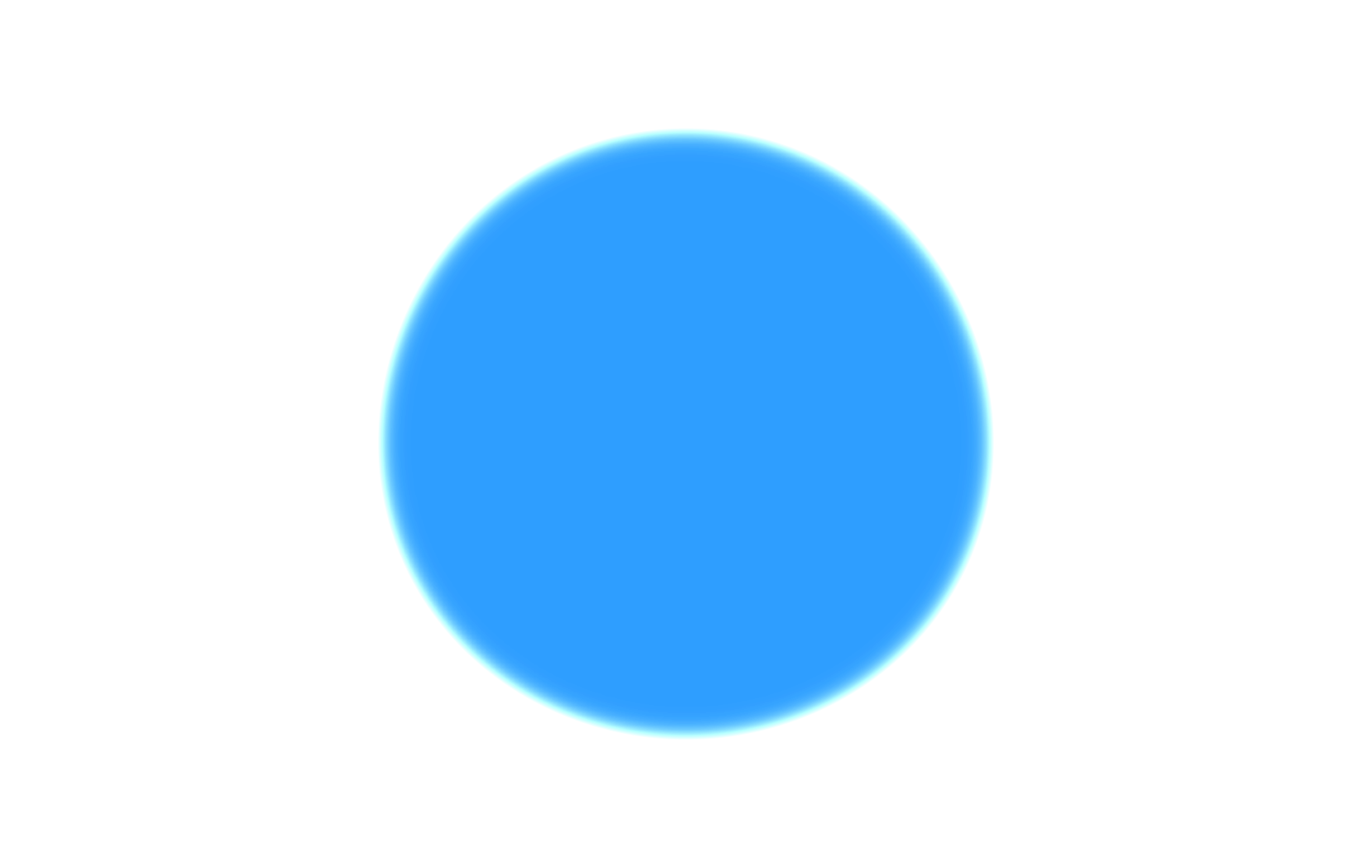} &
        \includegraphics[width=\linewidth,trim=300 170 300 170,clip]{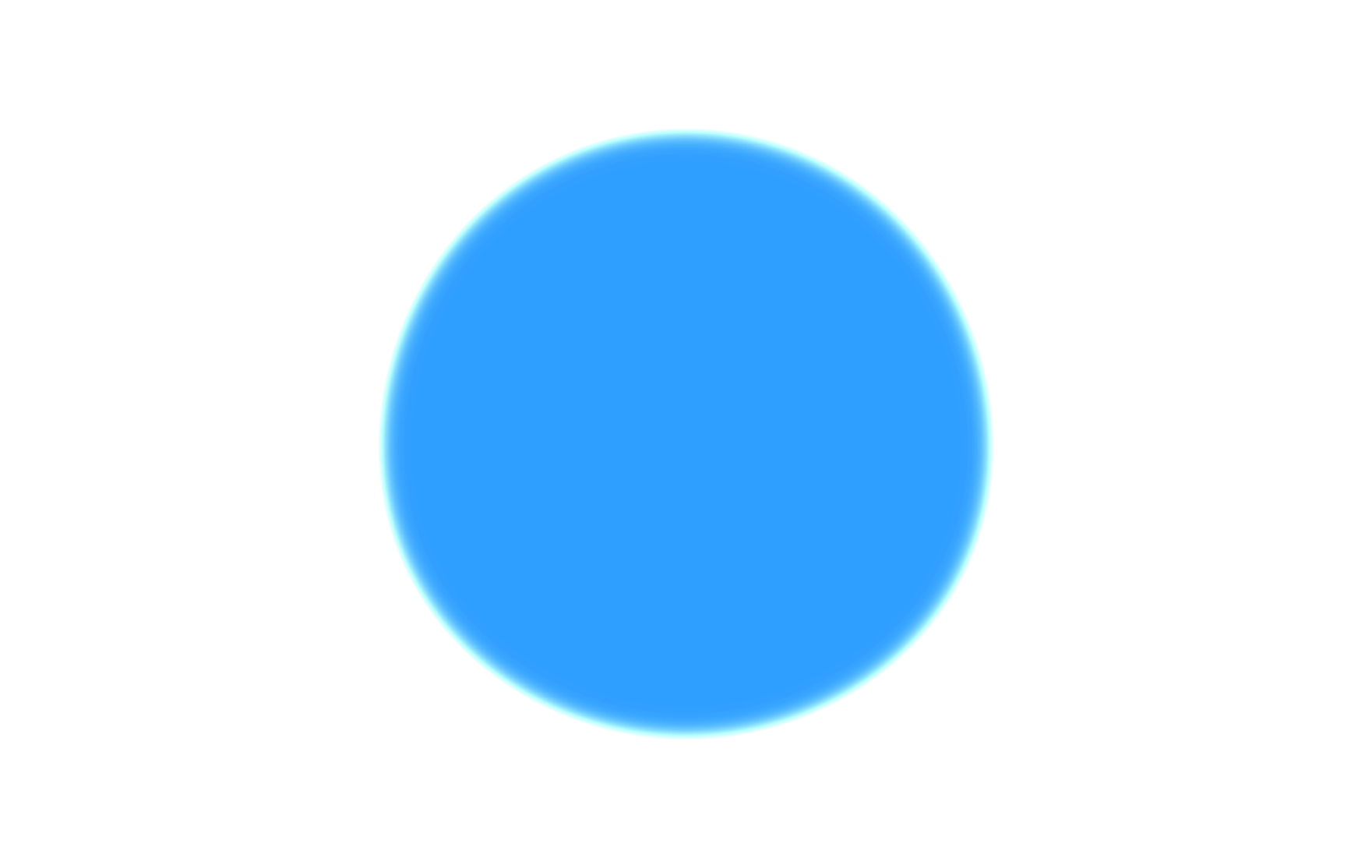} \\
        \hline
        \end{tabular}
        \caption{Rising Bubble}
        \label{2D_contours_bubble}
    \end{subfigure}
    \hspace{0.1\linewidth}
    \begin{subfigure}{0.4\linewidth}
        \centering
        \setlength\extrarowheight{3pt}
        \begin{tabular}{>{\centering\arraybackslash}m{0.05\linewidth}>{\centering\arraybackslash}m{0.28\linewidth}>{\centering\arraybackslash}m{0.28\linewidth}>{\centering\arraybackslash}m{0.28\linewidth}}
        \hline
         & D1 & D2 & D3 \\
        \hline
        % [trim={left bottom right top},clip]
        t1 &
        \includegraphics[width=\linewidth,trim=300 0 300 0,clip]{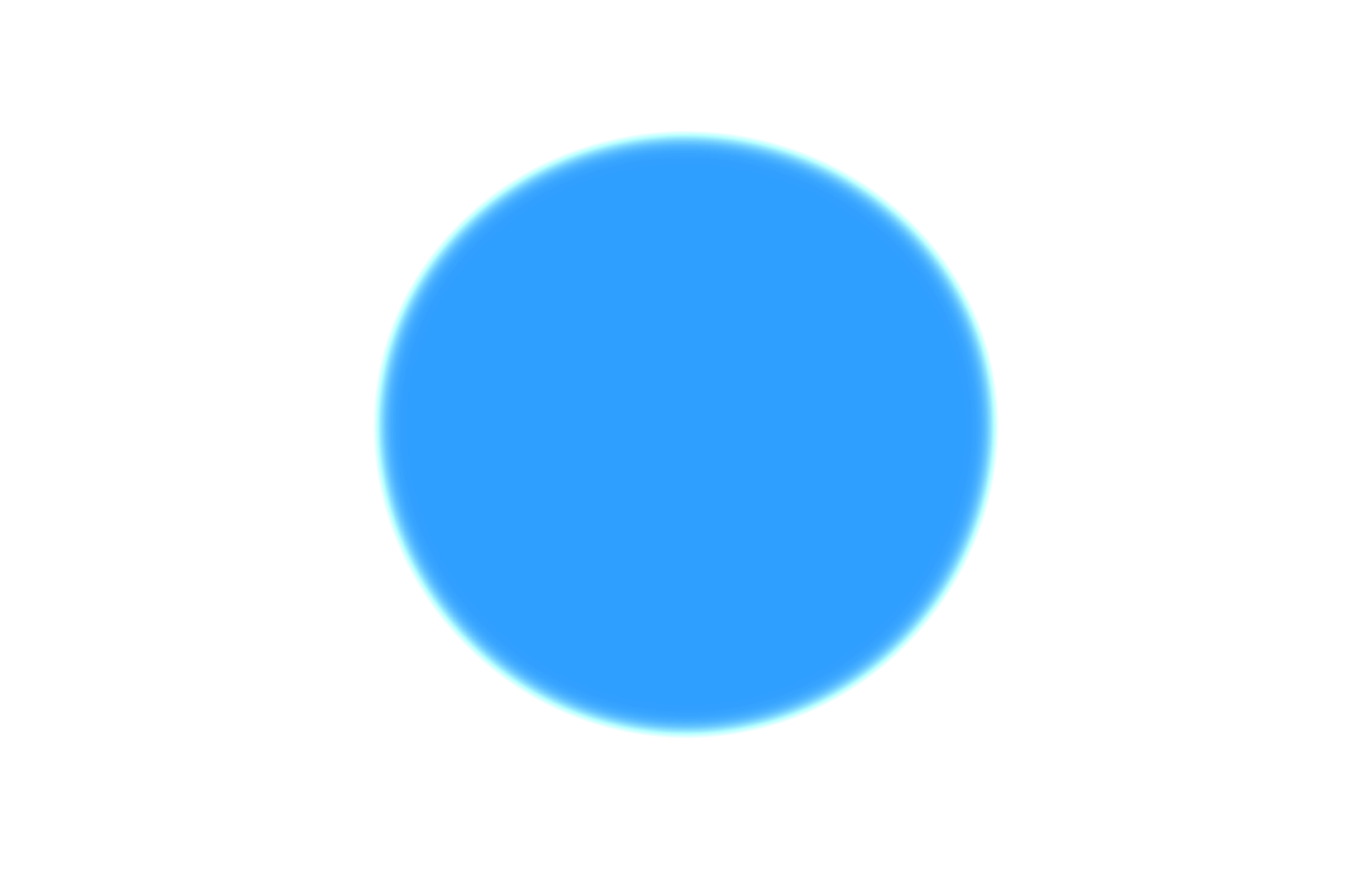} &
        \includegraphics[width=\linewidth,trim=300 0 300 0,clip]{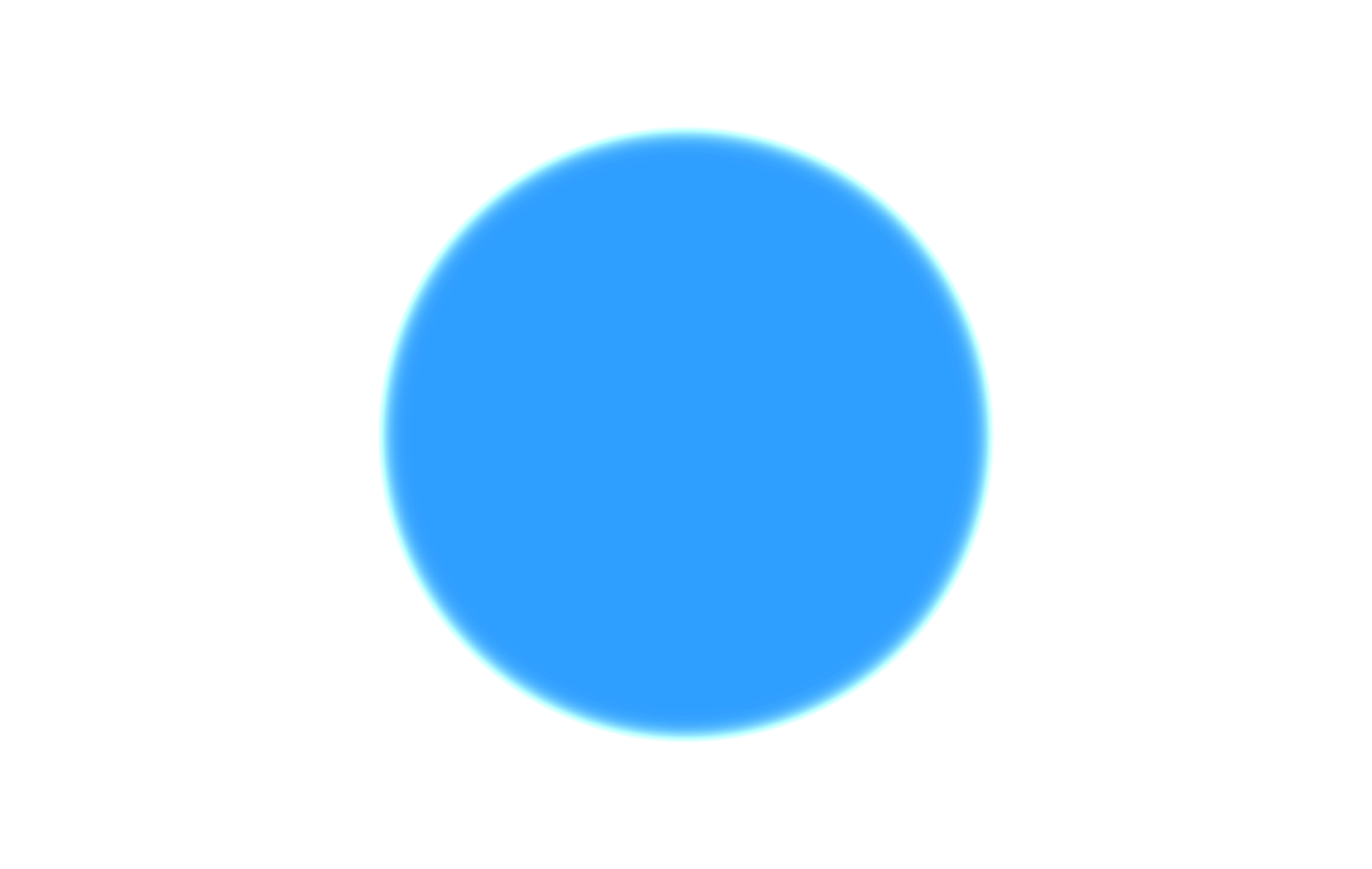} &
        \includegraphics[width=\linewidth,trim=300 0 300 0,clip]{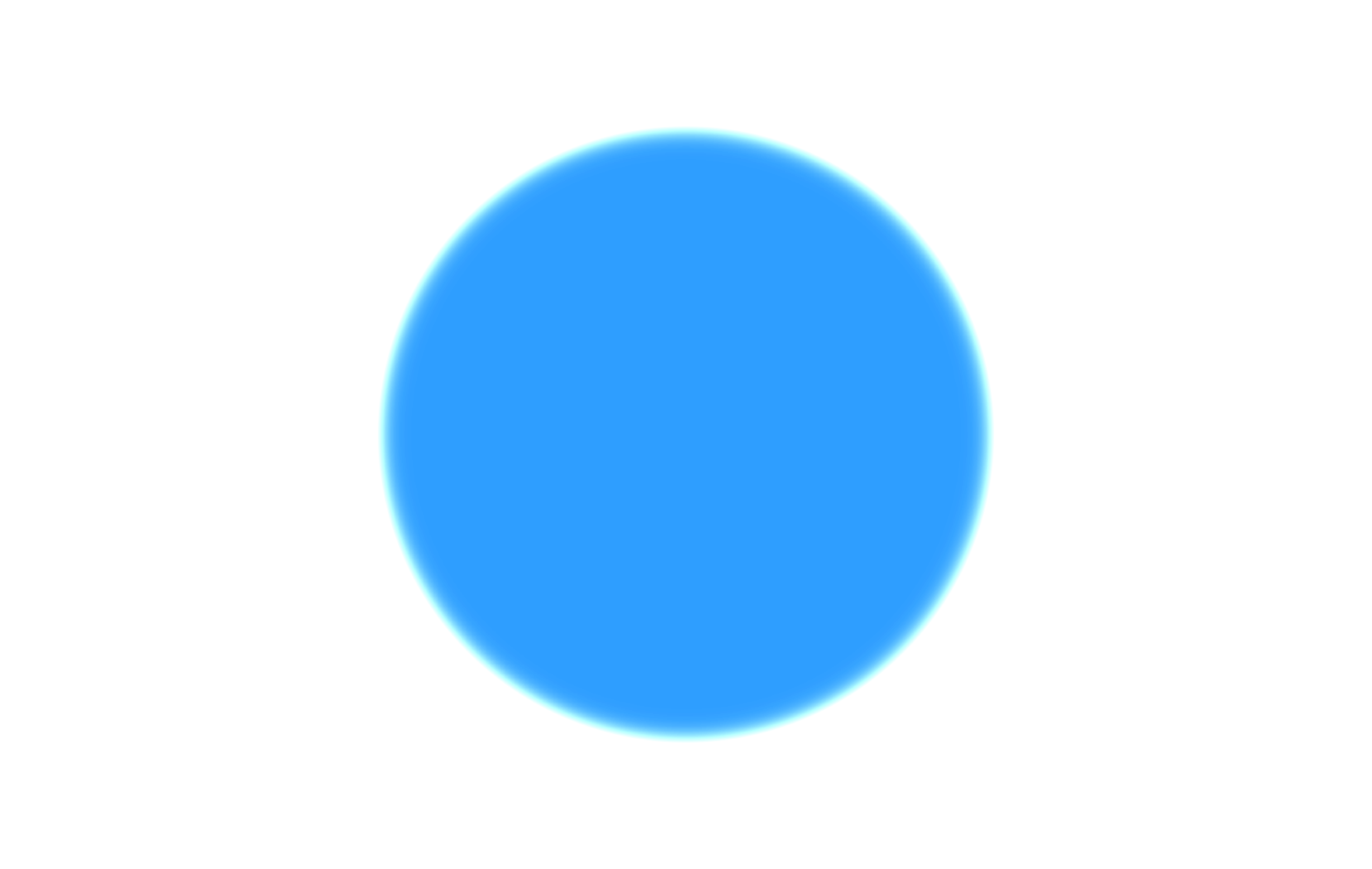} \\
        \hline
        t2 &
        \includegraphics[width=\linewidth,trim=300 170 300 170,clip]{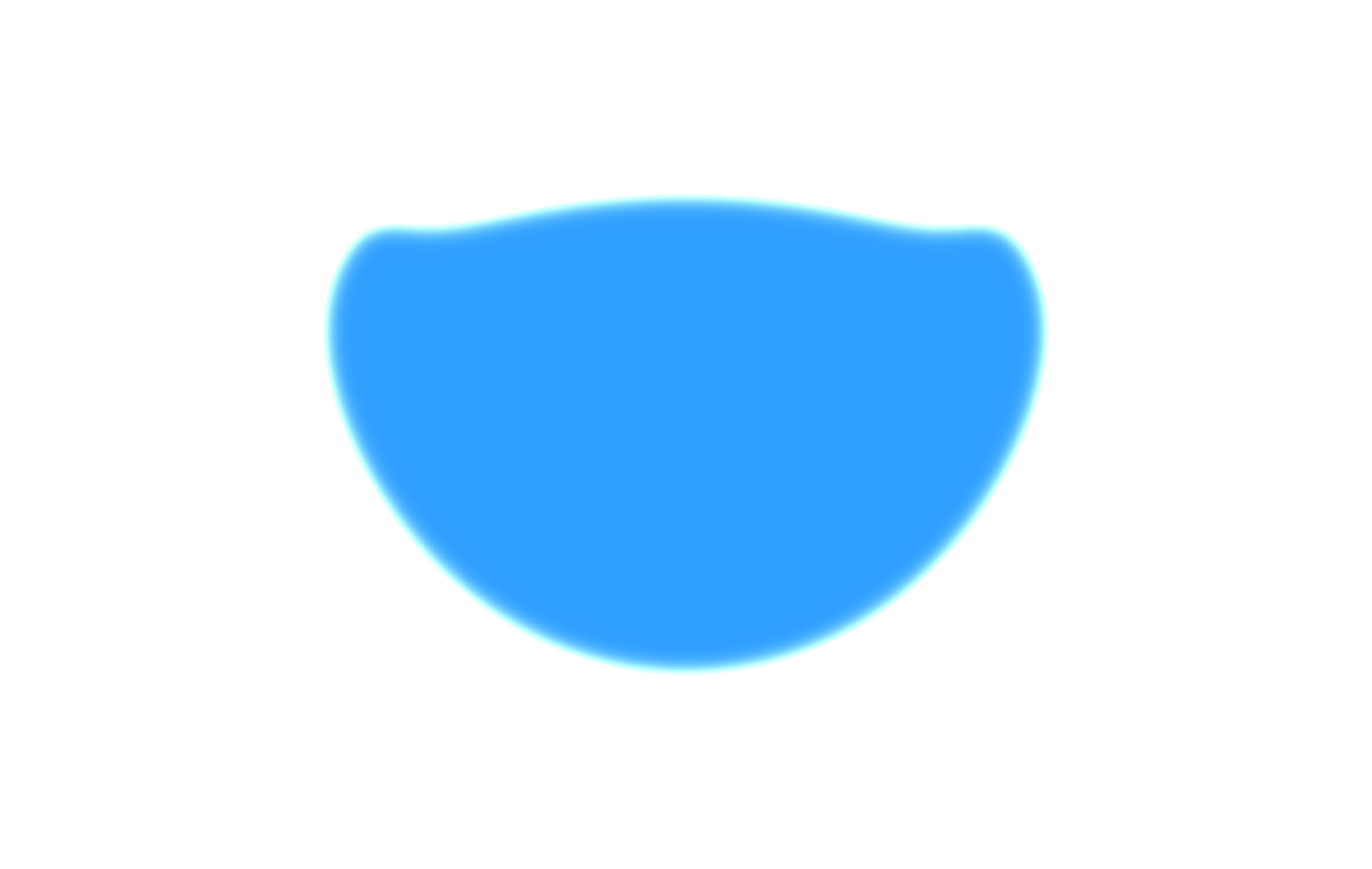} &
        \includegraphics[width=\linewidth,trim=300 170 300 170,clip]{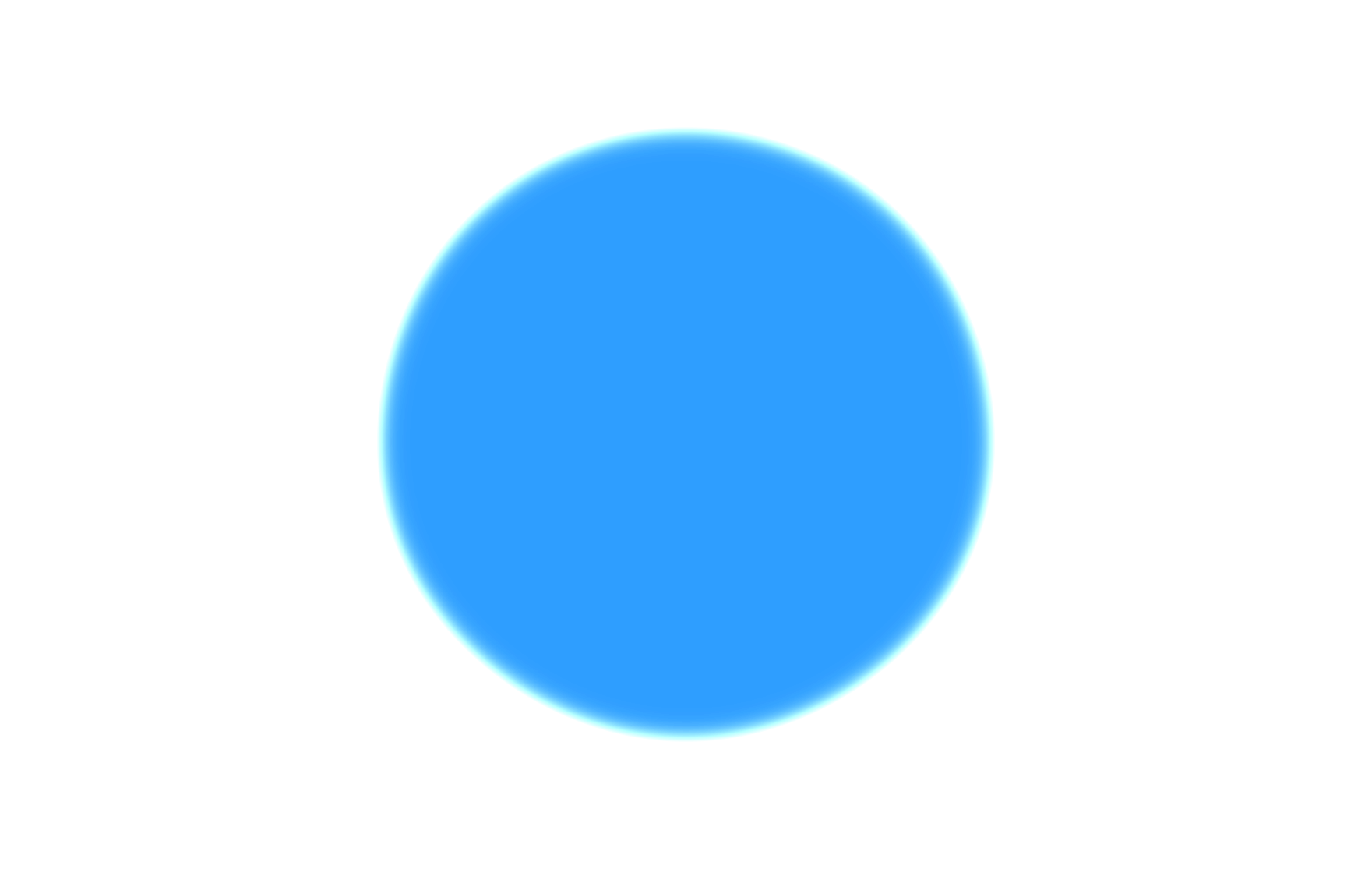} &
        \includegraphics[width=\linewidth,trim=300 170 300 170,clip]{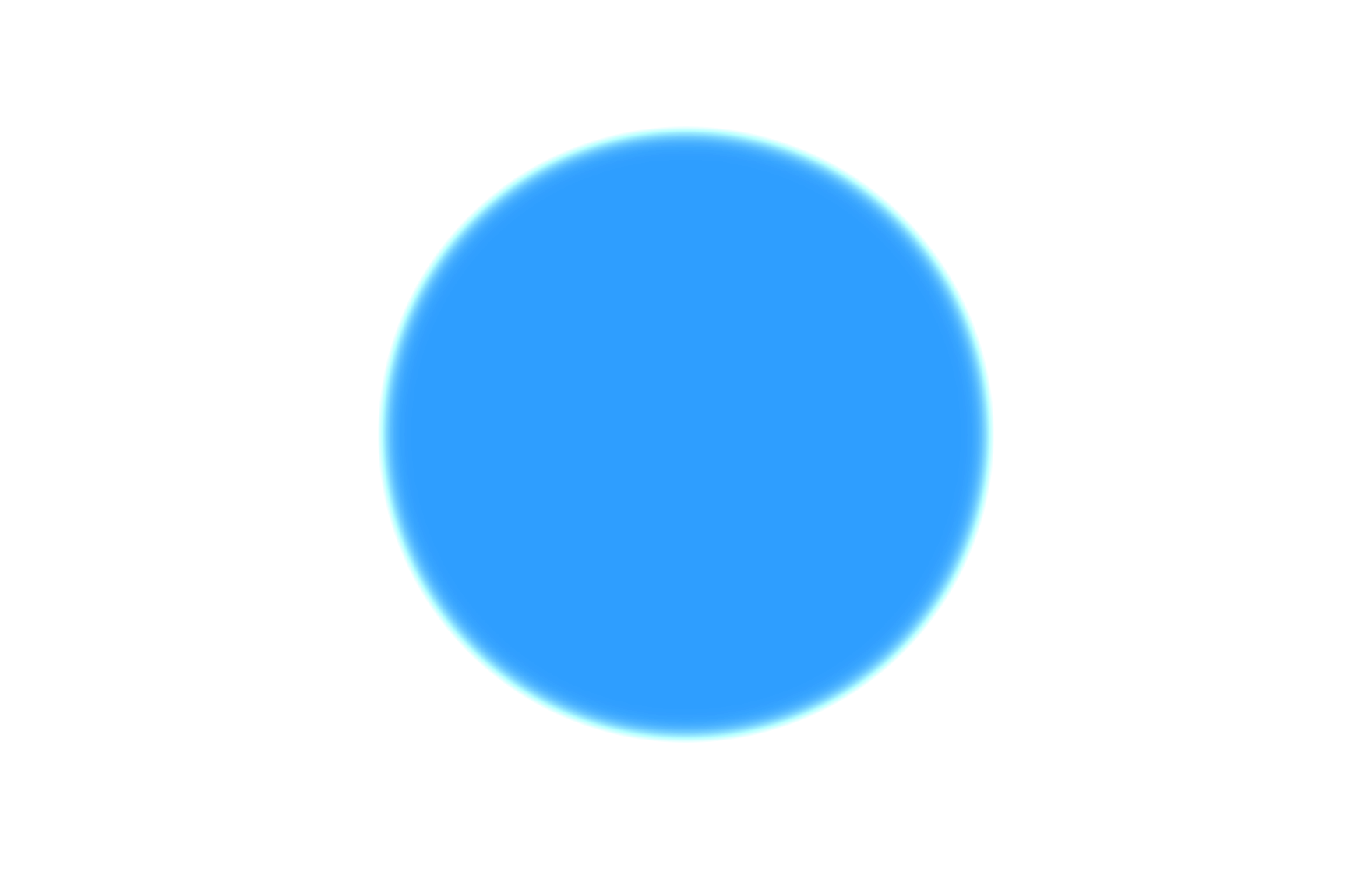} \\
        \hline
        t3 &
        \includegraphics[width=\linewidth,trim=300 170 300 170,clip]{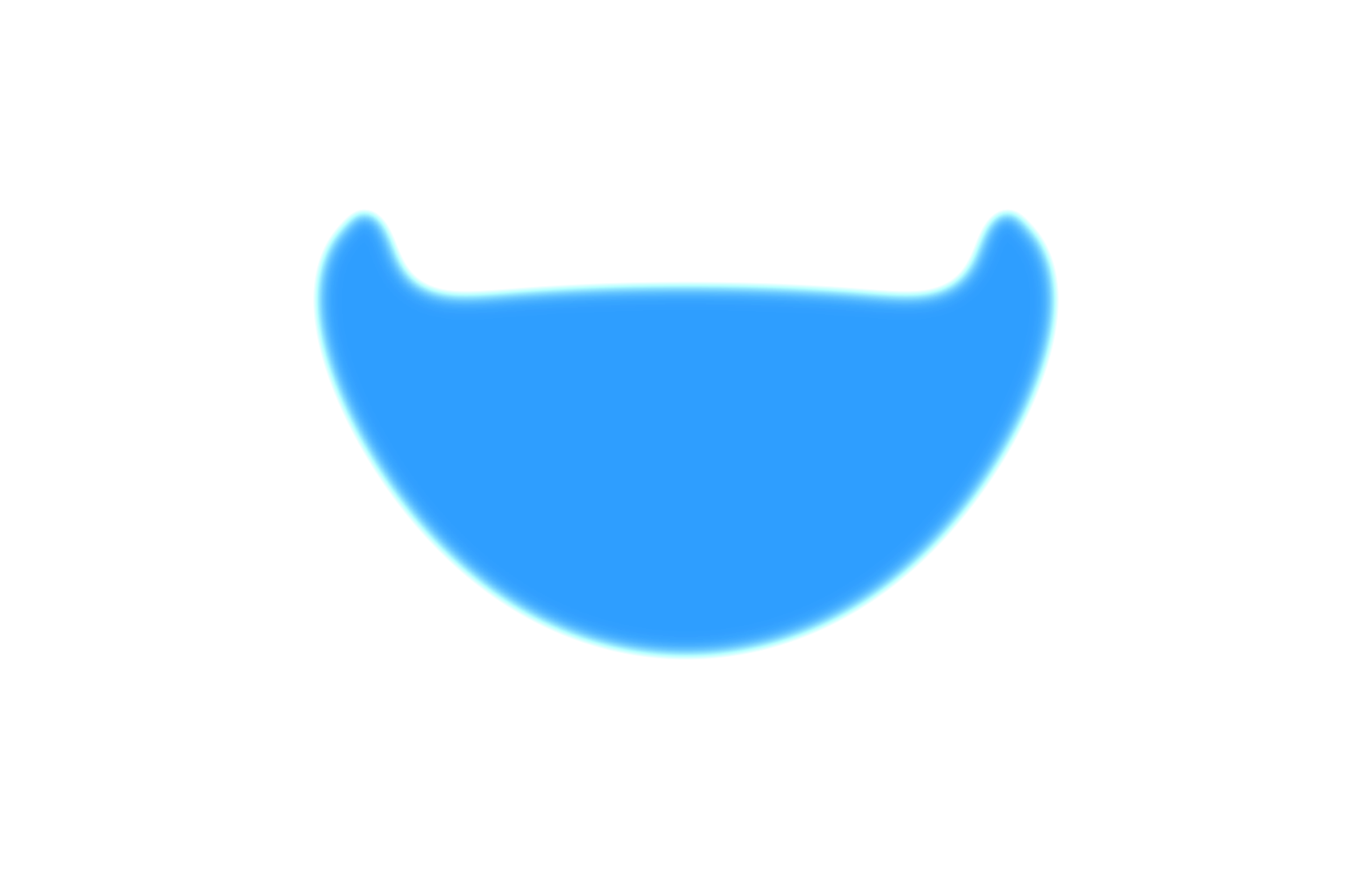} &
        \includegraphics[width=\linewidth,trim=300 170 300 170,clip]{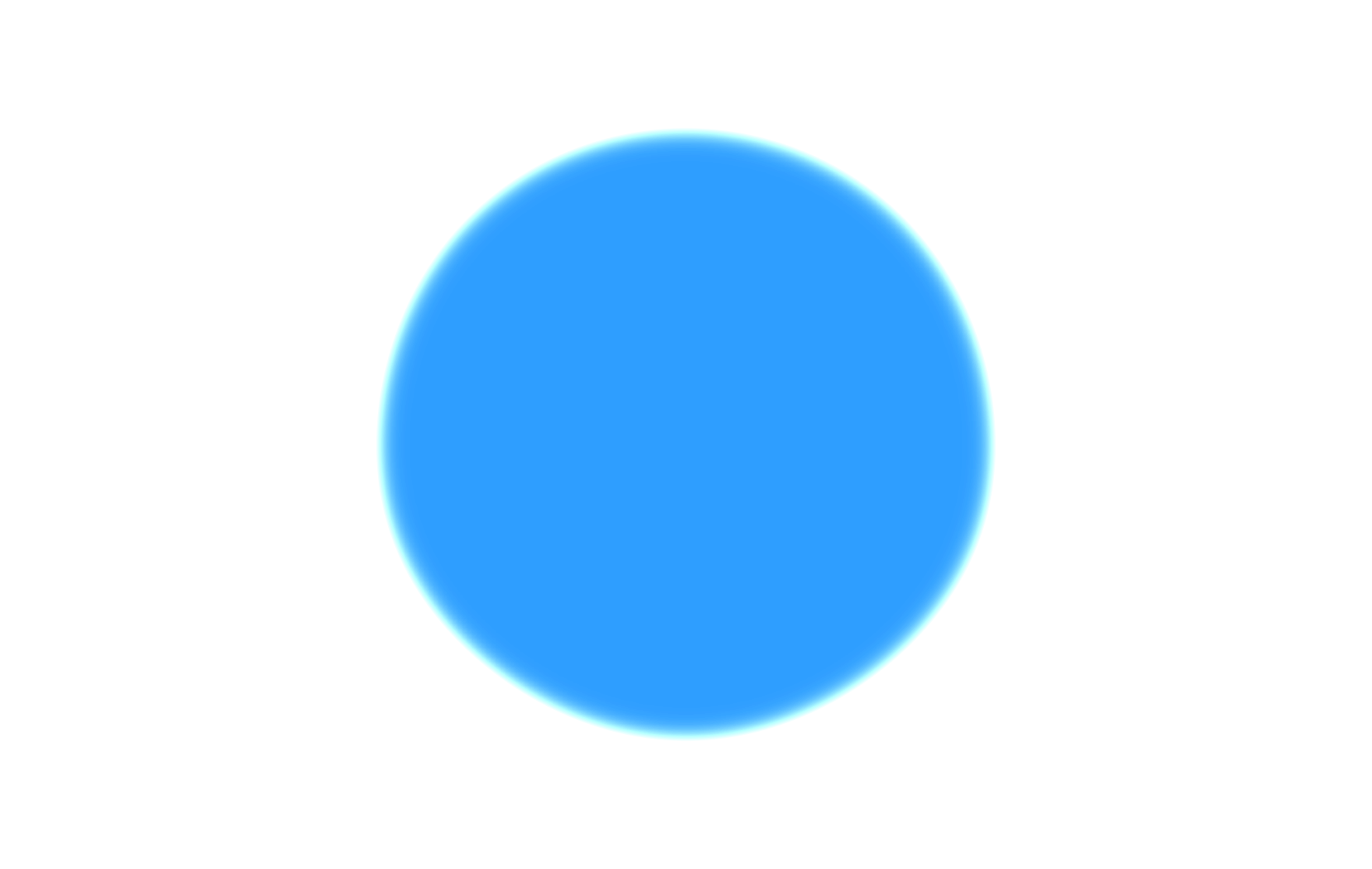} &
        \includegraphics[width=\linewidth,trim=300 170 300 170,clip]{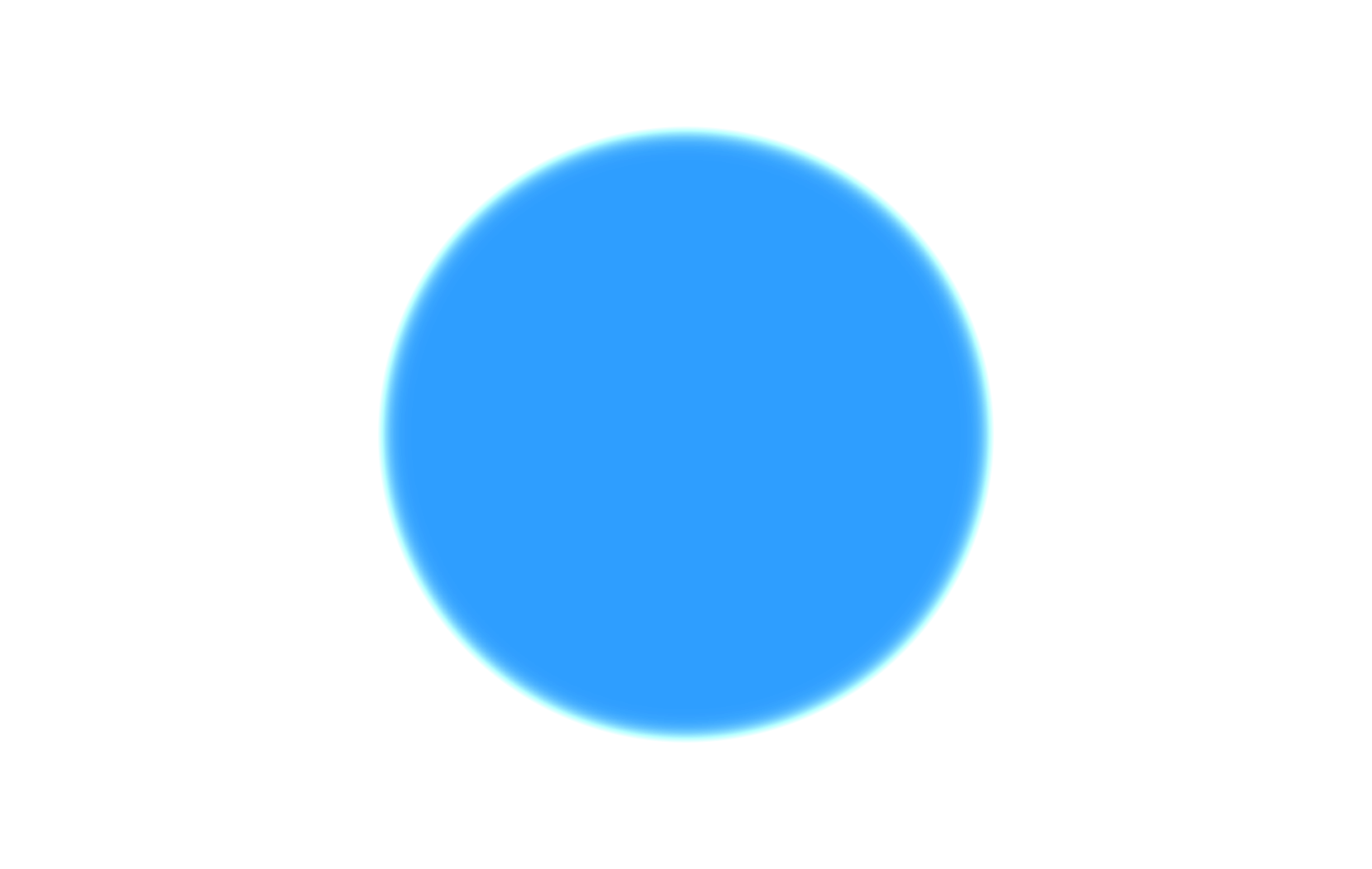} \\
        \hline
        t4 &
        \includegraphics[width=\linewidth,trim=300 170 300 170,clip]{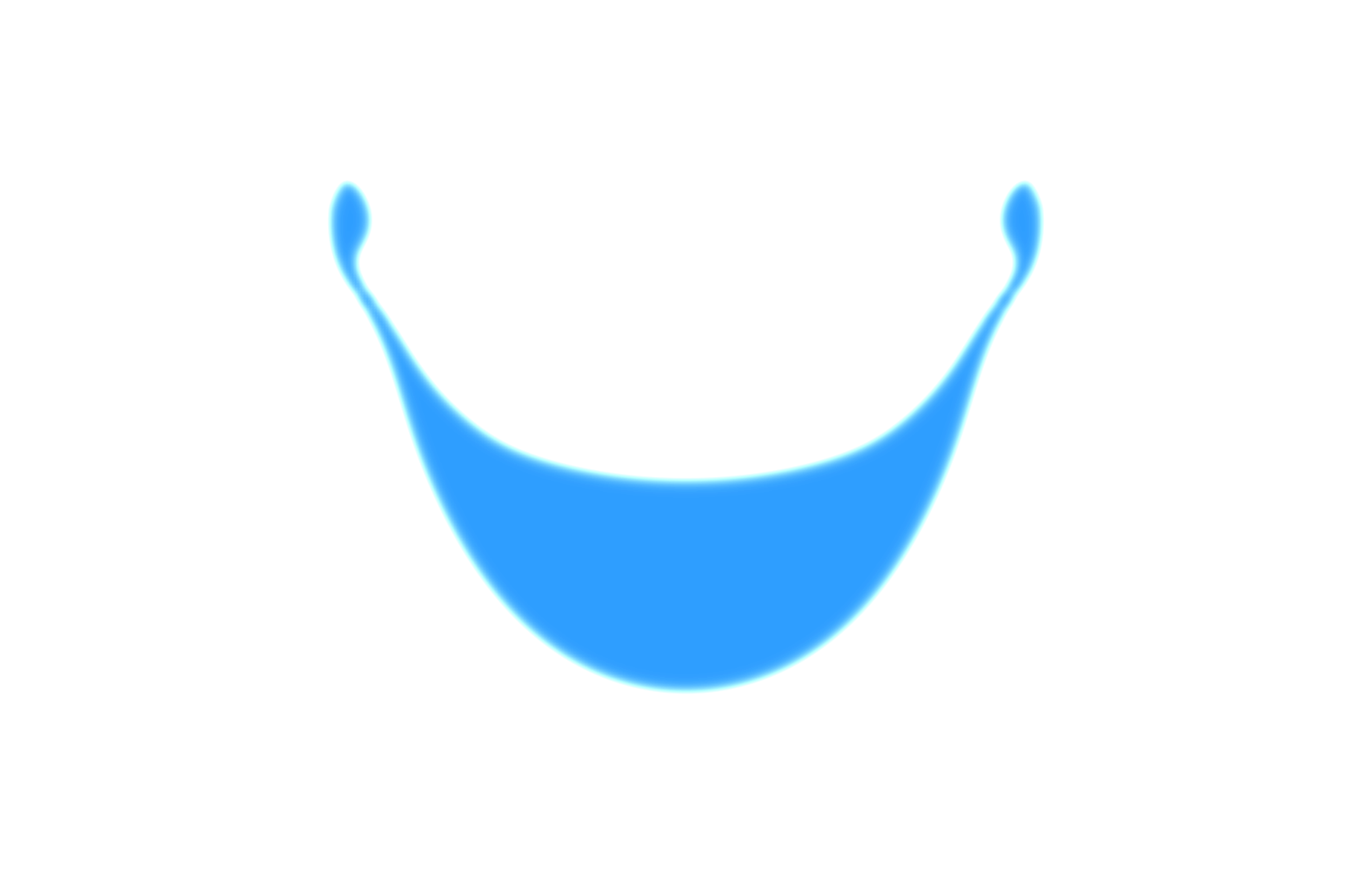} &
        \includegraphics[width=\linewidth,trim=300 170 300 170,clip]{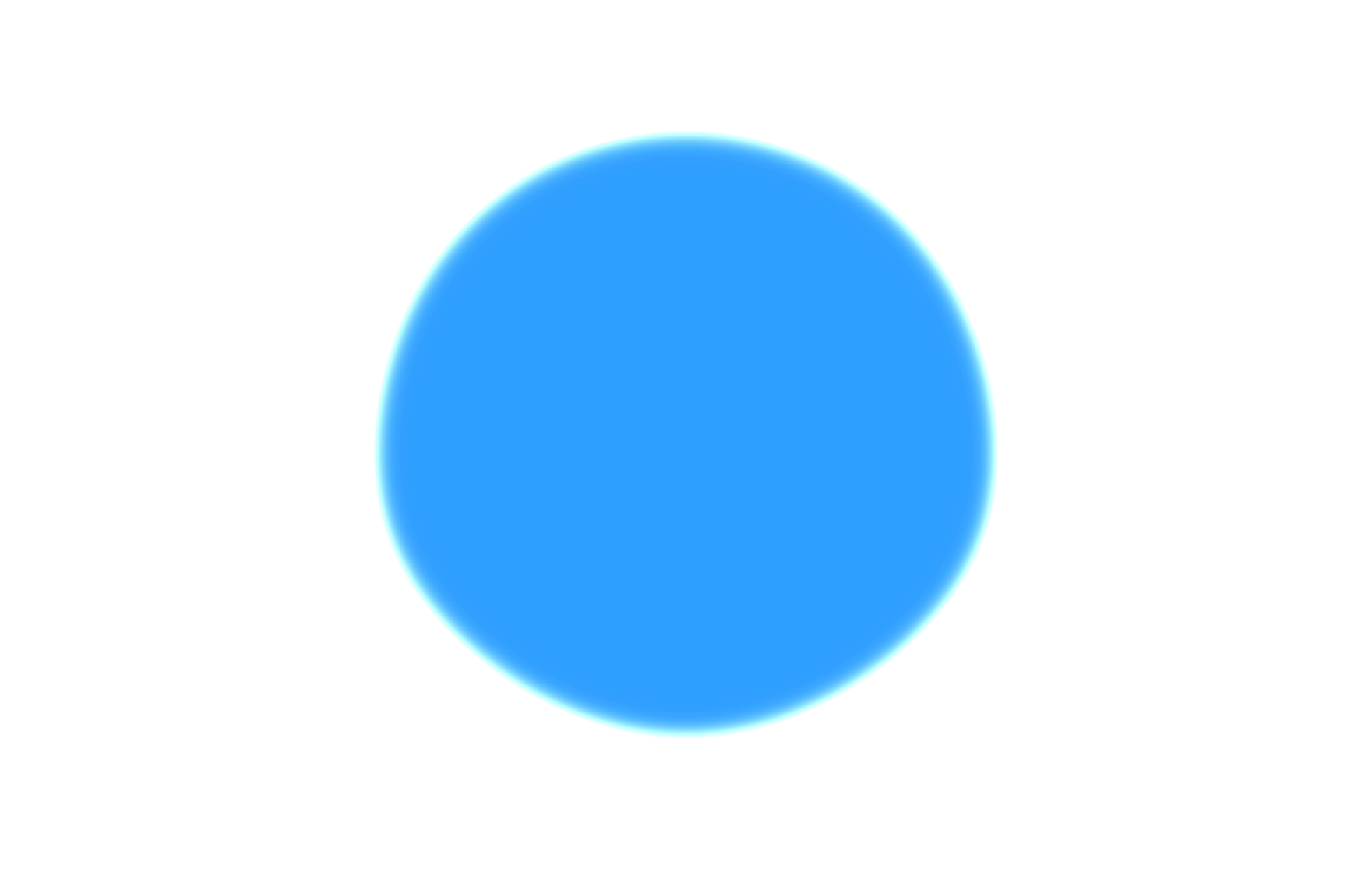} &
        \includegraphics[width=\linewidth,trim=300 170 300 170,clip]{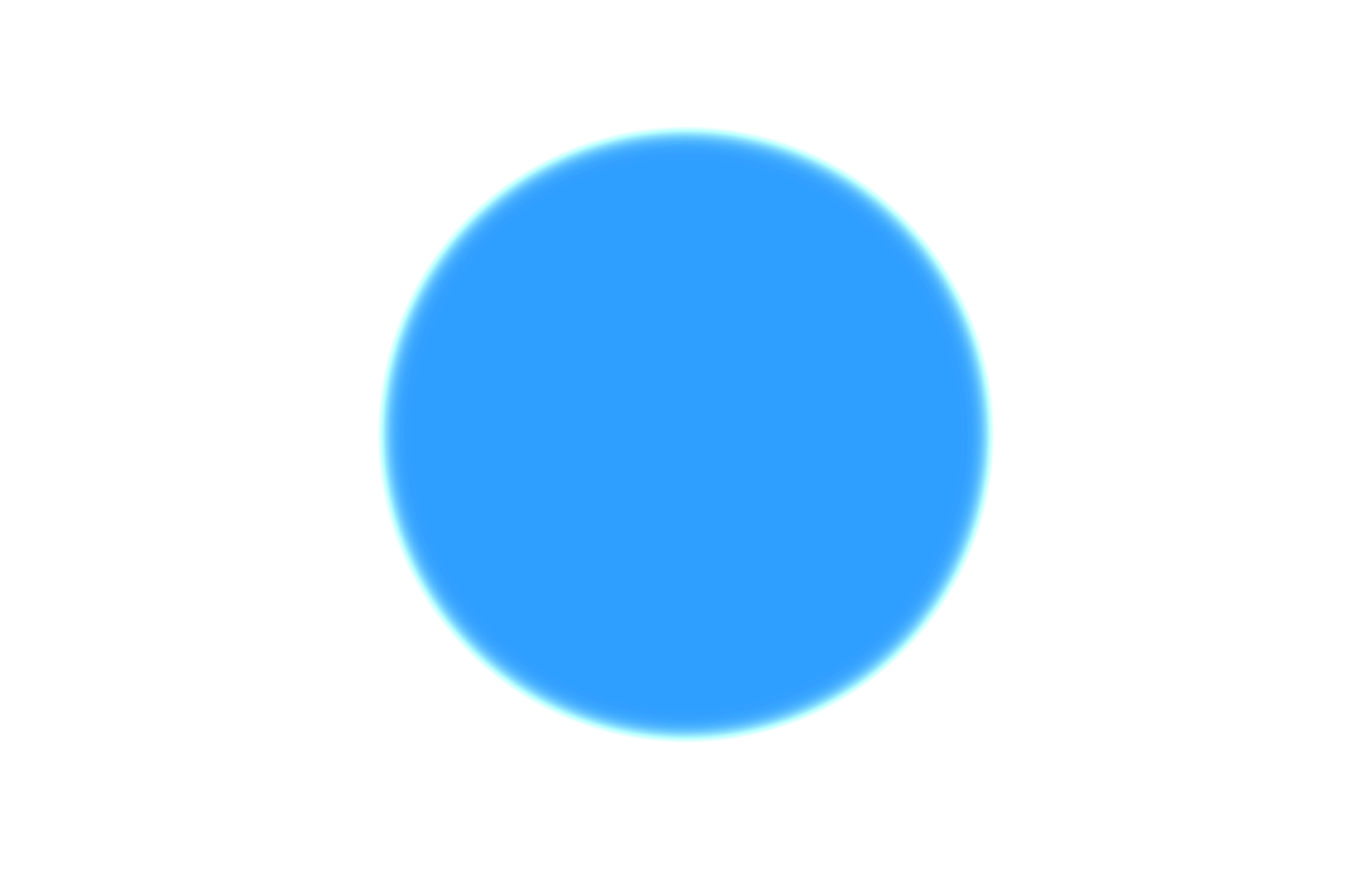} \\
        \hline
        t5 &
        \includegraphics[width=\linewidth,trim=300 170 300 170,clip]{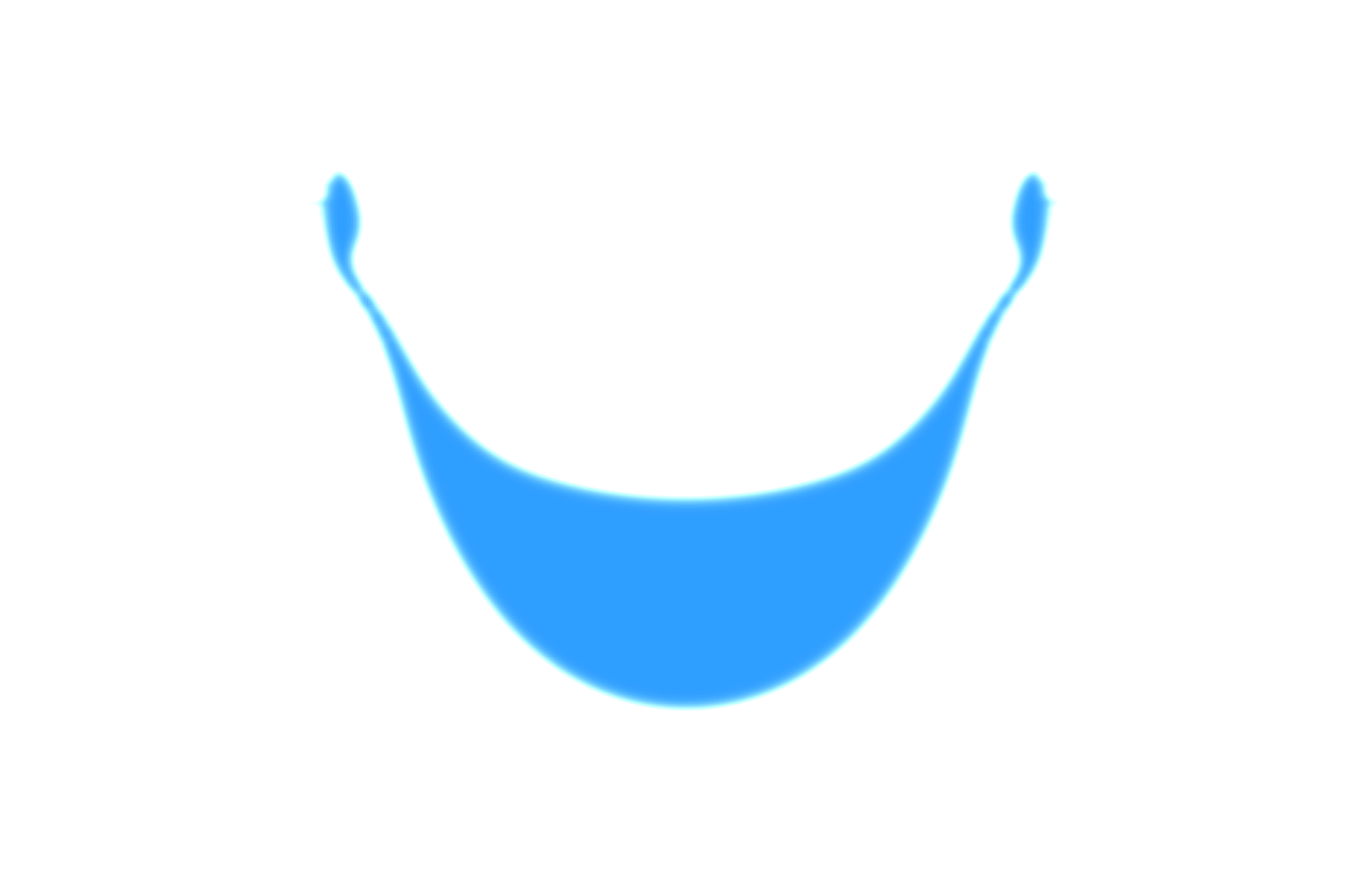} &
        \includegraphics[width=\linewidth,trim=300 170 300 170,clip]{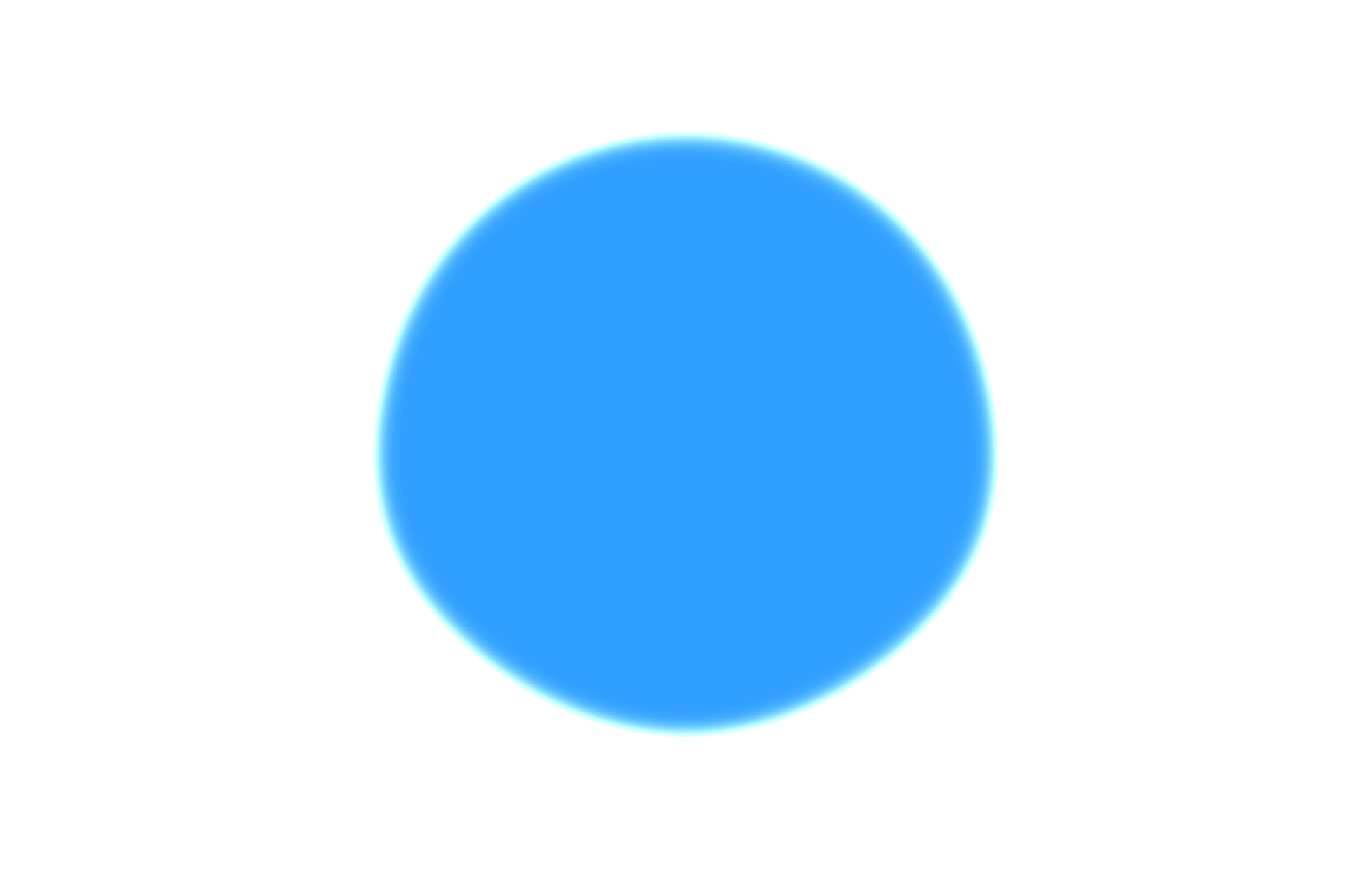} &
        \includegraphics[width=\linewidth,trim=300 170 300 170,clip]{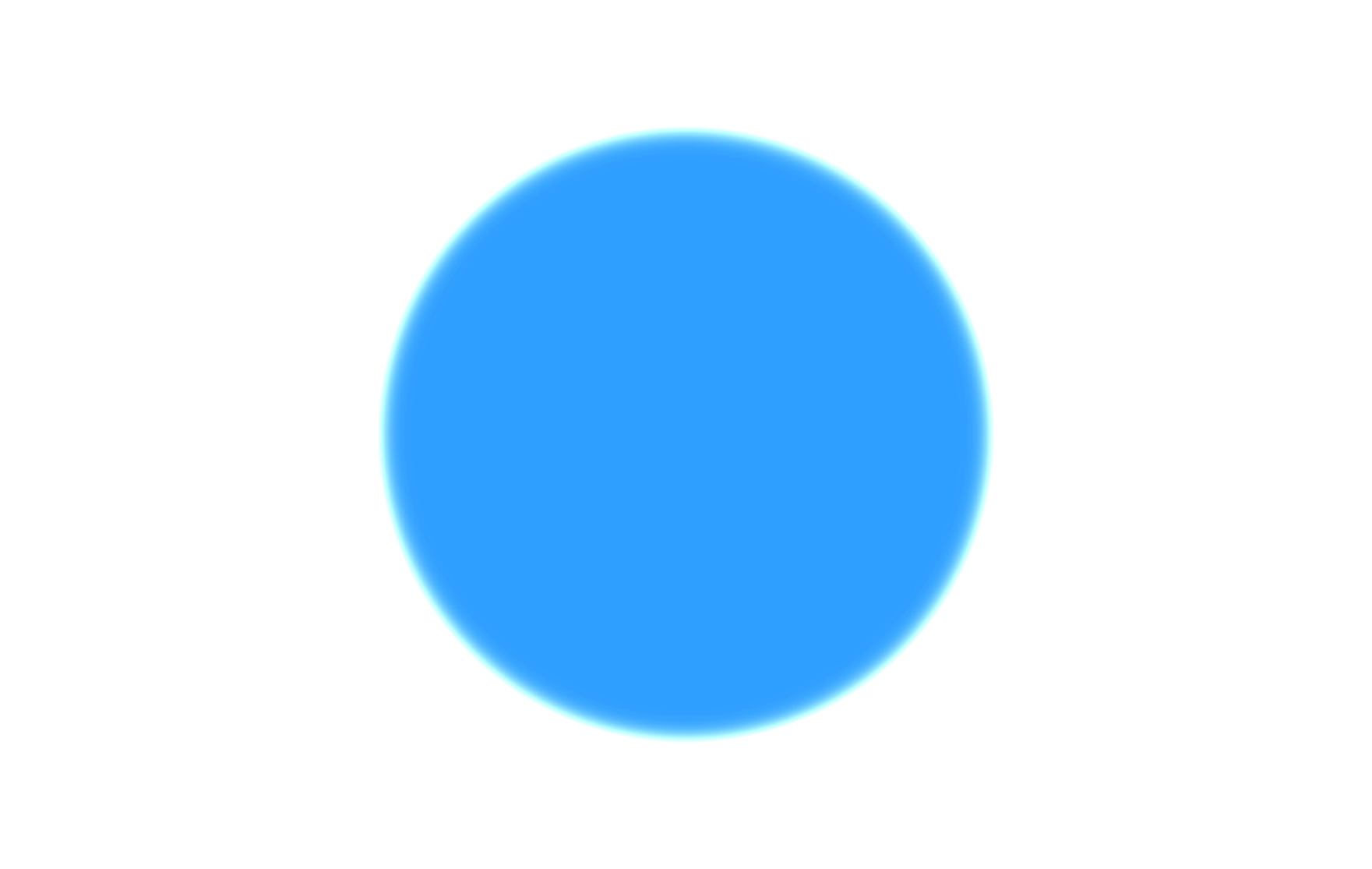} \\
        \hline
        \end{tabular}
        \caption{Falling Droplet}
        \label{2D_contours_drop}
    \end{subfigure}
    \caption{(a) Snapshot of a 2D rising bubble and (b) snapshot of a 2D falling droplet. The properties of the fluids for each case are detailed in \tabref{tab:properties}.\label{tab:2D-dropping}}
\label{2D_contours}
\end{figure}

\begin{figure}[t!]
    \centering

    \begin{subfigure}[t]{0.45\textwidth}
        \centering
        \includegraphics[width=\linewidth,trim=0 30 300 0 ,clip]{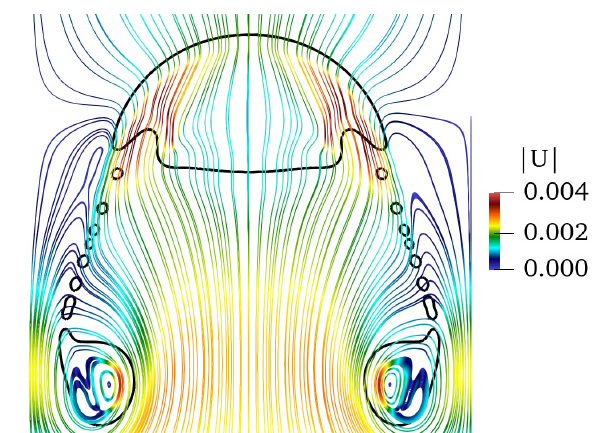} 
        \caption{Rising Bubble (B1)}
    \end{subfigure}
    \hspace{0.1in}
    \begin{subfigure}[t]{0.45\textwidth}
        \centering
        \includegraphics[width=\linewidth,trim=300 30 0 0 ,clip]{Figures/2D/S-2D.pdf} 
        \caption{Falling Droplet (D1)}
    \end{subfigure}    
    \caption{Streamlines of a 2D rising bubble (a) and a 2D falling droplet (b), with colors indicating the magnitude of velocity.    \label{tab:2D-LIC}}
\end{figure}

We selected a few representative cases from our dataset to illustrate the key physics of droplet and bubble dynamics (see \tabref{tab:properties}). As shown in \figureref{fig:3D-shape} and \figureref{tab:2D-dropping}, these cases highlight how variations in Bond number, Reynolds number, and density ratio affect droplet deformation and breakup patterns. Each case reveals distinct fluid behaviors, enhancing our understanding of the complex, nonlinear dynamics. The streamlines around the bubble and droplet, depicted in \figureref{tab:3D-LIC} and \figureref{tab:2D-LIC}, further illustrate how these physical parameters influence droplet breakup and stability across 3D and 2D flows."

\begin{table} [ht]
    \centering
    \vspace{0.2in}
    \caption{Material properties and nondimensional numbers of three bubble rise simulations (B1, B2, B3) and three droplet fall simulations (D1, D2, D3). The table shows the density ratio, viscosity ratio, Reynolds number, and Bond number of all six simulations.}
    \label{tab:properties}
\setlength\extrarowheight{3pt}
\begin{tabular}{cccccccc} 
    \hline
      case & B1 & B2 & B3 & D1 & D2 & D3 \\
    \hline
    Density Ratio ($\rho^*$) & $10^3$ & $10^3$ & $10^3$ & $10$ & $10^3$ & $10^3$ & \\
    Viscosity Ratio ($\mu^*$) & $10^2$ & $10^2$ & $10^2$ & $1$ & $10^2$ & $10^2$ & \\
    $Re$ & $5 \times 10^2$ & $10$ & $10$ & $10^3$ & $10^3$ & $10$ & \\
    $Bo$ & $5 \times 10^2$ & $10$ & $5 \times 10^2$ & $5 \times 10^2$ & $5 \times 10^2$ & $10$ & \\
    \hline
\end{tabular}
\end{table}

% \begin{figure}[ht]
%     \centering
%     \label{tab:2D-LIC}
    
%     \begin{subfigure}[t]{0.49\textwidth}
%         \centering
%         \includegraphics[width=\linewidth,trim=600 450 600 200,clip]{Figures/2D/Rising/B1-LIC.png} 
%         \caption{Bubble rising (B1)}
%     \end{subfigure}
%     \begin{subfigure}[t]{0.49\textwidth}
%         \centering
%         \includegraphics[width=\linewidth,trim=600 50 600 600,clip]{Figures/2D/Dropping/D1-LIC.png} 
%         \caption{Bubble dropping (D1)}
%     \end{subfigure}

%     \caption{2D bubble deformations with streamlines.}
% \end{figure}

\subsection{Simulation Framework and Compute Effort} \label{sec:compute-effort}
Our simulation framework employs a highly parallel, in-house Lattice Boltzmann code, utilizing one of the most accurate two-phase models, the phase field model, to capture the complexities of the interface. The code has been rigorously tested across various problems, with validation results provided in \sectionref{subsec:validations}. For 2D simulations, we used a uniform lattice grid of $256 \times 512$, while for 3D simulations, the domain was set to $ 128\times 256 \times 128$. We achieved high parallelization by distributing the computation across 12 Nvidia A100-SXM4 80GB GPUs. The total computational cost for 2D and 3D cases was approximately 4,000 GPU hours. We use the ParaView tool \citep{Paraview2015} to visualize and understand our dataset.

\subsection{MetaData}

\textbf{Input Fields}: We have provisioned the following dimensionless quantities as inputs to our study as defined in \sectionref{subsec:problem}. These are the Density Ratio ($\rho^*$), Viscosity Ratio ($\mu^*$), Bond Number ($Bo$), and Reynolds Number ($Re$). Since these are scalar values, we feed them to the neural network by creating a constant field with a dimension consistent with the number of samples, in this case, 10,000 in 2D and 1000 in 3D.

\textbf{Output Fields}: In analyzing multiphase flow problems, we are interested in solving the governing PDEs to obtain solutions at every point in the domain's interior for certain cardinal fields. For a 2D solution domain, these are: $c$ - interface indicator, $u$ - velocity in $x$ direction, $v$ - velocity in $y$ direction, $p$ - pressure. Additionally, because this is a time-dependent problem, we have these cardinal fields or a sequence of these fields distributed uniformly over time (100 time steps).

\textbf{Resolution}: We maintained the original resolution of our datasets, matching the Lattice Boltzmann simulation domain. This ensures the complete physics is presented to the Neural Operator and allows direct comparison with Lattice Boltzmann method simulations. Our datasets are published at $256\times512$ resolution for 2D and $128\times256\times128$ for 3D simulations.

\textbf{Dataset Format}: For both the bubble and droplet datasets, we have released a single file for each sample. This decision was taken with the view to allow for maximum flexibility to the end user in deciding what and how many time steps they want to use to train their models, as these time-dependent problems often take the shape of sequence to sequence formulations. In 2D, the resulting \texttt{.npz} files take the form:
{\small
\[
    [\mathbf{number\_of\_time\_steps}][\mathbf{number\_of\_channels}][\mathbf{resolution\_y}][\mathbf{resolution\_x}]
\]
}whereas in 3D, \([\mathbf{resolution\_z}]\) incorporated as an additional dimension for depth. In this study, we have released a total of 11,000 samples spread across two families of datasets. \tabref{tab:dataset} provides a detailed formulaic description of the packaging of the input and output \texttt{numpy} tensors for both these families:

\textbf{Level of Difficulty}: We provide \tabref{tab:difficulty} to help users select datasets based on varying difficulty levels. The dataset includes key parameters like Reynolds number (Re), Density Ratio, Viscosity Ratio, and Bond Number (Bo), with a difficulty classification to guide users. This classification reflects the complexity of interface deformations, making it easier to choose suitable cases for model training and evaluation.

\begin{table}[t!]
\centering
\caption{\label{tab:dataset} Formulaic description of the input and output tensors. 5000 - sample size for the dataset. 101 - number of time steps in the simulation. $x,y$ - The $x,y$ dimension of a field. E.g., $Y[0,100,1,:,:]$ indicates the pointwise $v$ velocity over the entire grid of size $256\times512$ for the first sample at time step 100.}
\setlength\extrarowheight{3pt}
\begin{tabular}{ cccc} 
\hline
\textbf{Dataset} & \textbf{Dim.} &\textbf{Input Tensor} & \textbf{Output Tensor} \\
\hline
Droplet & 2 & $X[5000][\rho^*, \mu^*, Bo, Re]$ & $Y[5000][101][c,u,v,p,\rho][y][x]$\\
Bubble & 2 & $X[5000][\rho^*, \mu^*, Bo, Re]$ & $Y[5000][101][c,u,v,p,\rho][y][x]$\\
Droplet & 3 & $X[500][\rho^*, \mu^*, Bo, Re]$ & $Y[500][51][c,u,v,w,p,\rho][z][y][x]$\\
Bubble & 3 & $X[500][\rho^*, \mu^*, Bo, Re]$ & $Y[500][51][c,u,v,w,p,\rho][z][y][x]$\\
\hline
\end{tabular}
\end{table}

\begin{table}[t!]
\centering
\caption{\label{tab:difficulty} Dataset parameters with difficulty levels for selecting appropriate cases based on Reynolds number, Density Ratio, Viscosity Ratio, and Bond Number.}
\begin{tabular}{lllll}
\toprule
\textbf{Density Ratio} & \textbf{Viscosity Ratio} & \textbf{$Bo$ Number} & \textbf{$Re$ Number} & \textbf{Difficulty} \\ 
\midrule
High  & High  & High  & High   & Challenging \\ 
High  & High  & Low   & Low    & Easy        \\ 
High  & High  & Low   & Low    & Moderate    \\ 
High  & High  & Low   & High   & Moderate    \\ 
High  & High  & High  & High   & Challenging \\ 
Low   & Low   & High  & High   & Challenging \\ 
Low   & Low   & Low   & Low    & Easy        \\ 
Low   & Low   & High  & High   & Easy        \\ 
Low   & Low   & Low   & High   & Moderate    \\ 
\bottomrule
\end{tabular}
\end{table}

% \begin{itemize}
%     \item \textcolor{red}{We might have 3D results (perhaps 1000 or 2000 samples).}

%     \item \textcolor{red}{output fields, resolution, examples: use chatgpt for paraphrasing from Flowbench + edit for two-phase flows.}
% \end{itemize}

\subsection{Evaluation Metrics and Test Dataset Analysis}

We assess the performance of the trained neural operators and foundation models using two primary metrics: Mean Squared Error (MSE) and relative $L_2$ error. Our models are trained on a random selection of 1000 samples from the bubble dataset. To model the transient nature of the data, we employ sequence-to-sequence and sequence-to-field mappings, where the solution fields at various time steps are concatenated and fed sequentially as input into the models.

The dataset consists of 100 temporal snapshots of bubble dynamics. However, to capture stronger temporal variations, we choose every fourth-time step, thereby reducing our time series to 25 snapshots for each case. For instance, $t3$ in our evaluation aligns with the 12th time step in the original 100-snapshot dataset. This approach ensures the input sequences contain larger changes in the bubble’s shape and velocity fields, thereby increasing the learning challenge for the machine learning models. By skipping every fourth timesteps, we ensure the models learn from substantial temporal variations rather than minor changes, ultimately making the learning problem more challenging for capturing multi-phase flow dynamics.

The models are evaluated on six distinct input-output mappings ($S1$ through $S6$) as outlined below:

\begin{itemize}[itemsep=0pt, leftmargin=*]
    \item \textbf{Sequence-to-field:} We set up three different input scenarios for subsets $S1$, $S2$, and $S3$. In these cases, the goal is to predict a single future time step given past information. The input consists of solution fields at:
    \begin{itemize}
        \item $t1$, which represents the first time snapshot in the reduced dataset.
        \item A short sequence from $t1$ to $t3$.
        \item A longer sequence from $t1$ to $t5$.
    \end{itemize}
    
    For each case, the model predicts the immediate next timestep in the reduced dataset, i.e., $t2$ for $S1$, $t4$ for $S2$, and $t6$ for $S3$.

    \item \textbf{Sequence-to-sequence:} In this case, the output is not a single time snapshot but a sequence of solutions. The goal is to predict multiple future time steps, capturing the evolving dynamics of the bubble rise. We input solution sequences of:
    \begin{itemize}
        \item $t1$ to $t3$, corresponding to the first three selected snapshots.
        \item $t1$ to $t5$, an extended sequence covering more temporal context.
        \item $t1$ to $t8$, further increasing the sequence length to capture longer-term dependencies.
    \end{itemize}
    For each case, the model predicts a sequence of three future timesteps. Specifically, the outputs for $S4$, $S5$, and $S6$ are the next three consecutive time snapshots after the given input sequence.
\end{itemize}

 This evaluation framework allows us to test the models' ability to generalize across different temporal dependencies ranging from short-term extrapolations in sequence-to-field tasks to longer-term sequence predictions in sequence-to-sequence tasks.

 The dataset used for these tasks ($S1$--$S6$) is available at \href{https://huggingface.co/datasets/BGLab/mpf-bench/tree/main/2Dbubble/mpf_paper_dataset}{HuggingFace} under \texttt{2Dbubble/mpf\_paper\_dataset}. This folder contains the \texttt{d\#X\_train\_pad\_flat.npz}, \texttt{d\#Y\_train\_pad\_flat.npz}, \texttt{d\#X\_test\_pad\_flat.npz}, and \texttt{d\#Y\_test\_pad\_flat.npz} files corresponding to each scenario from $S1$ through $S6$.

\subsection{Dataset Format}
\label{subsec:datasetformat}

We store all training and testing data for the six input-output mappings (\(S1\) to \(S6\)) in \texttt{.npz} files. Each \texttt{.npz} file contains a single NumPy array named \texttt{Y} of shape \((N, C, H, W)\), where:

\begin{itemize}
    \item \(N\) is the number of samples (800 for training, 200 for testing).
    \item \(C\) is the total number of channels, determined by the number of time snapshots \(\times\) physical fields. For instance, in a sequence-to-sequence task with 8 input snapshots (each containing 4 fields), \(C = 8 \times 4 = 32.\)
    \item \(H\) and \(W\) are the spatial dimensions, fixed at 256\(\times\)256 for these simulations.
\end{itemize}

For example, \texttt{d6X\_train\_pad\_flat.npz} has shape \((800, 32, 256, 256)\) because \(S6\) requires an 8-snapshot input (each snapshot with 4 channels). Similarly, the corresponding output \texttt{d6Y\_train\_pad\_flat.npz} has shape \((800, 12, 256, 256)\), where 12 channels correspond to 3 future snapshots \(\times\) 4 fields each. The same mapping applies to \texttt{d1X\_train\_pad\_flat.npz}, \texttt{d1Y\_train\_pad\_flat.npz}, etc., with varying numbers of channels depending on how many time snapshots each mapping \((S1\) to \(S6)\) requires.

\section{Experiments} \label{sec:experiments}

Neural operators represent a novel class of deep learning designed to learn function spaces of solutions to partial differential equations (PDEs). Unlike traditional deep learning methods which focus on finding a parametric solution for a fixed problem, Neural Operators are capable of generalizing to solutions of PDEs. While these frameworks have demonstrated notable success in modeling single-phase fluid flow, there is limited research on their accuracy in capturing multi-phase flows. Multi-phase flows present additional challenges due to phenomena like bubble or droplet breakup, coalescence, and shape oscillations. In this context, we aim to evaluate the performance of several Neural Operators and foundation models in learning these intricate fluid dynamics.

We provide baseline results by training a range of well-established neural PDE solvers, commonly referred to as scientific machine learning (SciML) models. Specifically, we examined the following neural operators and transformer-based foundation models on the 2D bubble benchmark: (a) Fourier Neural Operator (FNO) \citep{li2021}, (b) Convolutional Neural Operators (CNO) \citep{raonic2023}, (c) DeepONet \citep{lu2020}, (d) UNet \citep{ronneberger2015}, (e) scOT (a randomly initialized vision-based transformer), (f) Poseidon (a pre-trained, vision-based foundation model) \citep{herde2024poseidon}. For training, we closely followed the published code examples, and the code used for training these models is available in the following GitHub repository: \href{https://github.com/baskargroup/GeometryMatters/}{GitHub}. All models were trained for 200 epochs on a single A100 80GB GPU using the Adam optimizer, and the hyperparameter settings for each model are detailed in \sectionref{subsec:model-hyperparams}. The validation loss for all models converged by 200 epochs, and the training and validation loss curves for two representative models are shown in \sectionref{subsec:loss-plots}.

\begin{table}[t!]
\centering
\caption{Comparison of mean squared error MSE and relative \(L_2\) Error for Sequence-to-Field predictions using UNet, DeepONet, FNO, CNO, scOT, and Poseidon on Bubble Datasets (S1-S3).}
\begin{scriptsize} 
\begin{tabular}{|c|c|cc|cc|cc|}
\hline
\multirow{2}{*}{Model} & \multirow{2}{*}{Channel} & \multicolumn{2}{c|}{S1} & \multicolumn{2}{c|}{S2} & \multicolumn{2}{c|}{S3} \\ \cline{3-8} 
                       &                          & MSE  & \(L_2\) & MSE  & \(L_2\) & MSE  & \(L_2\) \\ \hline
\multirow{4}{*}{UNet}  
                       & \(c\)                    & $\mathbf{2.60 \times 10^{-2}}$  & $\mathbf{2.59 \times 10^{-2}}$  & \(8.40 \times 10^{-3}\)  & \(8.07 \times 10^{-3}\)  & \(9.56 \times 10^{-3}\)  & \(9.04 \times 10^{-3}\)  \\ \cline{2-8} 
                       & \(u\)                    & \(8.80 \times 10^{-5}\)  & \(3.13 \times 10^{0}\)  & \(8.00 \times 10^{-6}\)  & \(2.04 \times 10^{0}\)  & \(1.00 \times 10^{-5}\)  & \(2.09 \times 10^{0}\)  \\ \cline{2-8} 
                       & \(v\)                    & \(1.00 \times 10^{-6}\)  & \(1.56 \times 10^{0}\)  & \(1.00 \times 10^{-6}\)  & \(7.84 \times 10^{-1}\)  & \(1.00 \times 10^{-6}\)  & \(8.76 \times 10^{-1}\)  \\ \cline{2-8} 
                       & \(p\)                    & \(1.00 \times 10^{-6}\)  & \(2.74 \times 10^{2}\)  & \(1.00 \times 10^{-6}\)  & \(3.23 \times 10^{2}\)  & \(1.00 \times 10^{-6}\)  & \(5.17 \times 10^{2}\)  \\ \hline

\multirow{4}{*}{DeepONet}  
                       & \(c\)                    & \(2.66 \times 10^{-2}\)  & \(2.65 \times 10^{-2}\)  & \(1.01 \times 10^{-1}\)  & \(1.01 \times 10^{-1}\)  & \(1.18 \times 10^{-1}\)  & \(1.18 \times 10^{-1}\)  \\ \cline{2-8} 
                       & \(u\)                    & \(9.10 \times 10^{-5}\)  & \(6.71 \times 10^{0}\)  & \(1.27 \times 10^{-3}\)  & \(1.24 \times 10^{0}\)  & \(1.71 \times 10^{-3}\)  & \(9.18 \times 10^{-1}\)  \\ \cline{2-8} 
                       & \(v\)                    & \(1.00 \times 10^{-6}\)  & \(8.68 \times 10^{-1}\)  & \(1.00 \times 10^{-6}\)  & \(5.35 \times 10^{-1}\)  & \(1.00 \times 10^{-6}\)  & \(8.89 \times 10^{-1}\)  \\ \cline{2-8} 
                       & \(p\)                    & \(1.00 \times 10^{-6}\)  & \(1.89 \times 10^{2}\)  & \(1.00 \times 10^{-6}\)  & \(2.43 \times 10^{2}\)  & \(1.00 \times 10^{-6}\)  & \(1.66 \times 10^{2}\)  \\ \hline

\multirow{4}{*}{FNO}    
                       & \(c\)                    & \(2.72 \times 10^{-2}\)  & \(2.68 \times 10^{-2}\)  & \(9.73 \times 10^{-3}\)  & \(8.97 \times 10^{-3}\)  & \(2.10 \times 10^{-2}\)  & \(1.98 \times 10^{-2}\)  \\ \cline{2-8} 
                       & \(u\)                    & \(9.30 \times 10^{-5}\)  & \(8.56 \times 10^{0}\)  & \(1.00 \times 10^{-5}\)  & \(2.77 \times 10^{0}\)  & \(4.80 \times 10^{-5}\)  & \(5.29 \times 10^{0}\)  \\ \cline{2-8} 
                       & \(v\)                    & \(1.00 \times 10^{-6}\)  & \(3.43 \times 10^{0}\)  & \(1.00 \times 10^{-6}\)  & \(1.09 \times 10^{0}\)  & \(2.00 \times 10^{-6}\)  & \(2.02 \times 10^{0}\)  \\ \cline{2-8} 
                       & \(p\)                    & \(1.00 \times 10^{-6}\)  & \(7.18 \times 10^{2}\)  & \(1.00 \times 10^{-6}\)  & \(7.44 \times 10^{2}\)  & \(2.00 \times 10^{-6}\)  & \(1.04 \times 10^{3}\)  \\ \hline

\multirow{4}{*}{CNO}   
                       & \(c\)                    & \(2.62 \times 10^{-2}\)  & \(2.60 \times 10^{-1}\)  & $\mathbf{5.89 \times 10^{-3}}$  & $\mathbf{5.65 \times 10^{-3}}$  & $\mathbf{9.41 \times 10^{-3}}$  & $\mathbf{9.00 \times 10^{-3}}$  \\ \cline{2-8} 
                       & \(u\)                    & \(8.80 \times 10^{-5}\)  & \(5.04 \times 10^{0}\)  & \(4.00 \times 10^{-6}\)  & \(1.60 \times 10^{0}\)  & \(1.00 \times 10^{-5}\)  & \(1.79 \times 10^{0}\)  \\ \cline{2-8} 
                       & \(v\)                    & \(1.00 \times 10^{-6}\)  & \(1.73 \times 10^{0}\)  & \(1.00 \times 10^{-6}\)  & \(5.07 \times 10^{-1}\)  & \(1.00 \times 10^{-6}\)  & \(9.19 \times 10^{-1}\)  \\ \cline{2-8} 
                       & \(p\)                    & \(1.00 \times 10^{-6}\)  & \(4.46 \times 10^{2}\)  & \(1.00 \times 10^{-6}\)  & \(2.24 \times 10^{2}\)  & \(1.00 \times 10^{-6}\)  & \(3.96 \times 10^{2}\)  \\ \hline

\multirow{4}{*}{scOT}   
                       & \(c\)                    & \(2.76 \times 10^{-2}\)  & \(2.68 \times 10^{-2}\)  & \(1.77 \times 10^{-2}\)  & \(1.68 \times 10^{-2}\)  & \(2.23 \times 10^{-2}\)  & \(2.17 \times 10^{-2}\)   \\ \cline{2-8} 
                       & \(u\)                    & \(1.29 \times 10^{1}\)  & \(1.29 \times 10^{1}\)  & \(3.50 \times 10^{-5}\)  & \(3.91 \times 10^{0}\)  & \(5.80 \times 10^{-5}\)  & \(2.82 \times 10^{0}\)   \\ \cline{2-8} 
                       & \(v\)                    & \(5.24 \times 10^{0}\)  & \(5.24 \times 10^{0}\)  & \(1.00 \times 10^{-6}\)  & \(2.31 \times 10^{0}\)  & \(1.00 \times 10^{-6}\)  & \(1.75 \times 10^{0}\)   \\ \cline{2-8} 
                       & \(p\)                    & \(9.65 \times 10^{2}\)  & \(9.65 \times 10^{2}\)  & \(2.00 \times 10^{-6}\)  & \(8.12 \times 10^{2}\)  & \(2.00 \times 10^{-6}\)  & \(7.80 \times 10^{2}\)   \\ \hline

\multirow{4}{*}{Poseidon}   
                       & \(c\)                    & \(2.87 \times 10^{-2}\)  & \(2.79 \times 10^{-1}\)  & \(3.34 \times 10^{-2}\)  & \(3.01 \times 10^{-2}\)  & \(2.49 \times 10^{-2}\)  & \(2.23 \times 10^{-2}\) \\ \cline{2-8} 
                       & \(u\)                    & \(1.00 \times 10^{-4}\)  & \(1.16 \times 10^{1}\)  & \(1.14 \times 10^{-4}\)  & \(1.31 \times 10^{1}\)  & \(6.10 \times 10^{-5}\)  & \(8.28 \times 10^{0}\) \\ \cline{2-8} 
                       & \(v\)                    & \(2.00 \times 10^{-6}\)  & \(5.86 \times 10^{0}\)  & \(1.10 \times 10^{-5}\)  & \(4.05 \times 10^{0}\)  & \(5.00 \times 10^{-6}\)  & \(2.83 \times 10^{0}\)  \\ \cline{2-8} 
                       & \(p\)                    & \(2.00 \times 10^{-6}\)  & \(1.04 \times 10^{3}\)  & \(6.00 \times 10^{-6}\)  & \(2.35 \times 10^{3}\)  & \(4.00 \times 10^{-6}\)  & \(1.94 \times 10^{3}\)  \\ \hline

\end{tabular}
\end{scriptsize}
\label{tab:seq2field_s1_s3}
\end{table}

\begin{table}[t!]
\centering
\caption{Comparison of mean squared error MSE and relative \(L_2\) Error for Sequence-to-Sequence predictions using UNet, DeepONet, FNO, CNO, scOT, and Poseidon on Bubble Datasets (S4-S6).}
\begin{scriptsize} 
\begin{tabular}{|c|c|cc|cc|cc|}
\hline
\multirow{2}{*}{Model} & \multirow{2}{*}{Channel} & \multicolumn{2}{c|}{S4} & \multicolumn{2}{c|}{S5} & \multicolumn{2}{c|}{S6} \\ \cline{3-8} 
                       &                          & MSE  & \(L_2\) & MSE  & \(L_2\) & MSE  & \(L_2\) \\ \hline
\multirow{4}{*}{UNet}  
                       & \(c\)                    & \(3.27 \times 10^{-2}\)  & \(2.87 \times 10^{-2}\)  & \(3.93 \times 10^{-2}\)  & \(3.31 \times 10^{-2}\)  & \(7.34 \times 10^{-2}\)  & \(6.58 \times 10^{-2}\)  \\ \cline{2-8} 
                       & \(u\)                    & \(1.35 \times 10^{-4}\)  & \(3.54 \times 10^{0}\)  & \(2.02 \times 10^{-4}\)  & \(3.10 \times 10^{0}\)  & \(6.65 \times 10^{-4}\)  & \(3.90 \times 10^{0}\)  \\ \cline{2-8} 
                       & \(v\)                    & \(1.00 \times 10^{-6}\)  & \(1.38 \times 10^{0}\)  & \(1.00 \times 10^{-6}\)  & \(1.45 \times 10^{0}\)  & \(1.00 \times 10^{-6}\)  & \(9.54 \times 10^{-1}\)  \\ \cline{2-8} 
                       & \(p\)                    & \(1.00 \times 10^{-6}\)  & \(6.76 \times 10^{2}\)  & \(1.00 \times 10^{-6}\)  & \(6.30 \times 10^{2}\)  & \(1.00 \times 10^{-6}\)  & \(7.79 \times 10^{2}\)  \\ \hline

\multirow{4}{*}{DeepONet}  
                       & \(c\)                    & \(1.64 \times 10^{-1}\)  & \(1.60 \times 10^{-1}\)  & \(2.03 \times 10^{-1}\)  & \(1.99 \times 10^{-1}\)  & \(1.94 \times 10^{-1}\)  & \(1.92 \times 10^{-1}\)  \\ \cline{2-8} 
                       & \(u\)                    & \(3.33 \times 10^{-3}\)  & \(3.15 \times 10^{0}\)  & \(5.04 \times 10^{-3}\)  & \(1.64 \times 10^{0}\)  & \(4.57 \times 10^{-3}\)  & \(1.32 \times 10^{0}\)  \\ \cline{2-8} 
                       & \(v\)                    & \(1.00 \times 10^{-6}\)  & \(1.04 \times 10^{0}\)  & \(1.00 \times 10^{-6}\)  & \(7.59 \times 10^{-1}\)  & \(1.00 \times 10^{-6}\)  & \(3.83 \times 10^{-1}\)  \\ \cline{2-8} 
                       & \(p\)                    & \(1.00 \times 10^{-6}\)  & \(7.30 \times 10^{2}\)  & \(1.00 \times 10^{-6}\)  & \(3.53 \times 10^{2}\)  & \(1.00 \times 10^{-6}\)  & \(3.17 \times 10^{2}\)  \\ \hline

\multirow{4}{*}{FNO}    
                       & \(c\)                    & $\mathbf{1.16 \times 10^{-2}}$  & $\mathbf{1.10 \times 10^{-2}}$  & \(2.33 \times 10^{-2}\)  & \(2.20 \times 10^{-2}\)  & \(4.24 \times 10^{-2}\)  & \(4.00 \times 10^{-2}\)  \\ \cline{2-8} 
                       & \(u\)                    & \(1.70 \times 10^{-5}\)  & \(1.00 \times 10^{0}\)  & \(6.50 \times 10^{-5}\)  & \(2.98 \times 10^{0}\)  & \(2.28 \times 10^{-4}\)  & \(1.13 \times 10^{0}\)  \\ \cline{2-8} 
                       & \(v\)                    & \(1.00 \times 10^{-6}\)  & \(4.50 \times 10^{-1}\)  & \(1.00 \times 10^{-6}\)  & \(1.28 \times 10^{0}\)  & \(1.00 \times 10^{-6}\)  & \(3.98 \times 10^{-1}\)  \\ \cline{2-8} 
                       & \(p\)                    & \(1.00 \times 10^{-6}\)  & \(1.38 \times 10^{2}\)  & \(1.00 \times 10^{-6}\)  & \(7.87 \times 10^{2}\)  & \(1.00 \times 10^{-6}\)  & \(2.69 \times 10^{2}\)  \\ \hline

\multirow{4}{*}{CNO}   
                       & \(c\)                    & \(1.74 \times 10^{-2}\)  & \(1.69 \times 10^{-2}\)  & $\mathbf{1.72 \times 10^{-2}}$  & $\mathbf{1.63 \times 10^{-2}}$  & $\mathbf{3.78 \times 10^{-2}}$  & $\mathbf{3.53 \times 10^{-2}}$  \\ \cline{2-8} 
                       & \(u\)                    & \(3.70 \times 10^{-5}\)  & \(1.51 \times 10^{0}\)  & \(3.60 \times 10^{-5}\)  & \(1.37 \times 10^{0}\)  & \(1.76 \times 10^{-4}\)  & \(1.74 \times 10^{0}\)  \\ \cline{2-8} 
                       & \(v\)                    & \(1.00 \times 10^{-6}\)  & \(6.87 \times 10^{-1}\)  & \(1.00 \times 10^{-6}\)  & \(6.85 \times 10^{-1}\)  & \(1.00 \times 10^{-6}\)  & \(5.98 \times 10^{-1}\)  \\ \cline{2-8} 
                       & \(p\)                    & \(1.00 \times 10^{-6}\)  & \(2.84 \times 10^{2}\)  & \(1.00 \times 10^{-6}\)  & \(2.93 \times 10^{2}\)  & \(1.00 \times 10^{-6}\)  & \(4.22 \times 10^{2}\)  \\ \hline

\multirow{4}{*}{scOT}   
                       & \(c\)                    & \(3.85 \times 10^{-2}\)  & \(3.69 \times 10^{-2}\)  & \(3.92 \times 10^{-2}\)  & \(3.79 \times 10^{-2}\)  & \(6.27 \times 10^{-2}\)  & \(6.09 \times 10^{-2}\)   \\ \cline{2-8} 
                       & \(u\)                    & \(1.73 \times 10^{-4}\)  & \(6.73 \times 10^{0}\)  & \(1.80 \times 10^{-4}\)  & \(5.73 \times 10^{0}\)  & \(4.85 \times 10^{-4}\)  & \(5.31 \times 10^{0}\)   \\ \cline{2-8} 
                       & \(v\)                    & \(3.00 \times 10^{-6}\)  & \(2.78 \times 10^{0}\)  & \(3.00 \times 10^{-6}\)  & \(2.12 \times 10^{0}\)  & \(2.00 \times 10^{-6}\)  & \(1.93 \times 10^{0}\)   \\ \cline{2-8} 
                       & \(p\)                    & \(3.00 \times 10^{-6}\)  & \(1.48 \times 10^{3}\)  & \(3.00 \times 10^{-6}\)  & \(1.34 \times 10^{3}\)  & \(4.00 \times 10^{-6}\)  & \(1.21 \times 10^{3}\)   \\ \hline

\multirow{4}{*}{Poseidon}   
                       & \(c\)                    & \(3.06 \times 10^{-2}\)  & \(2.84 \times 10^{-2}\)  & \(3.33 \times 10^{-2}\)  & \(3.17 \times 10^{-2}\)  & \(5.99 \times 10^{-2}\)  & \(5.79 \times 10^{-1}\) \\ \cline{2-8} 
                       & \(u\)                    & \(1.01 \times 10^{-4}\)  & \(8.26 \times 10^{0}\)  & \(1.26 \times 10^{-4}\)  & \(6.49 \times 10^{0}\)  & \(4.30 \times 10^{-4}\)  & \(7.79 \times 10^{0}\) \\ \cline{2-8} 
                       & \(v\)                    & \(5.00 \times 10^{-6}\)  & \(3.51 \times 10^{0}\)  & \(4.00 \times 10^{-6}\)  & \(2.47 \times 10^{0}\)  & \(5.00 \times 10^{-6}\)  & \(2.16 \times 10^{0}\)  \\ \cline{2-8} 
                       & \(p\)                    & \(5.00 \times 10^{-6}\)  & \(1.71 \times 10^{3}\)  & \(4.00 \times 10^{-6}\)  & \(1.48 \times 10^{3}\)  & \(5.00 \times 10^{-6}\)  & \(1.81 \times 10^{3}\)  \\ \hline

\end{tabular}
\end{scriptsize}
\label{tab:seq2seq_s4_s6}
\end{table}

\begin{figure}[t!]
\begin{center}
\includegraphics[width=0.99\linewidth,trim=0 0 0 0,clip]{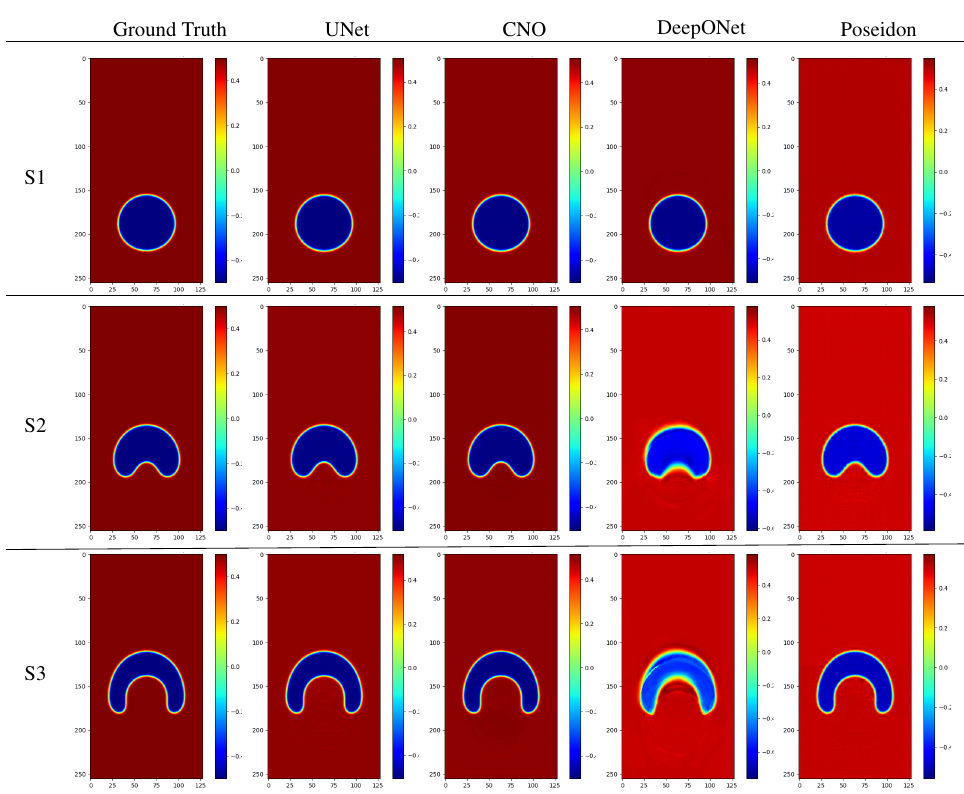} 
\end{center}
\caption{The figure presents a comparison of sequence-to-field predictions for the concentration field \(C\) against the ground truth. The predictions are generated by four models: UNet, Convolutional Neural Operator (CNO), DeepONet, and Poseidon, across three data subsets (S1, S2, and S3). Each row corresponds to a different subset (S1, S2, or S3), while each column displays the predictions made by the respective models.}
\label{fig:S2F}
\end{figure}

\begin{figure}[t!]
\begin{center}
\includegraphics[width=0.99\linewidth,trim=0 0 0 0,clip]{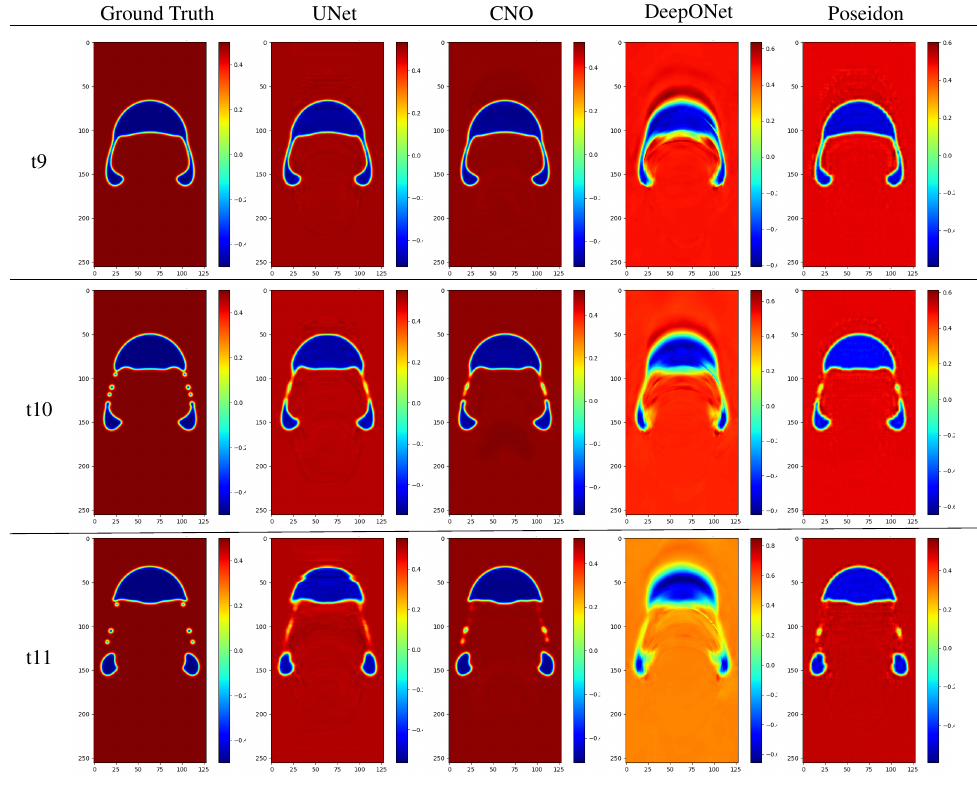} 
\end{center}
\caption{The figure presents sequence-to-sequence predictions for the concentration field \(C\) compared to the ground truth. Predictions are made by four models: UNet, Convolutional Neural Operator (CNO), DeepONet, and Poseidon. Each row represents different time steps (t9, t10, and t11) from dataset S6, while each column shows the predictions from the respective models.}
\label{fig:S2S}
\end{figure}

%Furthermore, \figureref{fig:S2F} and \figureref{fig:S2S} illustrate field predictions of the concentration field $C$ using UNet, CNO, DeepONet, and Poseidon for sequence-to-field and sequence-to-sequence scenarios, respectively. These figures show that DeepONet performs poorly in both the sequence-to-field and sequence-to-sequence scenarios. Also, UNet's accuracy declines as the prediction horizon extends to longer time sequences, as shown in \figureref{fig:S2S}. This may be attributed to UNet's architecture, which, unlike neural operators, is more adept at capturing local rather than global interface patterns. In contrast, CNO consistently delivers the best performance in both sequence-to-field and sequence-to-sequence predictions, reinforcing its capability in handling complex fluid dynamics over time. Also, \figureref{fig:S2S} shows that CNO can capture small bubble formation after breakup more accurately than other models.

\tabref{tab:seq2field_s1_s3} and \tabref{tab:seq2seq_s4_s6} compare the Mean Squared Error (MSE) and relative $L_2$ error for sequence-to-field and sequence-to-sequence predictions across various models on the six bubble rise datasets (S1-S6). These results highlight the performance differences between models in predicting the solution fields for different data subsets (S1-S6). All model predictions improve as more time snapshots are incorporated, reinforcing the importance of temporal context in learning transient phenomena. Additionally, CNO generally outperforms the other models in predicting the concentration field, making it particularly effective for modeling sharp gradients in the interface region. Another interesting observation is that scOT marginally outperforms the pre-trained version of Poseidon. This suggests that Poseidon, having been pretrained exclusively on single-phase flow, struggles to generalize to multi-phase phenomena, highlighting the limitations of transfer learning when faced with physics characterized by sharp gradients near the interface. This performance gap underscores the importance of dataset diversity in foundation models designed for physics-informed learning.

Furthermore, \figureref{fig:S2F} and \figureref{fig:S2S} illustrate field predictions of the concentration field $C$ using UNet, CNO, DeepONet, and Poseidon for sequence-to-field and sequence-to-sequence scenarios, respectively. These figures show that DeepONet performs poorly in both the sequence-to-field and sequence-to-sequence scenarios. Additionally, UNet's accuracy declines as the prediction horizon extends to longer time sequences, as shown in \figureref{fig:S2S}. This suggests that convolution-based architectures may not effectively capture long-range dependencies crucial for tracking evolving multi-phase interfaces over time. In contrast, CNO consistently delivers the best performance in both sequence-to-field and sequence-to-sequence predictions. Its strong performance may be attributed to its ability to blend convolutional representations with continuous function approximations, allowing it to better capture sharp interface dynamics compared to other models. Additionally, \figureref{fig:S2S} demonstrates that CNO is capable of capturing small bubble formation post-breakup, a key feature of multi-phase flow that is often difficult to learn with purely spectral or transformer-based architectures.

\section{Conclusions}\label{sec:limit_conclusion}
In summary, we have introduced a comprehensive time series dataset comprising 10,000 simulations in 2D and 1,000 simulations in 3D, focusing on bubble rise and droplet fall dynamics. This dataset captures a wide range of two-phase flow phenomena, including simulations with density ratios as high as 1,000, Reynolds numbers up to 1,000, and Bond numbers up to 500. Using a subsample of 1,000 samples from the bubble dataset, we successfully trained neural operators and foundation models, demonstrating encouraging results. By feeding in more time snapshots to models, they can more accurately predict the trajectory of bubble dynamics. Specifically, we found that CNO outperformed other models in capturing fine-scale interfacial details. We also concluded that the foundation model Poseidon pre-trained on single-phase phenomena might not be effective in learning multiphase flow, which demonstrates the need to train foundation models on multiphase flow data. 

\textbf{Limitations}: The dataset has the following constraints:
\begin{itemize}[topsep=0pt,left=0pt] 
\item \textbf{Different orders of magnitude for solution fields}: The dataset includes solution fields that span different orders of magnitude. This is evident in the large disparity between the mean squared error (MSE) and relative \(L_2\) errors for different solution fields.
\item \textbf{Limited 3D Simulations}: Due to the substantial computational cost, only a small number of 3D simulations were conducted, resulting in a more restricted set of 3D cases in the dataset. 
\item \textbf{Model fitting with a limited number of time steps}: GPU memory limitations constrained the number of time steps that could be fitted on a single GPU. As a result, we had to use a limited number of time snapshots. An alternative approach could involve using an auto-regressive model to model the time series for each sample, inputting the predictions of previous time steps to predict the future time solutions to capture the full trajectory of the bubble. 
\end{itemize}

\section*{Reproducibility Statement}

In this work, we introduce a dataset and provide detailed explanations of the methodology and mathematical framework used for data generation in the Appendix \sectionref{appendix}. To evaluate the dataset, we applied various neural operators and foundation models, and the code for these models is available on the GitHub page on our \href{https://github.com/baskargroup/GeometryMatters/}{GitHub}. The repository includes detailed instructions for easy reproducibility of our results. All experiments were conducted on Nvidia A100-SXM4 80GB. Please refer to the repository’s \texttt{README.md} for complete instructions on replicating the model evaluations.

\section*{CRediT authorship contribution statement}

\textbf{Mehdi Shadkhah}: Conceptualization, Data curation, Formal Analysis, Investigation, Methodology, Software, Validation, Visualization, Writing – original draft, Writing – review \& editing.
\textbf{Ronak Tali}: Data curation, Formal Analysis, Software, Writing – original draft, Writing – review \& editing. 
\textbf{Ali Rabeh}: Conceptualization, Data curation, Formal Analysis, Software, Writing – original draft, Writing – review \& editing. 
\textbf{ChengHau Yang}: Visualization, Writing – original draft, Writing – review \& editing. 
\textbf{Ethan Herron}: Software. 
\textbf{Abhisek Upadhyaya}: Software. 
\textbf{Adarsh Krishnamurthy}: Conceptualization, Validation, Writing – review \& editing. 
\textbf{Chinmay Hegde}: Conceptualization. 
\textbf{Aditya Balu}: Conceptualization, Validation, Writing – review \& editing. 
\textbf{Baskar Ganapathysubramanian}: Conceptualization, Funding acquisition, Project administration, Supervision, Writing – review \& editing.

\clearpage

\bibliography{Refs}
\bibliographystyle{iclr2025_conference}

\clearpage
\appendix
\section*{Appendix}
\section{Details of the CFD simulation framework}\label{appendix}
Our computational framework employs the CUDA platform to implement the algorithms necessary for the Lattice Boltzmann Method (LBM). We achieve significant computational performance enhancements by leveraging CUDA's parallel processing capabilities. The primary performance bottleneck in GPU architectures is often the data transfer between GPU memory and unified CPU memory. To mitigate this, we minimize such data transfers, conducting them only when necessary for convergence checks or final output retrieval.

We utilize a single one-dimensional array in conjunction with macro functions to handle the substantial data volumes intrinsic to LBM simulations. This method optimizes memory usage and computational efficiency on the GPU, ensuring that we fully exploit the GPU's computational power and memory bandwidth. This strategy allows for the high-performance execution of LBM algorithms, crucial for large-scale simulations and complex fluid dynamics problems.

\subsection{Formulation of Navier Stokes and Allen Cahn equations}
Several lattice Boltzmann models, such as the Cahn-Hilliard and Allen-Cahn models, utilize interface tracking equations and are thus categorized as phase-field models \citep{Penrose1990, Jacqmin1999}. These models describe multiphase flows using a diffuse interface, with the Allen-Cahn equation commonly employed for this purpose \citep{Allen1976}. In some studies, this approach is called the conservative phase-field LB model \citep{Fakhari2019}. The phase-field variable, $\phi$, which tracks the interface, ranges from 0 to 1, leading to the following expression for the phase-field equation \citep{Chiu2011}:
\begin{equation}
\frac{\partial\phi}{\partial t}+\mathrm{\nabla}.\left(\phi u\right)=\mathrm{\nabla}.\left[M(\mathrm{\nabla\phi}-\frac{1-4\left(\phi-\phi_0\right)^2}{\xi}\hat{n})\right],
\label{eq:1}
\end{equation}
where $t$ represents time, $u$ is the velocity, $M$ denotes a positive constant for the mobility parameter, $\xi$ is the interfacial thickness, and $\phi_0 = \frac{\phi_H + \phi_L}{2}$. \(\phi_H\) and \(\phi_L\) represent the interface indicator values for the heavy and light fluids, respectively, set to 1.0 for the heavy fluid and 0.0 for the light fluid. The unit normal vector $\hat{n}$ for the interface can be defined as:

\begin{equation}
\hat{n}=\frac{\mathrm{\nabla\phi}}{\left|\mathrm{\nabla\phi}\right|}.
\label{eq:2}
\end{equation}
Note, the interface location at $x_0$ is initialized as \cite{Yan2007}:
\begin{equation}
\phi\left(x\right)=\phi_0\pm\frac{\phi_H-\phi_L}{2}\tanh(\frac{\left|x-x_0\right|}{\xi/2}).
\label{eq:3}
\end{equation}
According to the phase-field model, the following equations exist for incompressible multiphase flows (\cite{Ding2007, Li2012}):
\begin{subequations}
\begin{equation}
\frac{\partial \rho}{\partial t} + \nabla \cdot (\rho u) = 0,
\label{eq:4a}
\end{equation}
\begin{equation}
\rho \left(\frac{\partial u}{\partial t} + u \cdot \nabla u \right) = -\nabla p + \nabla \cdot \left(\mu \left[\nabla u + (\nabla u)^T \right] \right) + F_s + F_b.
\label{eq:4b}
\end{equation}
\end{subequations}

In \eqnref{eq:4a}, $\rho$ represents the density of fluids, $p$ denotes the macroscopic pressure, $F_b$ is the body force, and $F_s$ corresponds to the surface tension force. The equation for calculating the surface tension force term is also expressed as \cite{Jamet2002}:
\begin{equation}
F_s=\mu_\phi\mathrm{\nabla\phi},
\label{eq:5}
\end{equation}
where
\begin{equation}
\mu_\phi=4\beta\phi\left(\phi-1\right)\left(\phi-1/2\right)-\kappa\mathrm{\nabla}^2\phi,
\label{eq:6}
\end{equation}
\vspace{0.2\baselineskip}% Remove extra line inserted by subfloat
denotes the chemical potential equation utilized for binary fluids \citep{JACQMIN2000}. \eqnref{eq:7} establishes a relation between the coefficients $\beta$ and $\kappa$, interface thickness $\xi$, and surface tension \(\sigma\), as;
\begin{equation}
\beta=12\sigma/\xi\ ,\quad \kappa=3\sigma\xi/2.
\label{eq:7}
\end{equation}
\vspace{0.2\baselineskip}% Remove extra line inserted by subfloat

\subsection{Lattice Boltzmann Method}\label{subsec:numerical}

Given that interfaces are typically of mesoscopic scale, the kinetic-based Lattice Boltzmann Method (LBM) presents a more effective approach for simulating multiphase flows compared to the traditional Navier-Stokes solvers \citep{Sukop2006, Huang2015}. The Chapman-Enskog analysis validates the consistency between the LBM and the Navier-Stokes equations \citep{Kruger2017}. In this study, we investigate hydrodynamic properties such as velocity and pressure using the standard form of the Lattice Boltzmann equation as outlined in \cite{Guo2002}:
\begin{equation}
f_a\left(x+e_a\delta t,\ t+\delta t\right)=f_a\left(x,t\right)+\mathrm{\Omega}_a(x,t)+F_a(x,t),
\label{eq:8}
\end{equation}
In this context, \( f_a \) denotes the velocity-based hydrodynamic distribution function for incompressible fluids, \( \Omega_a \) represents the collision operator, and \( F_a \) signifies the force term. This study employs the two-dimensional nine-velocity (D2Q9) model for 2D simulations and the three-dimensional nineteen-velocity (D3Q19) model for 3D simulations. 

To define the interface between phases, we employed the following Lattice Boltzmann Equation (LBE) to accurately determine the interface between fluid phases \citep{Geier2015F}:

\begin{equation}
g_a\left(x + e_a \delta t, t + \delta t\right) = g_a\left(x, t\right) 
- \frac{g_a\left(x, t\right) - \bar{g}_a^{eq}\left(x, t\right)}{\tau_\phi + 1/2} 
+ F_a^\phi(x, t).
\label{eq:28}
\end{equation}

Here, \( g_a \) represents the distribution function for the phase-field, and \( \tau_\phi \) denotes the dimensionless phase-field relaxation time. The forcing term is calculated as follows:
\begin{equation}
F_a^\phi\left(x,t\right)=\delta t\frac{\left[1-4{(\phi-\phi_0)}^2\right]}{\xi}\omega_ae_a\ \cdot \ \frac{\mathrm{\nabla\phi}}{\left|\mathrm{\nabla\phi}\right|}.
\label{eq:29}
\end{equation}

\begin{figure}[ht]
    \centering
    \includegraphics[width=0.6\textwidth]{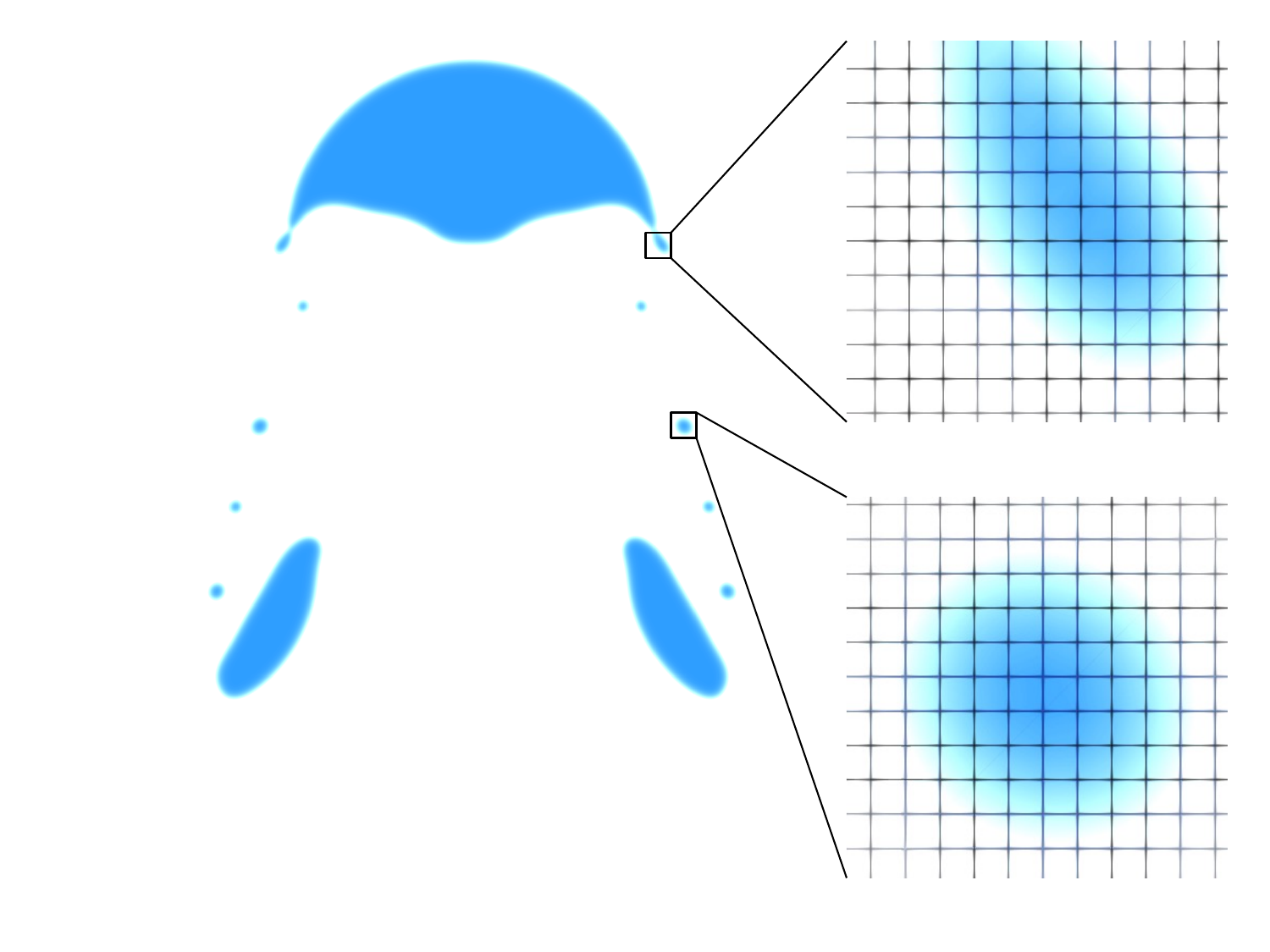}
\caption{Illustration of the interface region captured by the computational mesh. The magnified views show the distribution of lattice nodes within the interface, ensuring precise resolution of interfacial dynamics and transitions. The careful selection of the interface thickness parameter \(\xi\) ensures that the mesh adequately represents the gradients and variations in the interface region.}
    \label{fig:mesh}
\end{figure}

In \eqnref{eq:29}, $\omega_a$ and $e_a$ denote the weight coefficient and the mesoscopic velocity set, respectively. Here, \(\xi\) denotes the thickness of the interface. As illustrated in \figureref{fig:mesh}, we carefully selected this parameter to ensure adequate lattice nodes within the interface. This choice is critical for accurately capturing the complex physics in the rapid change of material properties across the interface. The appropriate selection of \(\xi\) ensures that the computational mesh can effectively represent the gradients and variations within the interface, thus enhancing the overall stability and accuracy of the simulation.

\subsection{Validation}\label{subsec:validations}
In this section, we validate our numerical model through benchmark tests covering a range of two-phase flow phenomena. We include four distinct validation cases to comprehensively assess the accuracy and robustness of our approach: (1) the capillary wave problem, which evaluates the model's capability to handle surface tension-driven flows; (2) the bubble rising dynamics, which tests the interaction between buoyancy and viscous forces; (3) the falling droplet dynamics, which examines the breakup mechanisms of liquid droplets under gravity; and (4) the Rayleigh-Taylor instability, which explores the interfacial instability between fluids of differing densities under gravitational influence. Each subsection compares our simulation results and established experimental or numerical data, demonstrating the model's fidelity across various flow regimes.

\subsubsection{Capillary Wave}
To validate our Lattice Boltzmann Method (LBM) simulations of two-phase flow, we focus on the dynamic behavior of capillary waves at the interface between two immiscible fluids. In our study, a sinusoidal perturbation with a small amplitude $\eta_0$ and wave number $k$ is applied to the initially quiescent interface. This setup provides a rigorous test for the LBM framework, as it has a well-established analytical solution for cases with identical kinematic viscosities $\nu$ but differing densities of the two fluids. The temporal evolution of the interface amplitude $\eta(t)$ is utilized as a benchmark for our simulations. The analytical expression for the decay of the wave amplitude, $\eta(t)$, is given by ~\cite{Prosperetti19811217}:

\begin{align}
\frac{\eta(t)}{\eta_0} &= \frac{4(1-4\gamma)\nu^2 k^4}{8(1-4\gamma)\nu^2 k^4 + \omega_0} \, \text{erfc}(\sqrt{\nu k^2 t}) + \sum_{i=1}^{4} \frac{z_i}{Z_i} \frac{\omega_0^2}{z_i^2 - \nu k^2} e^{(z_i^2 - \nu k^2)t} \, \text{erfc}(z_i\sqrt{\nu t})
\label{eq:decay_wave_amplitude}
\end{align}

where $\omega_0 = \sqrt{\frac{\sigma k^3}{\rho_H + \rho_L}}$ is the angular frequency, $\gamma = \frac{\rho_H \rho_L}{(\rho_H + \rho_L)^2}$ and $Z_i = \prod_{\substack{1 \le j \le 4 \\ j \ne i}} (z_j - z_i)$. The evaluation of the complementary error function $\text{erfc}(z_i)$ can be done by solving the following algebraic equation:

\begin{equation}
z^4 - 4\gamma \sqrt{\nu k^2} z^3 + 2 (1 - 6\gamma) \nu k^2 z^2 + 4 (1 - 3\gamma) (\nu k^2)^{3/2} z + (1 - 4\gamma) \nu k^2 + \omega_0^2 = 0.
\label{eq:algebraic_equation}
\end{equation}

\begin{figure}[ht]
    \centering
    \includegraphics[width=0.50\textwidth]{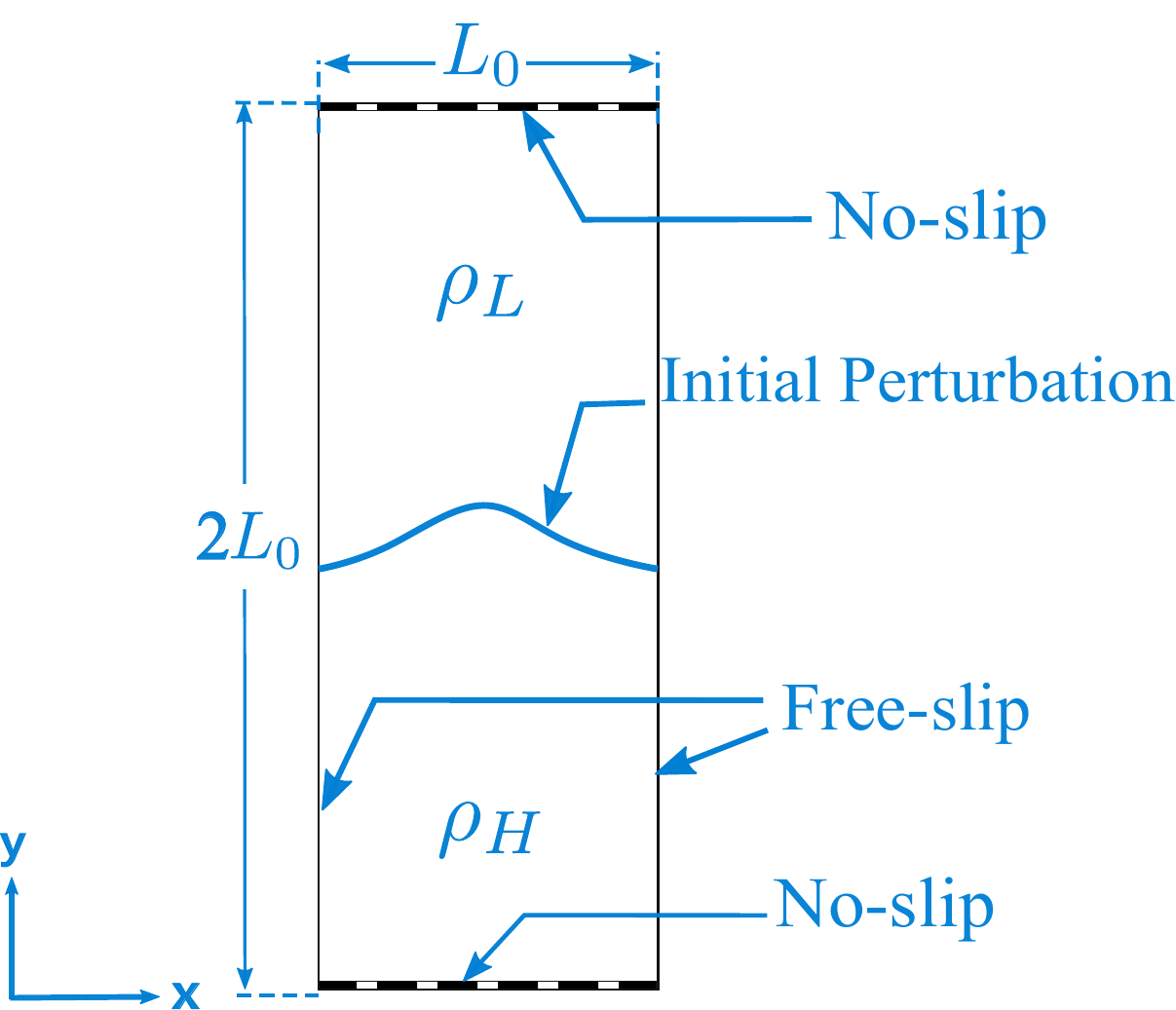}
    \caption{Schematic diagram of the capillary wave problem setup.}
    \label{fig:capillary_wave_setup}
    \label{fig:CWBC}
\end{figure}

Our validation involves analyzing the propagation of capillary waves, an inherently transient process that tests the model's ability to accurately capture key physical parameters such as density and viscosity ratios, along with surface tension effects. By varying these parameters and the wavelength, we compare the simulation results with predictions from linear theory. According to \figureref{fig:CWBC}, the lighter fluid with density $\rho_L$ overlays the heavier fluid with density $\rho_H$, with the initial interface described by $y = L + \eta_0 \cos(2 \pi x)$, where $\eta_0$ is the initial perturbation amplitude. The decay of this wavy profile to a flat interface, driven by viscosity and surface tension, without external forces like gravity, serves as a critical validation test for our LBM approach.
The computational domain is discretized into a grid of 256 by 512 lattice nodes. Free-slip boundary conditions are applied in the direction of wave propagation, while no-slip conditions are imposed at the top and bottom boundaries. The simulation parameters are set as follows: $\eta_0 = 0.02$, $\sigma = 10^{-4}$, $\xi = 4$, and $M_\phi = 0.02$. Since the interface may not align exactly with the grid points, the values of $\eta(t)$ are interpolated from $\phi$ values using the following relationship:

\begin{equation}
\eta(t) = y - \frac{\phi(x_{L0/2}, y)}{\phi(x_{L0/2}, y) - \phi(x_{L0/2}, y - 1)}, \quad \phi(x_{L0/2}, y)\phi(x_{L0/2}, y - 1) < 0.
\end{equation}

The length ($\eta$) and time scales ($t$) are normalized by the initial amplitude $a_0$ and the angular frequency $\omega_0$, respectively, denoted as $\eta^* = \eta/\eta_0$ and $t^* = t\omega_0$.

It is worth noting that angular frequency is crucial for any wave system. It depends on surface tension, viscosity, wave number, and density values.  The equation is derived assuming that both fluids have the same viscosity, set to $\nu = 0.005, 0.0005$. Note that the wavelength magnitude matches the grid size $L_0 = 256$.

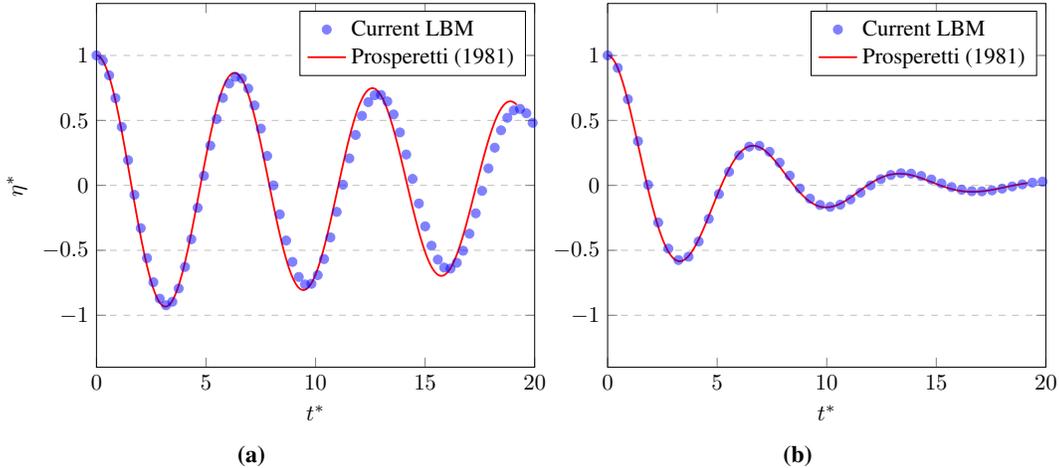
\begin{figure}[ht]
    \centering
    \begin{subfigure}[b]{0.48\textwidth}
        \centering
        \begin{tikzpicture}[scale=0.85]
        \begin{axis}[
            xlabel={$t^*$},
            ylabel={$\eta^*$},
            xmin=0, xmax=20,
            ymin=-1.4, ymax=1.4,
            ymajorgrids=true,
            grid style=dashed,
            legend cell align={left}
        ]
        
        \addplot[
            only marks,
            mark=*,
            mark options={fill=blue, draw=blue, opacity=0.5},
            mark size=2pt
        ]
        table[x=X,y=Y,col sep=comma] {Validation/Capillary0005.csv};
        \addlegendentry{Current LBM}
        
        \addplot[
            color=red,
            solid,
            thick
        ]
        table[x=t,y=eta,col sep=comma] {Validation/CapillaryAnalytical0005.csv};
        \addlegendentry{Prosperetti (1981)}
        
        % \node at (axis cs: 1,0.95) {\textcolor{red}{(c)}};
        
        \end{axis}
        \end{tikzpicture}
        \caption{}
    \end{subfigure}
    \hfill
    \begin{subfigure}[b]{0.48\textwidth}
        \centering
        \begin{tikzpicture}[scale=0.85]
        \begin{axis}[
            xlabel={$t^*$},
            xmin=0, xmax=20,
            ymin=-1.4, ymax=1.4,
            ymajorgrids=true,
            grid style=dashed,
            legend cell align={left}
        ]
        
        \addplot[
            only marks,
            mark=*,
            mark options={fill=blue, draw=blue, opacity=0.5},
            mark size=2pt
        ]
        table[x=X,y=Y,col sep=comma] {Validation/Capillary005.csv};
        \addlegendentry{Current LBM}
        
        \addplot[
            color=red,
            solid,
            thick
        ]
        table[x=t,y=eta,col sep=comma] {Validation/CapillaryAnalytical005.csv};
        \addlegendentry{Prosperetti (1981)}
        
        % \node at (axis cs: 1,0.95) {\textcolor{red}{(c)}};
        
        \end{axis}
        \end{tikzpicture}
        \caption{}
    \end{subfigure}
    \caption{Comparison of the normalized interface amplitude $\eta^*$ as a function of normalized time $t^*$ between the current LBM simulation and the analytical solution by Prosperetti (1981). (a) corresponds to a viscosity of $\nu = 0.0005$, and (b) corresponds to a viscosity of $\nu = 0.005$. The LBM results (blue circles) closely match the analytical results (red line).}
    \label{fig:main}
\end{figure}

\subsubsection{Rise of a single bubble in quiescent fluid}\label{subsec:rising_bubble}

The dynamics of a rising bubble have been extensively studied due to their significance in various natural and industrial processes. When a bubble rises through a liquid, it is subjected to several forces, including buoyancy, drag, and surface tension, which influences its shape, velocity, and trajectory ~\citep{Bhaga_Weber_1981, AMAYABOWER20101191, HUA2007769, KHANWALE2023111874}. Our investigation focuses on the dynamics of a bubble rising within a rectangular channel. The simulation begins with a circular bubble of diameter $D = L_0/5$ placed at the coordinates $(L_0/2, L_0/2)$ within a domain with a length of $L_0$ and a height of $4L_0$. Boundary conditions are set such that the no-slip is applied at the top and bottom, while free-slip boundary conditions are used for the lateral boundaries. The fluids experience a volumetric buoyancy force $F_b = -(\rho - \rho_h) g_y \mathbf{j}$, where $g_y$ represents the gravitational acceleration in the $y$-direction. This study highlights four crucial dimensionless parameters: the density ratio $\rho_h/\rho_l$, the viscosity ratio $\mu_h/\mu_l$, the gravity Reynolds number, and the Eötvös (Bond) number.

The gravity Reynolds number is defined as:
\begin{equation}
\text{Re}_h = \frac{\sqrt{g_y \rho_h (\rho_h - \rho_l) D^3}}{\mu_h}
\end{equation}

The Eötvös (Bond) number is defined as:
\begin{equation}
\text{Eo} = \frac{g_y (\rho_h - \rho_l) D^2}{\sigma}
\end{equation}

In many studies, the Morton number is also considered, defined as:
\begin{equation}
\text{Mo} = \frac{g_y (\rho_h - \rho_l) \mu_h^4}{\sigma^3 \rho_h^2}
\end{equation}
The dimensionless time is also defined by:

\begin{equation}
t^* = t \sqrt{\frac{g_y}{D}}
\end{equation}

\begin{figure}[ht]
    \centering
    
    \begin{tabular}{ccccc}
    \toprule
        & A1 & A2 & A3 & A4 \\
        & $Bo = 17.7$ & $Bo = 243$ & $Bo = 115$ & $Bo = 115$ \\
        & $Mo = 711$ & $Mo = 266$ & $Mo = 1.31$ & $Mo = 0.001$ \\
        \midrule
        \multirow{-6}{*}{\rotatebox{90}{Experiment}} &
        \subfloat{\includegraphics[width=0.20\textwidth]{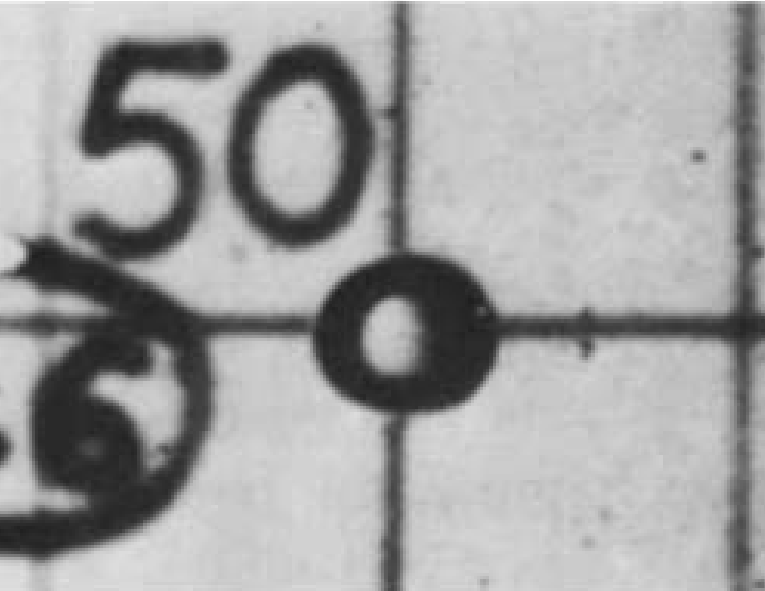}} & 
        \subfloat{\includegraphics[width=0.20\textwidth]{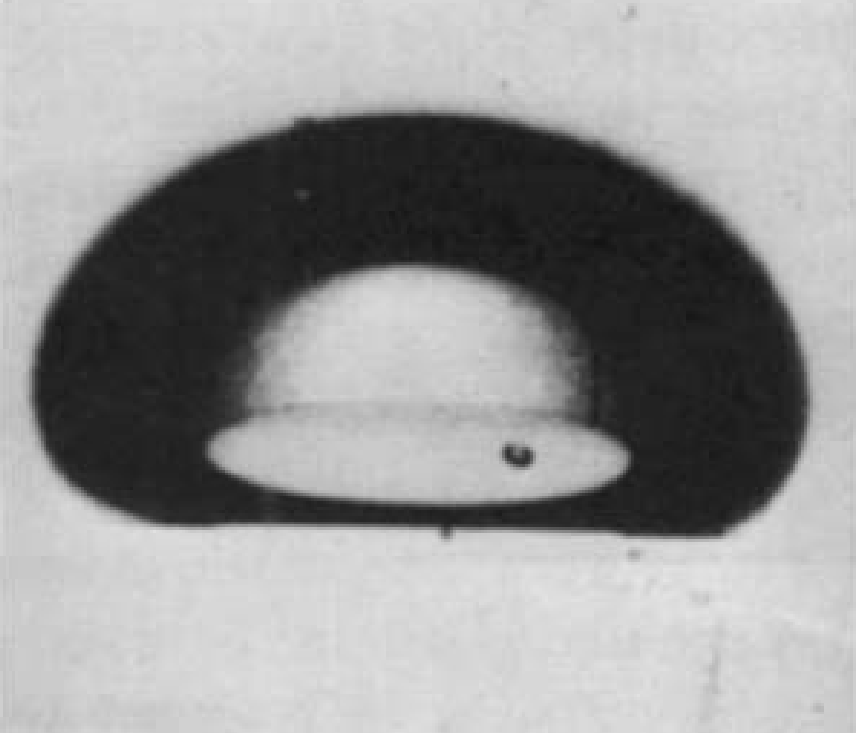}} & 
        \subfloat{\includegraphics[width=0.20\textwidth]{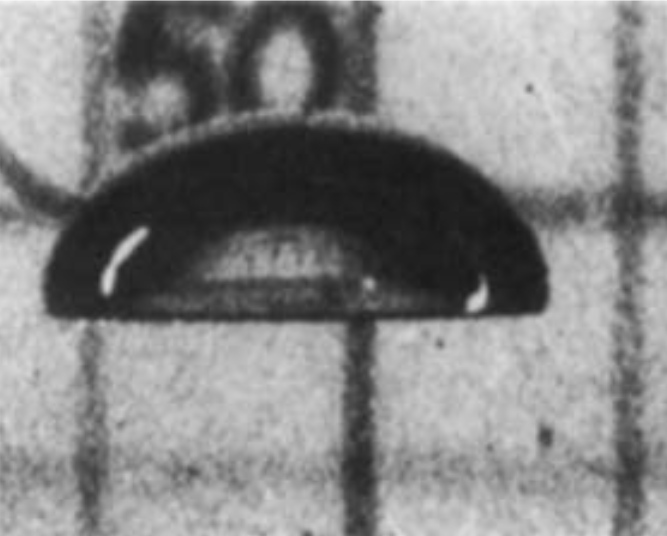}} & 
        \subfloat{\includegraphics[width=0.20\textwidth]{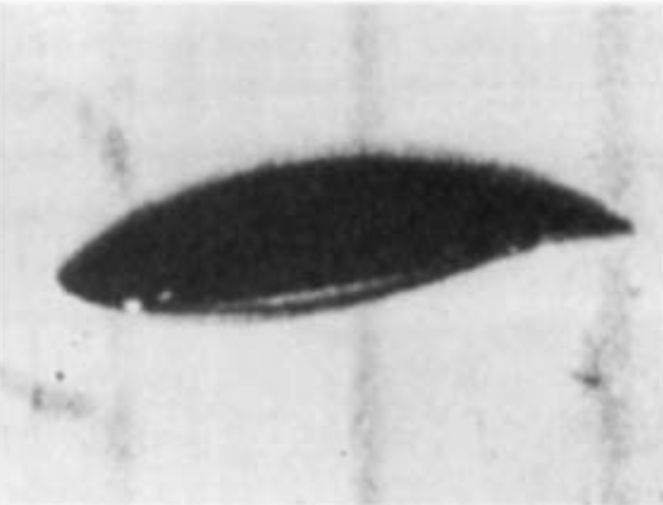}} \\
        \midrule
        \multirow{-6}{*}{\rotatebox{90}{FVM (3D)}} &
        \subfloat{\includegraphics[width=0.20\textwidth]{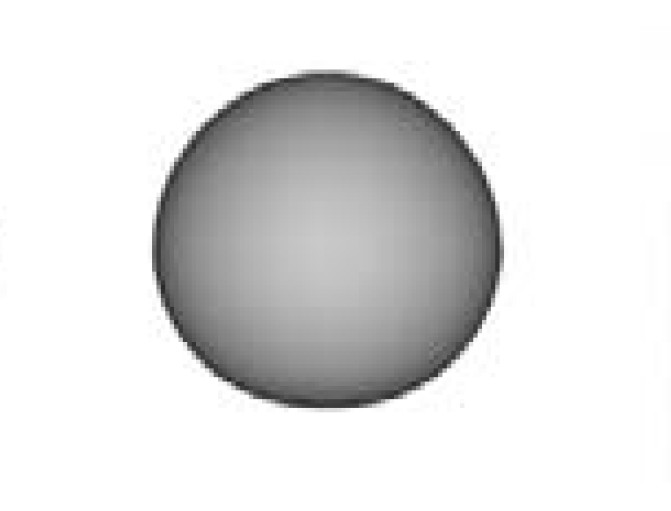}} & 
        \subfloat{\includegraphics[width=0.20\textwidth]{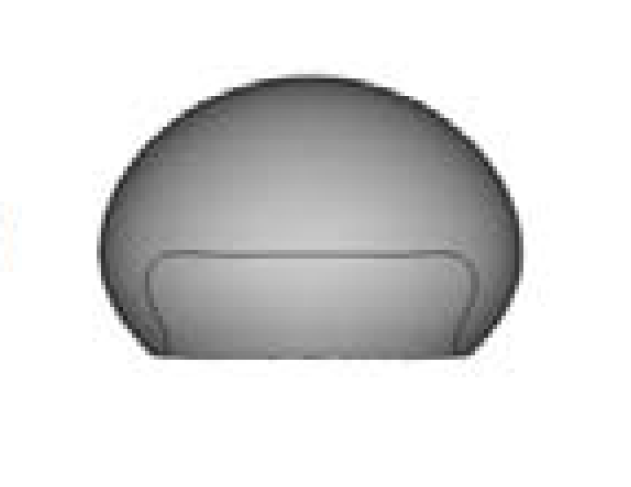}} & 
        \subfloat{\includegraphics[width=0.20\textwidth]{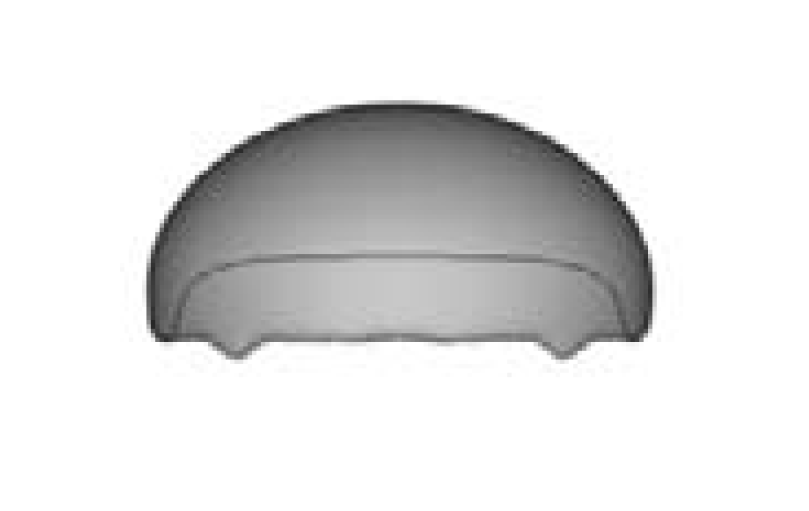}} & 
        \subfloat{\includegraphics[width=0.20\textwidth]{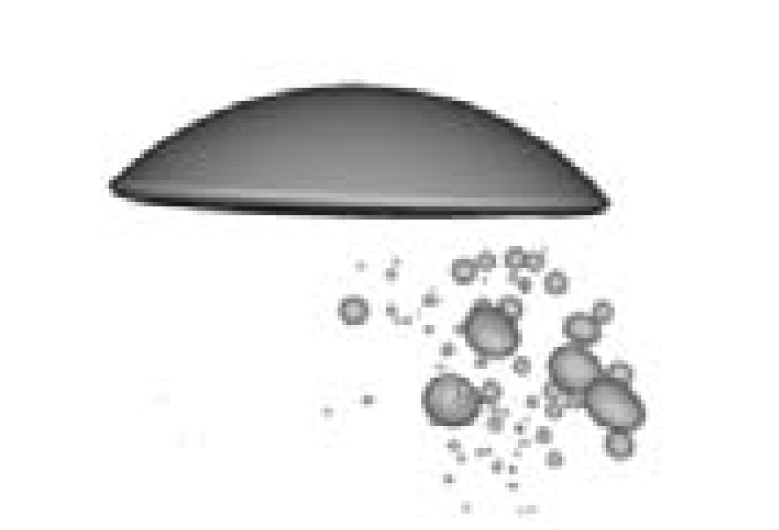}} \\
        \midrule
        \multirow{-6}{*}{\rotatebox{90}{LBM (2D)}} &
        \subfloat{\includegraphics[width=0.20\textwidth]{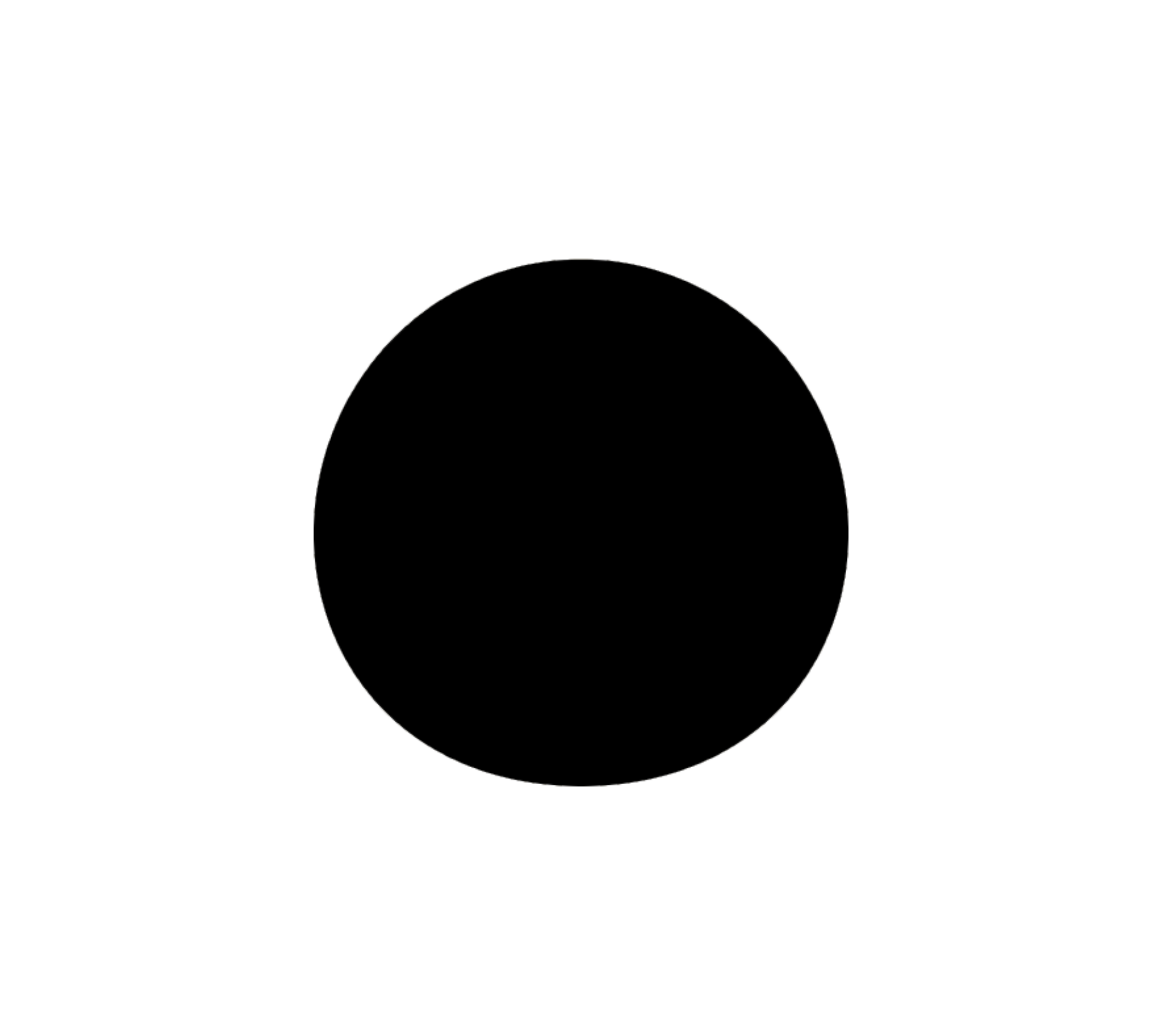}} & 
        \subfloat{\includegraphics[width=0.20\textwidth]{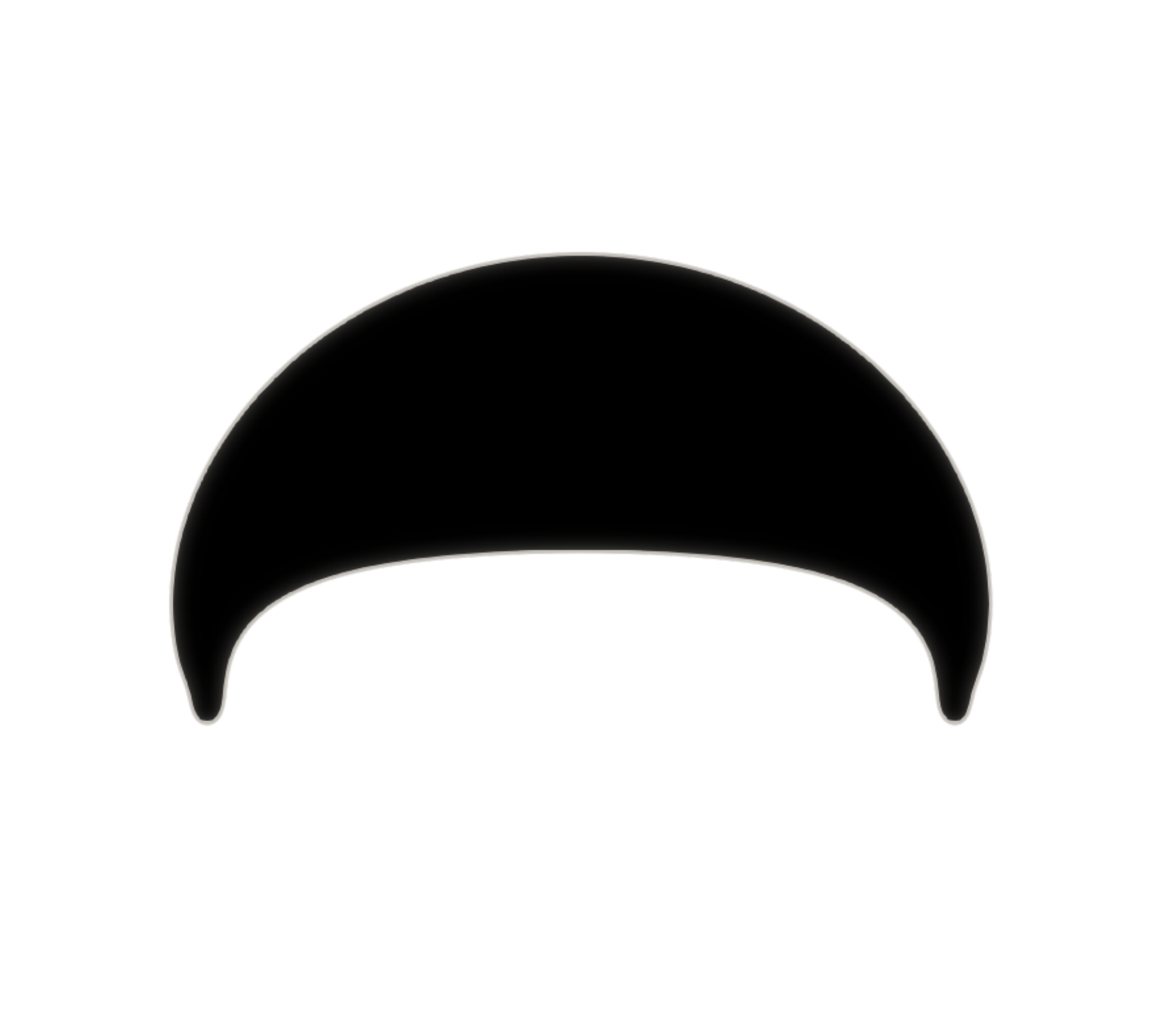}} & 
        \subfloat{\includegraphics[width=0.20\textwidth]{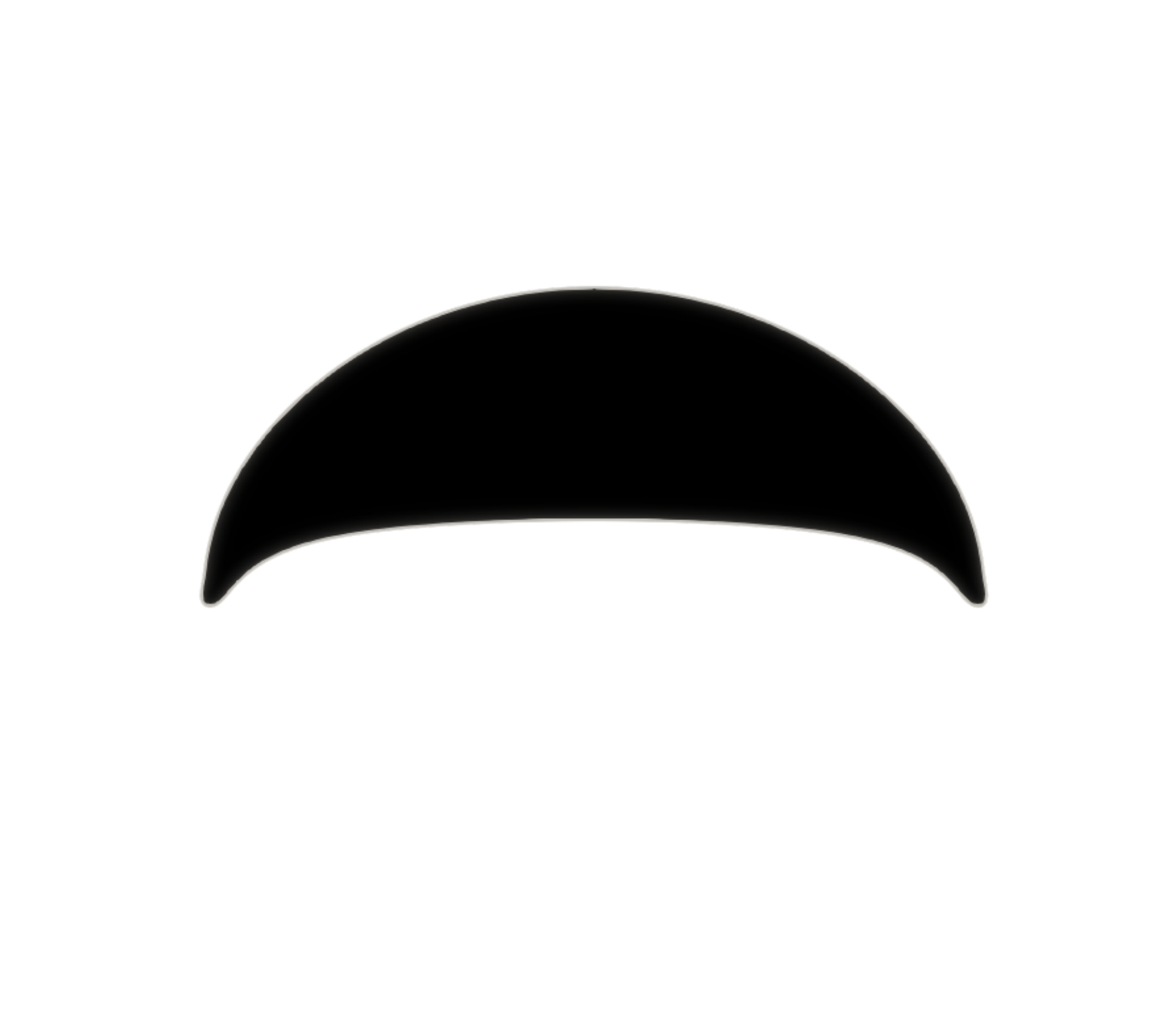}} & 
        \subfloat{\includegraphics[width=0.20\textwidth]{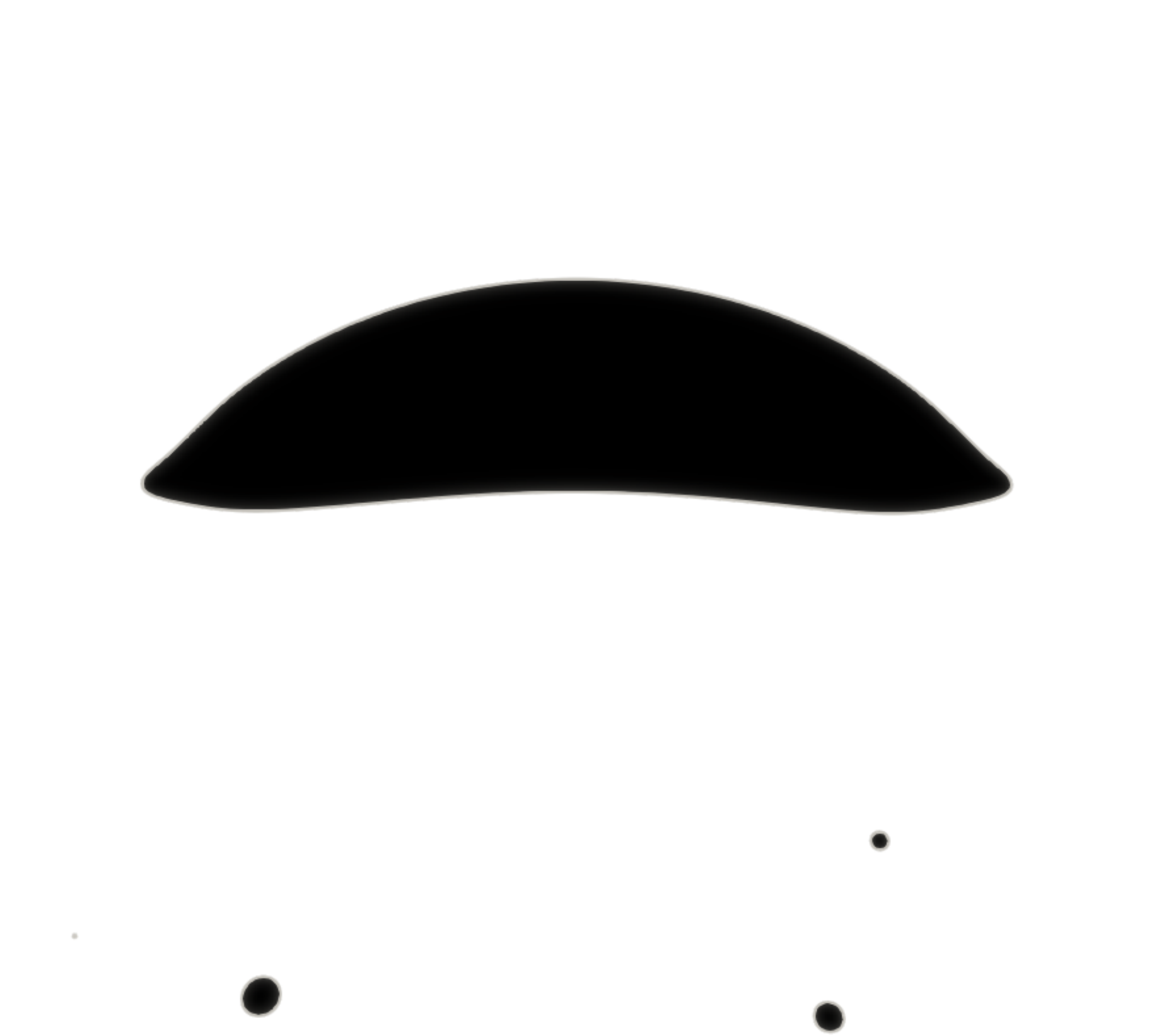}} \\
        \bottomrule
    \end{tabular}

\caption{Comparison of bubble shapes at constant rise velocity: Experimental results by~\cite{BhagaWeber1981}, LBM results, and FVM results by~\cite{GUMULYA2016298} for various Bond numbers ($Bo$) and Morton numbers ($Mo$).}
    \label{fig:Bubblecomparison}
\end{figure}

The reference velocity scale needed in the Péclet number can be chosen for gravity-driven flows as \( U_0 = \sqrt{g_y D} \). Four sets of simulations are conducted at Four different Eötvös and Morton numbers. The density and viscosity ratios are fixed at 1000 and 100, respectively. The numerical parameters are \( L_0 = 512 \), \( \text{Pe} = 25 \) and \( \text{Cn} = 0.010 \), and the LBM simulation results are shown in \figureref{fig:Bubblecomparison}.

To evaluate the accuracy and reliability of the proposed LBM, a comparison is made between the results obtained from the LBM approach and those from the experiments and FVM, as illustrated in \figureref{fig:Bubblecomparison}. In the spherical regime, surface tension dominates, resulting in small bubbles that maintain a nearly spherical shape due to the strong cohesive forces at the interface. As the bubble size increases, the shape transitions to an ellipsoidal form. In this ellipsoidal regime, the inertial forces become more significant, causing the bubble to deform. This deformation is influenced by the surrounding liquid's viscosity and the interface's surface tension. The dynamics of this regime can be described using correlations that account for the balance between inertial and surface tension forces~\citep{AMAYABOWER20101191}. In the spherical cap regime, the bubbles are large enough that inertia forces dominate, leading to further deformation into a cap shape. This regime is characterized by a significant increase in terminal velocity, which is proportional to the size of the bubble~\citep{BhagaWeber1981}. These patterns are consistent among all results.

\subsubsection{Falling Droplet}

The dynamics of a falling droplet under gravity is another fascinating two-phase flow phenomenon that has been extensively studied in the literature \citep{YANG2021103561,JALAAL2012115}. In this study, a liquid droplet with diameter $D = L_0/5$ is initially placed at $(L_0/2, 6L_0/2)$ within a rectangular computational domain of length $L_0$ and height $3L_0$. The same boundary conditions are applied as in the bubble rising simulations: the no-slip boundary condition is applied at the top and bottom, while free-slip boundary conditions are imposed at the lateral boundaries. The volumetric buoyancy force $F_b = -(\rho - \rho_l) g_y \mathbf{j}$, where \(\mathbf{j}\) is unit vector in $y$-direction and $g_y$ represents the gravitational acceleration in the $y$-direction, acts on the fluids.

The dimensionless analysis identifies several key parameters that characterize the flow: the density ratio $\rho_h / \rho_l$, the viscosity ratio $\mu_h / \mu_l$, the gravity Reynolds number, and the Eötvös (Bond) number. The gravity Reynolds number is defined as:

\begin{equation}
\text{Re}_h = \frac{\sqrt{g_y \rho_h (\rho_h - \rho_l) D^3}}{\mu_h}
\end{equation}

Similarly, the Eötvös number, which represents the ratio of gravitational forces to surface tension forces, is given by:

\begin{equation}
\text{Eo} = \frac{g_y (\rho_h - \rho_l) D^2}{\sigma}
\end{equation}

Another important dimensionless group in the literature is the Morton number, which characterizes the fluid properties affecting the bubble and droplet dynamics:

\begin{equation}
\text{Mo} = \frac{g_y (\rho_h - \rho_l) \mu_h^4}{\sigma^3 \rho_h^2}
\end{equation}

The Ohnesorge number (Oh) is a dimensionless number that characterizes the relative importance of viscous forces compared to inertial and surface tension forces in a fluid. It is particularly relevant in the study of droplet dynamics and is defined as:

\begin{equation}
\text{Oh} = \frac{\mu_h}{\sqrt{\rho_h \sigma D}}
\end{equation}

The simulation is conducted at a moderate density ratio to capture the breakup mechanisms of the falling droplet, allowing for comparisons with the VOF model. The simulation considers an Eötvös number: $Eo = 288$, with density and viscosity ratios fixed at 10 and 1, respectively, and the Oh number set to 0.05. The numerical parameters are $Pe = 5$ and $Cn = 0.010$. As mentioned in \sectionref{subsec:rising_bubble}, the reference velocity scale needed for the Péclet number can be chosen as $U_0 = \sqrt{g_y D}$ for gravity-driven flows. Also, dimensionless time can be defined by:

\begin{equation}
t^* = t \sqrt{\frac{g_y}{D}}
\end{equation}

\begin{figure}[ht]
    \centering
    
    \begin{tabular}{ccccc}
    \toprule
        \multirow{-6}{*}{\rotatebox{90}{VOF (3D)}} &
        \subfloat{\includegraphics[width=0.20\textwidth]{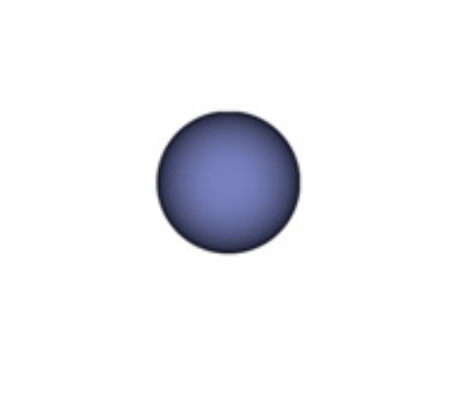}} & 
        \subfloat{\includegraphics[width=0.20\textwidth]{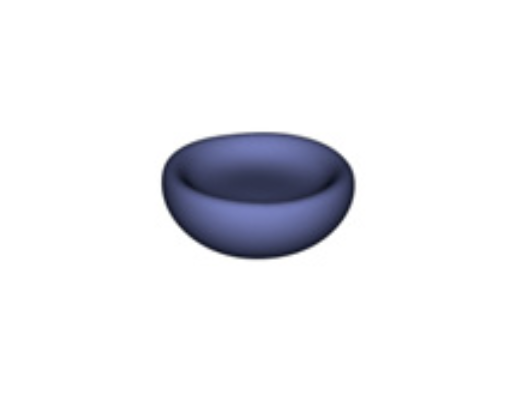}} & 
        \subfloat{\includegraphics[width=0.20\textwidth]{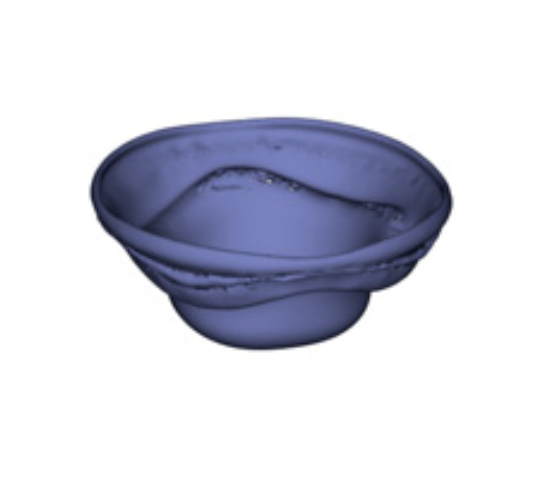}} & 
        \subfloat{\includegraphics[width=0.20\textwidth]{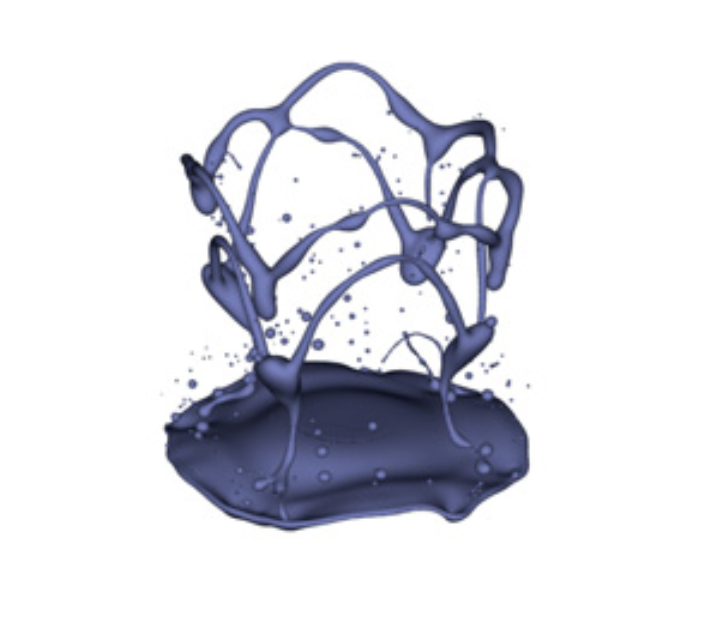}} \\
        \midrule
        \multirow{-6}{*}{\rotatebox{90}{LBM (2D)}} &
        \subfloat{\includegraphics[width=0.20\textwidth]{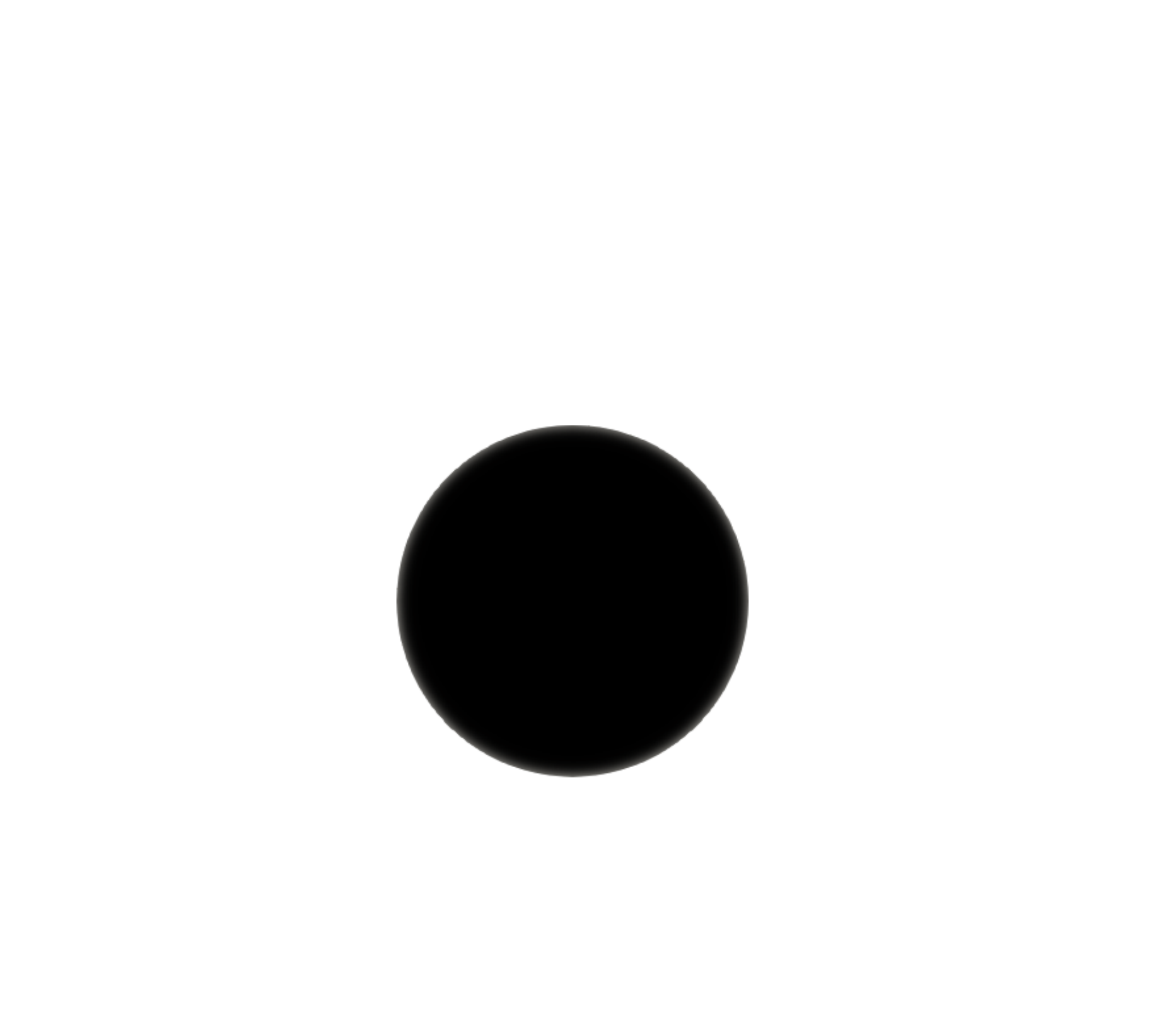}} & 
        \subfloat{\includegraphics[width=0.20\textwidth]{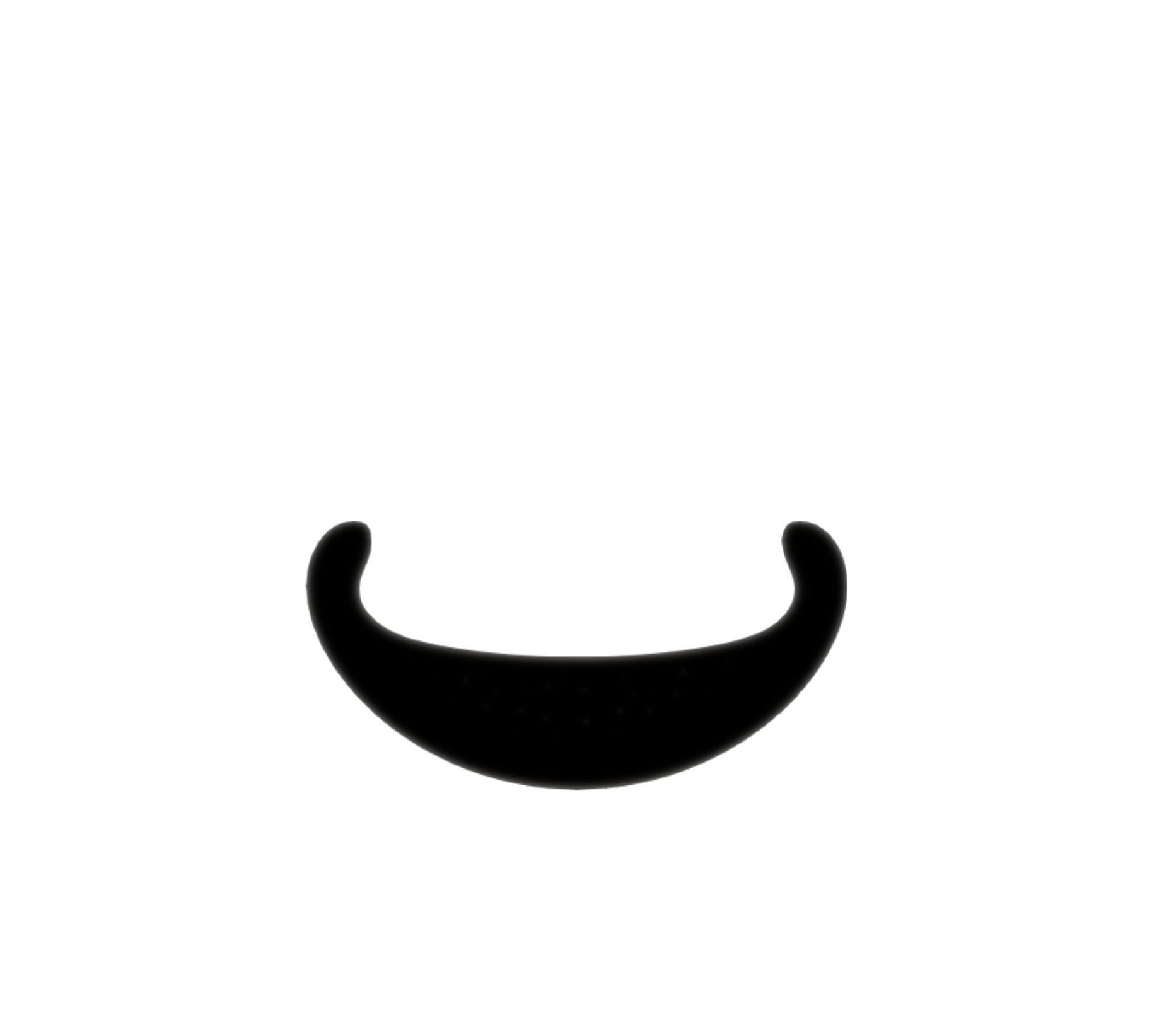}} & 
        \subfloat{\includegraphics[width=0.20\textwidth]{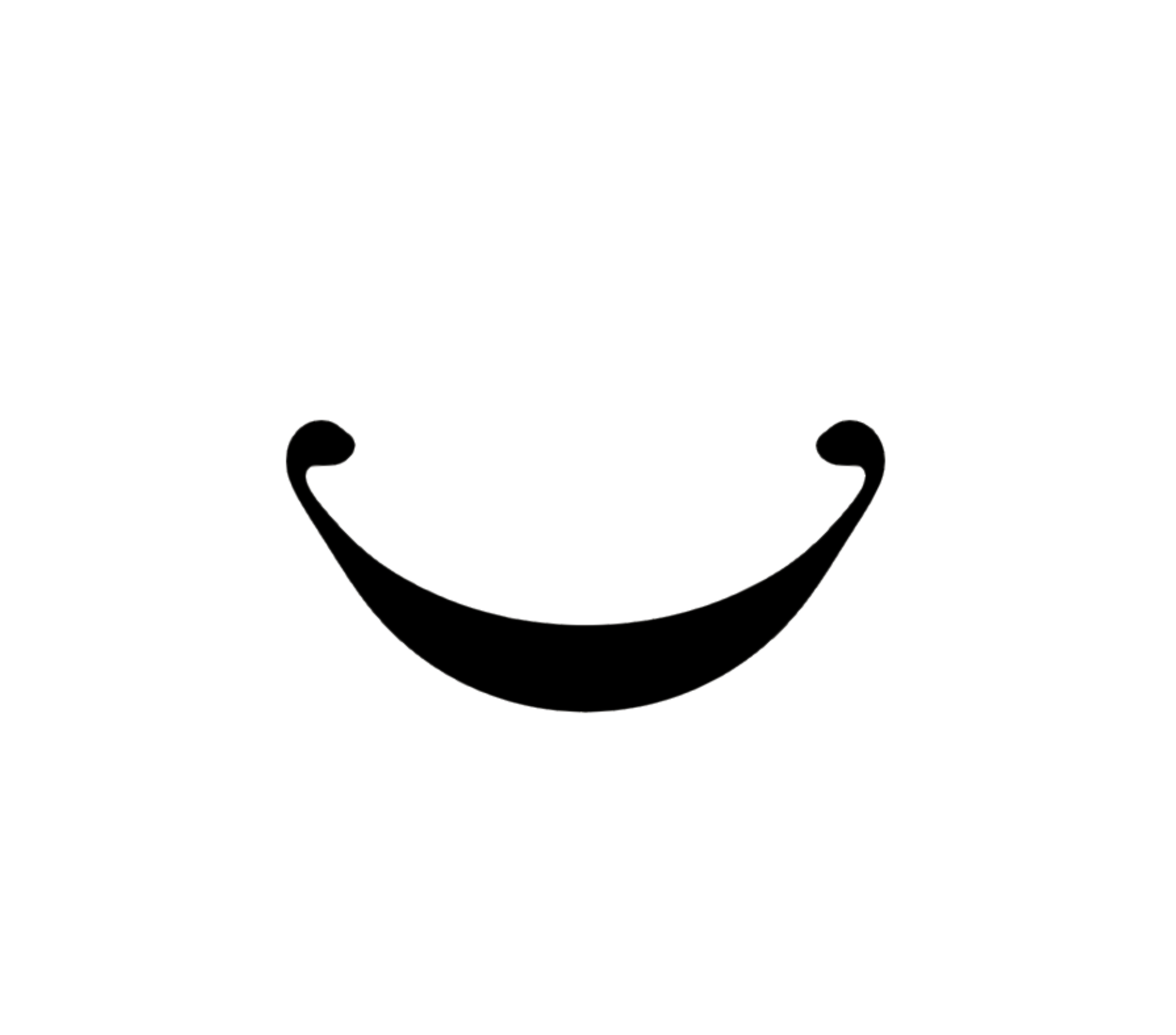}} & 
        \subfloat{\includegraphics[width=0.20\textwidth]{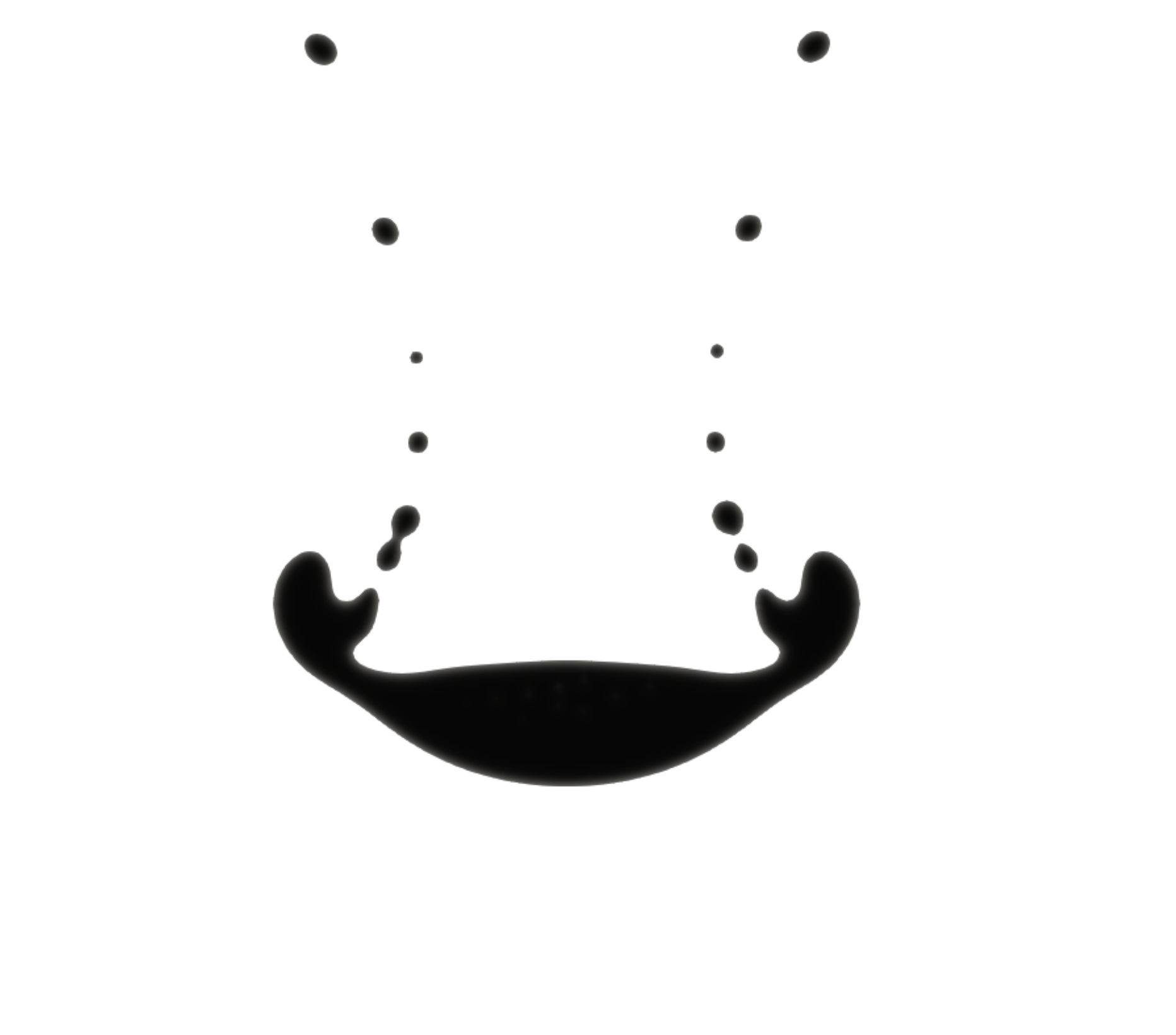}} \\
        \midrule
        & $t^* = 0.0$ & $t^* = 0.1647$ & $t^* = 0.2642$ & $t^* = 0.3575$ \\
        \bottomrule
    \end{tabular}

\caption{Comparison of the deformation of a liquid drop using the LBM and VOF methods: Current results and those of Jalaal et al. (2012)~\citep{JALAAL2012115} for $Eo = 288$, $Oh_h = Oh_l = 0.05$, and $\rho^* = 10$.}
    \label{fig:Dropletcomparison}
\end{figure}

Our simulation results exhibit excellent agreement with the findings of \citet{JALAAL2012115}. As shown in \figureref{fig:Dropletcomparison}, the comparison of the deformation of a liquid drop using both the Lattice Boltzmann Method (LBM) in 2D and the Volume of Fluid (VOF) method in 3D demonstrates that the evolution of the drop shapes over time is remarkably similar. For instance, at $t^* = 0.1647$, both methods capture the formation of a curved interface, and at $t^* = 0.3575$, the drop breakup into smaller droplets is observed in both approaches. This consistency across different numerical methods, with parameters set at $Eo = 288$, $Oh_h = Oh_l = 0.05$, and $\rho^* = 10$, validates the robustness and accuracy of our LBM simulations in replicating complex two-phase flow phenomena.

Overall, the dynamics of falling droplets involve complex interactions between buoyancy, inertia, and surface tension forces, leading to various deformation and breakup patterns, such as forming bags, ligaments, and secondary droplets. These phenomena are influenced significantly by the Eötvös number, with higher values leading to more pronounced deformations and faster breakup processes \citep{JALAAL2012115}.

\subsubsection{Rayleigh-Taylor Instabilities}

The Rayleigh-Taylor instability (RTI) arises when a denser fluid is positioned above a less dense fluid in the presence of a gravitational field, causing the interface between the two fluids to become unstable. This phenomenon has been extensively studied due to its relevance in various natural and engineering contexts ~\citep{KHANWALE2023111874,Ren2016,Zu2013}.

We consider a computational domain of size $[0, L_0] \times [0, 4L_0]$ with $L_0 = 256$ for our simulations. The initial interface is defined as $y_0(x) = 2L_0 + 0.1L_0 \cos(2\pi x/L_0)$. Periodic boundary conditions are applied on the left and right boundaries, while no-slip conditions are enforced at the top and bottom boundaries. The dimensionless numbers characterizing the RTI include the Atwood number, Reynolds number, Capillary number, and Peclet number:

\begin{equation}
\text{At} = \frac{\rho_H - \rho_L}{\rho_H + \rho_L},
\end{equation}

\begin{equation}
Re = \frac{\rho_H U_0 L_0}{\mu_H},
\end{equation}

where $U_0 = \sqrt{g_yL_0}$,

\begin{equation}
Ca = \frac{\mu_H U_0}{\sigma},
\end{equation}

\begin{equation}
Pe = \frac{U_0 L_0}{M}.
\end{equation}

In our study, we used a density ratio $\rho^* = 3$, viscosity ratio $\mu^* = 1$, Reynolds number $Re = 128$, Atwood number $At = 0.5$, Peclet number $Pe = 744$, and interface width $\xi = 5$. The results are compared with the findings from ~\citet{Ren2016} and ~\citet{Zu2013} . The dimensionless time is defined as $t^* = t / t_0$, where $t_0 = \sqrt{L_0 / (g At)}$ .

\begin{figure}[ht]
    \centering
    \begin{subfigure}[b]{0.48\textwidth}
        \centering
        \includegraphics[width=\textwidth]{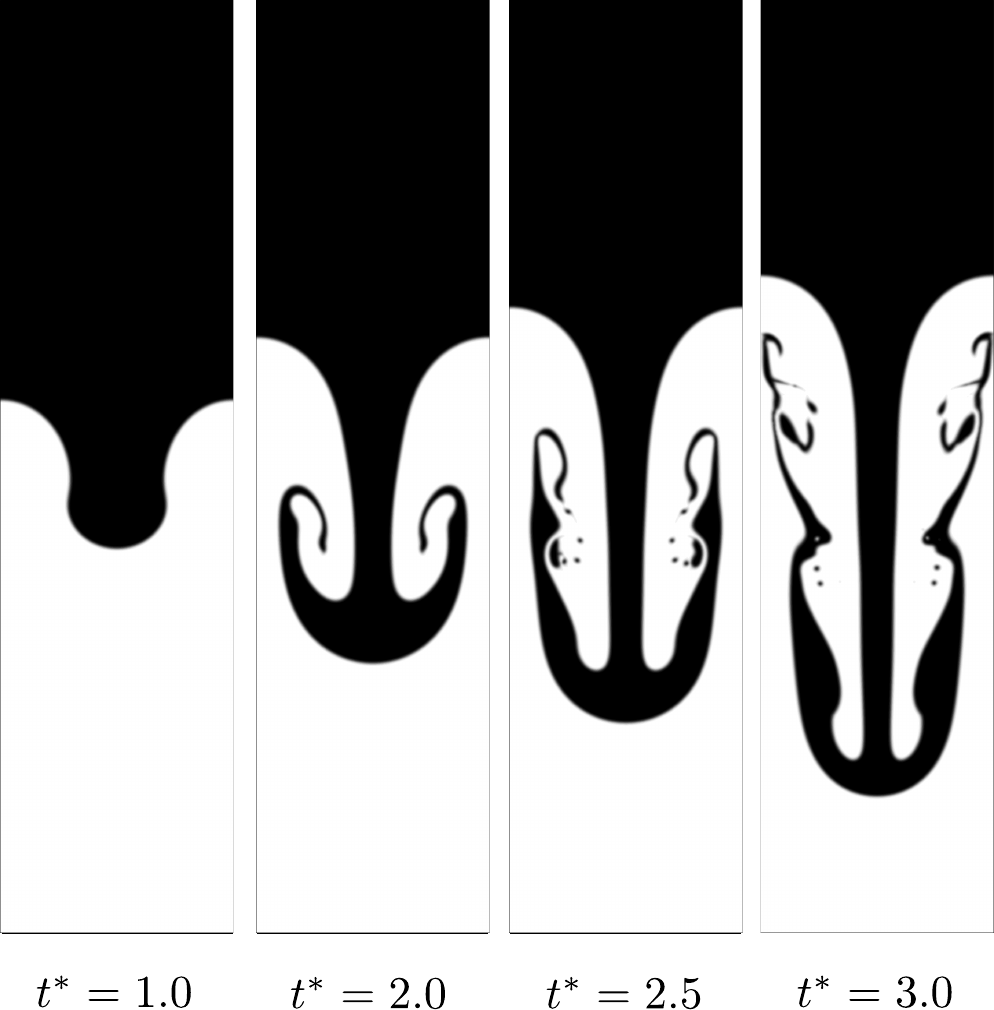} % Replace with the actual path to your image
        \caption{}
    \end{subfigure}
    \hfill
    \begin{subfigure}[b]{0.48\textwidth}
        \centering
        \includegraphics[width=\textwidth]{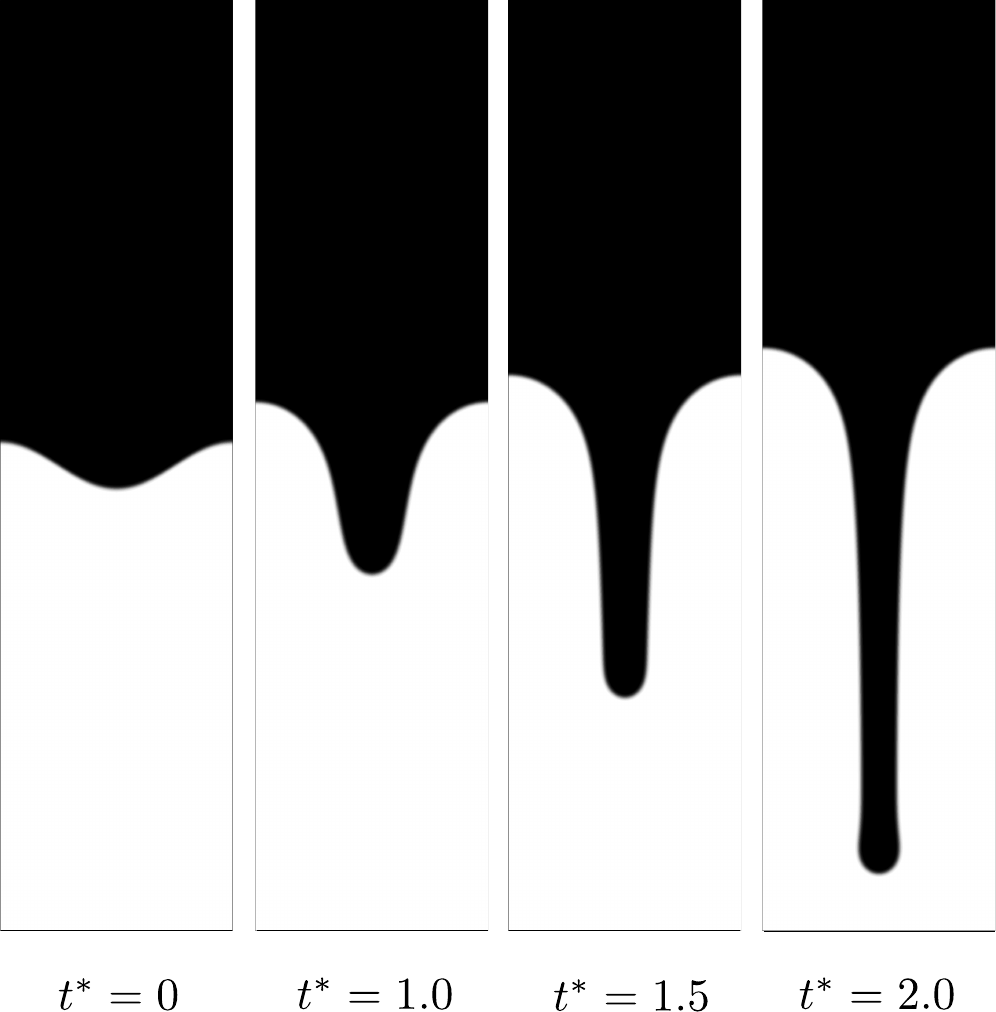} % Replace with the actual path to your image
        \caption{}
    \end{subfigure}
\caption{Evolution of the interface pattern of the 2D Rayleigh-Taylor instability for two scenarios: (a) $\rho^\ast=3$, $\mu^\ast=1$, $Re=128$, $At=0.500$, $Pe=744$,  $\xi=5$; (b) $\rho^\ast=1000$, $\mu^\ast=100$, $Re=3000$, $At=0.998$, $Pe=200$, $Ca=8.7$, $\xi=5$.}

    \label{fig:RTcomparison}
\end{figure}

Snapshots of the interface evolution for the 2D Rayleigh-Taylor instability at different times are shown in \figureref{fig:RTcomparison}. Initially, the interface undergoes a symmetrical penetration of the heavier fluid into the lighter fluid, forming counter-rotating vortices. As time progresses, the heavier fluid rolls up into mushroom-like shapes, and secondary vortices form at the tails of the roll-ups. Our simulations' interface patterns and vortex structures are consistent with those reported in previous studies ~\citep{Zu2013, Ren2016}.

\begin{figure}[ht]
    \centering
    \begin{subfigure}[b]{0.38\textwidth}
        \centering
        \includegraphics[width=\textwidth]{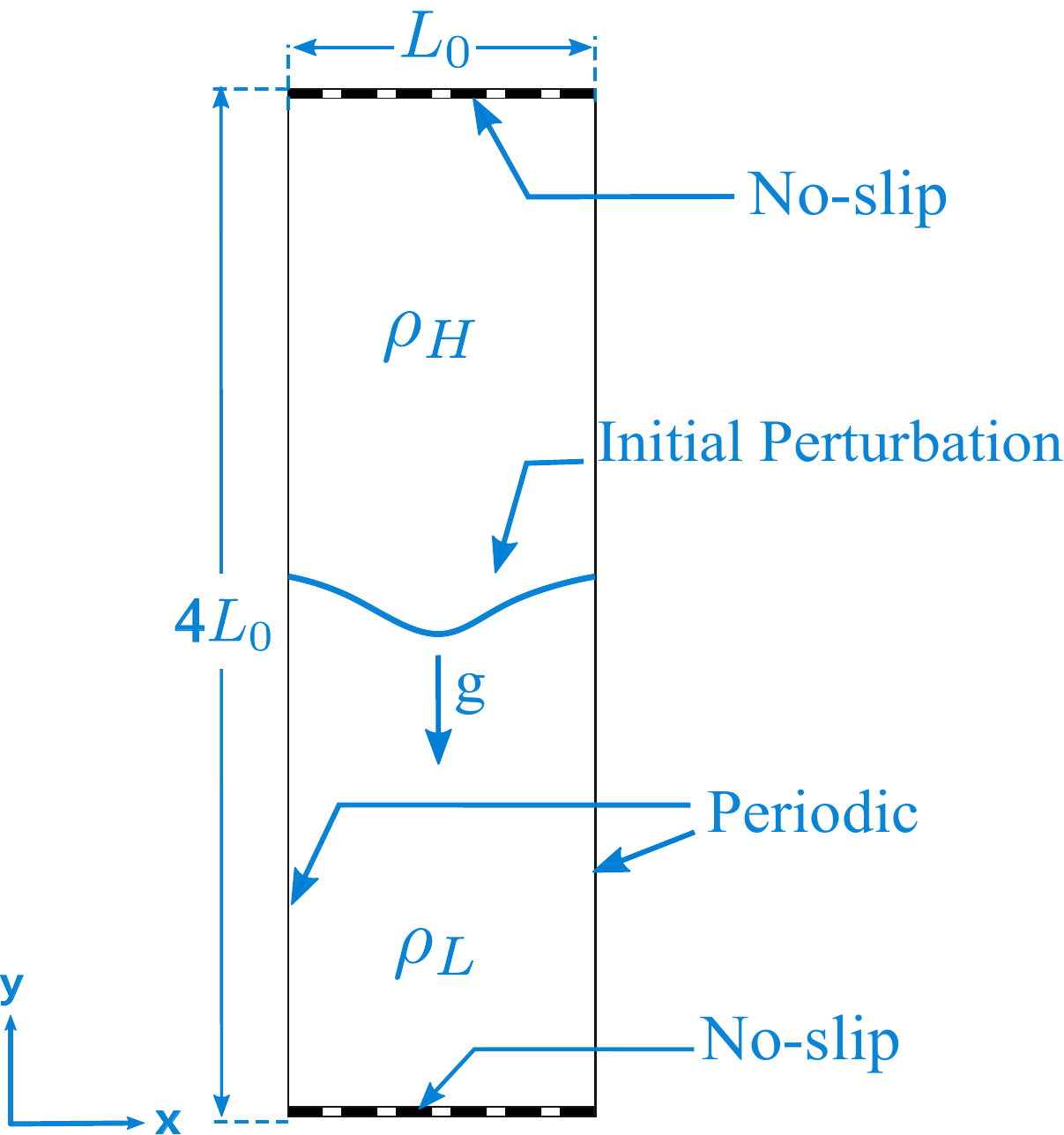}
        \caption{}
        \label{fig:image_subfigure}
    \end{subfigure}
    \hspace{0.01\textwidth}
    \begin{subfigure}[b]{0.40\textwidth}
        \centering
        \begin{tikzpicture}[scale=0.8]
        \begin{axis}[
            xlabel={$t^*$},
            ylabel={$y/L$},
            xmin=0, xmax=2.5,
            ymin=-1.6, ymax=1.6,
            ymajorgrids=true,
            grid style=dashed,
            legend cell align={left},
        ]

        \addplot[
            color=black,
            solid,
            thick
        ]
        table[x=X,y=Y1,col sep=comma] {Validation/RT.csv};
        \addlegendentry{Current LBM}
        
        \addplot[
            only marks,
            mark=o,
            color=red,
            mark options={solid},
            mark size=3pt
        ]
        table[x=X,y=Y2,col sep=comma] {Validation/RT.csv};
        \addlegendentry{Ren et al. (2016)}
        
        \addplot[
            only marks,
            mark=triangle*,
            mark options={fill=blue, draw=blue, opacity=0.5},
            mark size=3pt
        ]
        table[x=X,y=Y3,col sep=comma] {Validation/RT.csv};
        \addlegendentry{Zu et al. (2013)}

        \addplot[
            color=black,
            solid,
            thick
        ]
        table[x=X,y=Y1,col sep=comma] {Validation/Spike.csv};
        
        \addplot[
            only marks,
            mark=o,
            color=red,
            mark options={solid},
            mark size=3pt
        ]
        table[x=X,y=Y2,col sep=comma] {Validation/Spike.csv};
        
        \addplot[
            only marks,
            mark=triangle*,
            mark options={fill=blue, draw=blue, opacity=0.5},
            mark size=3pt
        ]
        table[x=X,y=Y3,col sep=comma] {Validation/Spike.csv};

        \node at (axis cs: 0.7,0.95) {\textcolor{black}{(Bubble Front)}};
        \node at (axis cs: 0.7,-0.95) {\textcolor{black}{(Spike Tip)}};

        \end{axis}
        \end{tikzpicture}
        \caption{}
        \label{fig:rayleigh_taylor_instability}
    \end{subfigure}
    \caption{(a) Schematic of the initial setup for the Rayleigh-Taylor instability simulation, showing the boundary conditions and initial perturbation. (b) Comparison of the bubble front and spike tip positions over time for the Rayleigh-Taylor instability case with parameters $\rho^\ast = 3$, $\mu^\ast = 1$, $Re = 128$, $At = 0.500$, $Pe = 744$, and $\xi = 5$. The current LBM results (solid line) are compared with the results of ~\citet{Ren2016} (red circles) and ~\citet{Zu2013} (blue triangles), showing excellent agreement in capturing the evolution of the instability.}
    \label{fig:combined_figure}
\end{figure}
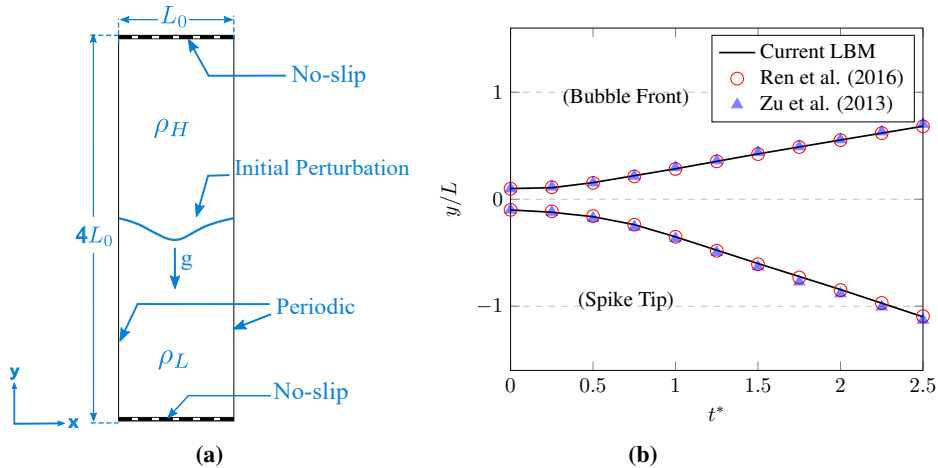

\section{SciML Training} 

\subsection{Model Hyperparameters} \label{subsec:model-hyperparams}

This section details the hyperparameters used for training the scientific machine learning (SciML) models. These parameters were carefully selected through extensive tuning to optimize model performance for the given datasets.  

\begin{itemize}

    \item \textbf{Fourier Neural Operator (FNO)}  
    \begin{itemize}
        \item Number of Fourier modes in both directions: \(64\)  
        \item Number of hidden channels: \(32\)  
        \item Projection channels: \(32\)  
        \item Number of layers: \(10\)  
        \item Learning rate: \(0.0005\)  
    \end{itemize}

    \item \textbf{Convolutional Neural Operator (CNO)}  
    \begin{itemize}
        \item Number of layers: \(4\)  
        \item Number of residual blocks (\(N_{\mathrm{res}}\)): \(6\)  
        \item Learning rate: \(0.0005\)  
    \end{itemize}

    \item \textbf{DeepONet}  
    \begin{itemize}  
        \item Branch network layers: \([512, 512, 512]\)  
        \item Trunk network layers: \([256, 256, 256]\)  
        \item Number of modes: \(128\)  
        \item Learning rate: \(0.0005\)  
    \end{itemize}
    
    \item \textbf{U-Net}  
    \begin{itemize}  
        \item Encoder channels: \([16, 32, 64, 128]\)  
        \item Decoder channels: \([128, 64, 32, 16]\)  
        \item Learning rate: \(0.0005\)  
    \end{itemize}
    \item \textbf{Poseidon}  
    
    \begin{itemize}
        \item Depths: \([4, 4, 4, 4]\)  
        \item Embedding dimension: \(48\)  
        \item Pretrained path: \(\texttt{camlab-ethz/Poseidon-T}\)  
        \item Learning rate: \(0.0005\)  
    \end{itemize}

    \item \textbf{scOT}  
    \begin{itemize} 
        \item Depths: \([4, 4, 4, 4]\)  
        \item Embedding dimension: \(48\)  
        \item Learning rate: \(0.0005\)  
    \end{itemize}
\end{itemize}

\subsection{Training and Validation Loss} \label{subsec:loss-plots}

\begin{figure}[b!]
    \centering
    \foreach \s in {1,3,4,6} {
        \begin{subfigure}[b]{0.48\textwidth}
            \centering
            \begin{tikzpicture}
            \begin{semilogyaxis}[
                width=0.9\textwidth,
                height=0.6\textwidth,
                xlabel={Epoch},
                ylabel={Loss},
                title={CNO (S\s)},
                legend pos=north east,
                grid=major,
                xmin=0, xmax=200,
                ymin=2e-6, ymax=8e-1
            ]
            \addplot[blue, thick] table[x=epoch, y=cno_S\s_train, col sep=comma] {Figures/training_loss/training-val_loss.csv};
            \addlegendentry{Training}
            \addplot[orange, thick] table[x=epoch, y=cno_S\s_val, col sep=comma] {Figures/training_loss/training-val_loss.csv};
            \addlegendentry{Validation}
            \end{semilogyaxis}
            \end{tikzpicture}
        \end{subfigure}
        \hfill
        \begin{subfigure}[b]{0.48\textwidth}
            \centering
            \begin{tikzpicture}
            \begin{semilogyaxis}[
                width=0.9\textwidth,
                height=0.6\textwidth,
                xlabel={Epoch},
                ylabel={Loss},
                title={Poseidon-T (S\s)},
                legend pos=north east,
                grid=major,
                xmin=0, xmax=200,
                ymin=2e-6, ymax=8e-1
            ]
            \addplot[blue, thick] table[x=epoch, y=poseidon_S\s_train, col sep=comma] {Figures/training_loss/training-val_loss.csv};
            \addlegendentry{Training}
            \addplot[orange, thick] table[x=epoch, y=poseidon_S\s_val, col sep=comma] {Figures/training_loss/training-val_loss.csv};
            \addlegendentry{Validation}
            \end{semilogyaxis}
            \end{tikzpicture}
        \end{subfigure}
        \\
    }
    \caption{Training and validation loss (semi-log scale) for CNO (left) and Poseidon-T (right) across four different input-output mappings $S1$, $S3$, $S4$, and $S6$.}
    \label{fig:cno-poseidon-loss}
\end{figure}
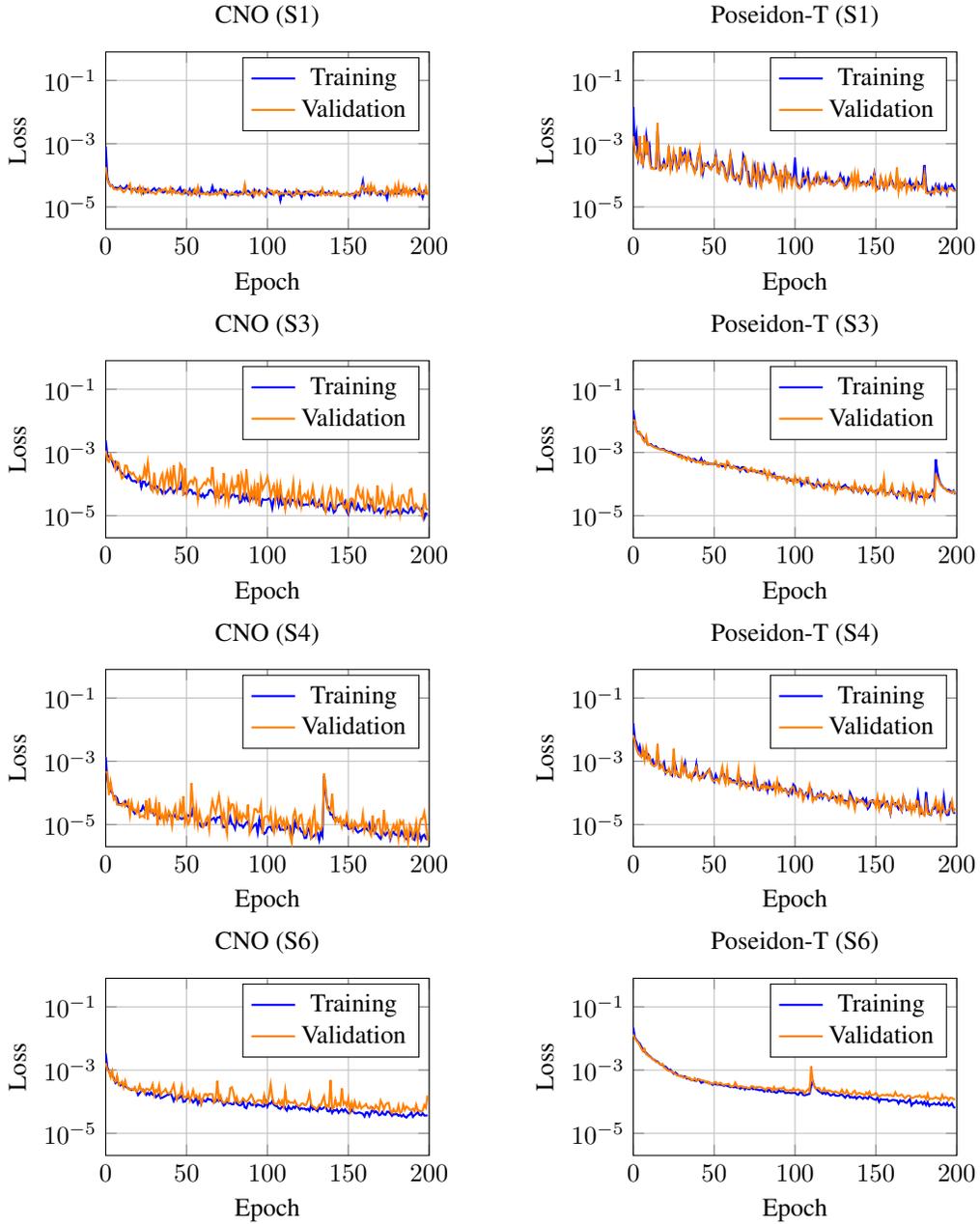

To further analyze the training performance, the evolution of training and validation loss for four representative models (Poseidon-T and CNO) are shown in~\figureref{fig:cno-poseidon-loss}. 

\end{document}

%% file: math_commands.tex
\usepackage{amsmath,amsfonts,bm}

\def\eqref#1{equation~\ref{#1}}
\def\1{\bm{1}}

\DeclareMathAlphabet{\mathsfit}{\encodingdefault}{\sfdefault}{m}{sl}
\SetMathAlphabet{\mathsfit}{bold}{\encodingdefault}{\sfdefault}{bx}{n}